\documentclass[acmtog]{acmart}

\usepackage{booktabs} %
\usepackage{diagbox}
\usepackage{subfig}
\usepackage{multirow}
\usepackage{mathrsfs}
\usepackage{graphicx}
\usepackage{enumitem}
\usepackage{soul}
\usepackage[normalem]{ulem}

\setcitestyle{square} %

\usepackage[ruled]{algorithm2e} %

\SetAlFnt{\small}
\SetAlCapFnt{\small}
\SetAlCapNameFnt{\small}
\SetAlCapHSkip{0pt}

\newcommand{\new}[1]{\textcolor{black}{#1}}

\acmJournal{TOG}

\begin{document}

\author{Yiwei Hu}
\affiliation{%
	\institution{Yale University}
	\country{USA}
}
\email{yiwei.hu@yale.edu}

\author{Chengan He}
\affiliation{%
	\institution{Yale University}
	\country{USA}
}

\author{Valentin Deschaintre}
\affiliation{%
	\institution{Adobe Research}
	\country{UK}
}
\affiliation{%
	\institution{Imperial College London}
	\country{UK}
}

\author{Julie Dorsey}
\affiliation{%
	\institution{Yale University}
	\country{USA}
}
\email{julie.dorsey@yale.edu}

\author{Holly Rushmeier}
\affiliation{%
	\institution{Yale University}
	\country{USA}
}
\email{holly.rushmeier@yale.edu}

\title{An Inverse Procedural Modeling Pipeline for SVBRDF Maps}

\begin{abstract}
Procedural modeling is now the de facto standard of material modeling in industry. Procedural models can be edited and are easily extended,
unlike pixel-based representations of captured materials. In this paper, we present a semi-automatic pipeline for general material proceduralization. 
Given Spatially-Varying Bidirectional Reflectance Distribution Functions (SVBRDFs) represented as  sets of pixel maps, our pipeline decomposes them into a tree of \emph{sub-materials} whose spatial distributions are encoded by their associated mask maps. This semi-automatic decomposition of material maps progresses hierarchically, driven by our new spectrum-aware material matting  and instance-based decomposition methods. Each decomposed sub-material is proceduralized by a novel multi-layer noise model to capture local variations at different scales. Spatial distributions of these sub-materials are modeled either by a by-example inverse synthesis method recovering Point Process Texture Basis Functions (PPTBF) \cite{Guehl20} or via random sampling. To reconstruct procedural material maps, we propose a differentiable rendering-based optimization that recomposes all generated procedures together to maximize the similarity between our 
procedural models
and the input material pixel maps. We evaluate our pipeline on a variety of synthetic and real materials. We demonstrate our method's capacity to process a wide range of material types, eliminating the need for artist designed material graphs required in previous work~\cite{hu2019, Shi20}. As fully procedural models, our results expand to arbitrary resolution and enable high level user control of appearance.
\end{abstract}



\begin{teaserfigure}
\includegraphics[width=\textwidth]{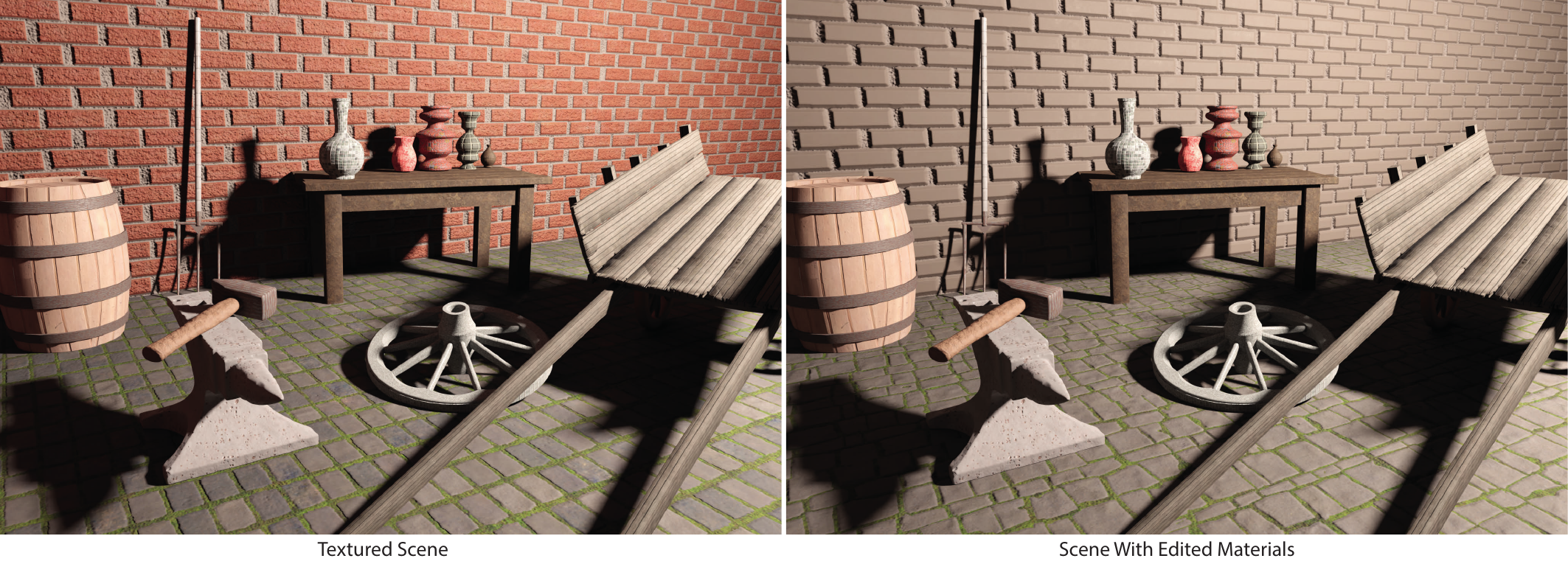}\vspace{-2mm}
\caption{On the left we show a virtual scene textured only with procedural materials generated from pixel map exemplars with our method. On the right we show examples of the editing made possible by our procedural representation on the wall and floor. We change the brick color and make it deeper and smoother. We also modify the pavement color variation, its regularity and show broken tiles. These edits are made solely through the parameters made available by our method. All of the input exemplars used to the generate procedural models are presented in  supplemental material.}
\label{fig:newTeaser}
\end{teaserfigure}

\maketitle

\section{Introduction}
The appearance of a scene is defined by the interaction between lighting, geometries and materials. Materials define the way the light 
is scattered and absorbed on and within geometries.
Despite progress in material authoring tools, creating a realistic material remains time consuming, even for expert artists. To reproduce existing materials, the \emph{Material Acquisition} field %
defines methods for
extracting properties through measurements sampling the material reflectance at different light and camera positions~\cite{Foo97}. Leveraging the recent progress in lightweight material acquisition, 
we propose the first method aiming at generating a \emph{procedural material representation},
allowing higher level editability 
as well as arbitrary scale and resolution.%

Procedural modeling is now common in
professional material modeling with tools such as Blender or Substance Designer~\cite{SubstanceDes}. Procedural models are often represented as graphs of operations, defining sub-elements of the targeted material. Each of these operations is procedural, allowing infinite resolution or scale, and high-level control through modification of parameters.

Recent methods for simplified material acquisition~\cite{dong19, Deschaintre18, li_kal18, Deschaintre19, Gao19} target the reconstruction of analytic material maps, which compactly represent  materials, but lack the editability and resolution/scale increase possibilities offered by procedural representations.

Inverse procedural modeling of materials focuses on the creation of a procedural material representation from images, or in our case, from an existing set of analytic material maps. We postulate that this ill-posed challenge requires division of a material into meaningful components which can be represented as procedures. Recently, several methods \cite{hu2019, Shi20} took a first step by proposing different inverse modeling frameworks to select a procedural graph among \emph{existing models} and optimize the nodes parameters to best match an input texture.

In this paper, we target the challenging task of creating entirely new procedural models from a set of SVBRDF pixel maps as input, eliminating the need for pre-existing artist-designed procedural models. We propose inverting the material modeling process in a sequential way by hierarchically breaking down the material into several sub-materials, then fitting their spatial distributions and material properties with procedures.

Given a material, we hierarchically decompose it into a tree of atomic \emph{sub-materials}. The different sub-materials are segmented to encourage uniform statistical variation through either a user-guided, Fourier spectrum aware KNN matting approach, or an automatic instance-based segmentation algorithm. Segmented regions provide information about local texture statistics and the properties of different sub-materials. We represent the global spatial distributions of sub-materials with mask maps. Each component of the tree is then automatically converted to a full procedural model. We present a multi-layer procedural noise model based on random phase noise~\cite{Galerne2011} to model the appearance of sub-materials to capture local variations at multiple scales. To procedurally model global spatial distributions, we propose an optimization-based inverse modeling method based on a procedural texture basis function~\cite{Guehl20} and random sampling to convert mask maps to procedures. With the proceduralized sub-materials and their procedural mask maps organized in a tree, we build a small material graph by adding optimizable operators. The material graph is optimized in a differentiable fashion based on a rendering loss to better match the input material.

Our  material representation relies solely on procedural components, enabling high level editing and arbitrary scale/resolution. We use an analytic  representation as input to reduce the uncertainty inherent in single-picture material acquisition. This allows our method to seamlessly benefit from progress in lightweight material acquisition methods~\cite{DDB20, Gao19}.

We show application of our pipeline on a wide range of materials and define a new taxonomy for material decomposition complexity, highlighting the existing challenges of this ill-posed task. In summary, this paper presents a novel research direction to create a procedural representation of general analytical materials without relying on pre-existing material graphs. %
Specifically we present the following contributions:
\begin{itemize}
    \item We present the first pipeline for semi-automatic \new{\emph{generation} of procedural representation of materials. As such, our method does \emph{not} rely on a pre-existing library of material node graphs}.
    \item We propose a new spectrum-aware, hierarchical segmentation method for guided material segmentation.
    \item We define procedures and their corresponding inverse approaches to proceduralize sub-materials and their distributions.
    \item We offer a differentiable rendering-based optimization routine to match our procedural material to the input material during reconstruction.
    \item \new{We define a procedural representation which allows for complex multi-scale edits and arbitrary scale and/or resolution}
\end{itemize}
Our implementation will be publicly available upon acceptance.

\section{Related Work} \label{sec:related-work}
\subsection{Material Acquisition}
Material acquisition seeks to recover the reflectance properties of existing surfaces or objects. Our method complements this body of work since acquired materials are used as input for our pipeline to create procedural representations. While material acquisition has challenged researchers for decades \new{as discussed in the excellent survey by} Guarnera et al. ~\shortcite{Guarnera16}, significant progress was achieved in recent years by leveraging deep learning. In particular, different methods were proposed to acquire materials from one~\cite{li17, li_kal18, Li18b, Deschaintre18} or multiple~\cite{Deschaintre19, Gao19, Guo20, Boss20, Deschaintre21} pictures of surfaces or objects. While our approach also targets the creation of a material representation, we do not seek to recovering material properties from captured sample(s), but rather to proceduralize an existing material. This difference not only allows our method to benefit from material acquisition and its progress, but also to handle any existing analytical material.

\subsection{Texture Synthesis}
In our method, we leverage texture synthesis, which creates new textures with larger scale or higher resolution from an exemplar. Several surveys provide a comprehensive overview of example-based texture synthesis~\cite{wie09, akl18, raad18}.
Texture synthesis can be classified into three families: patch re-arrangement based,  arranging patches available in the original texture to synthesize a new texture \cite{Efros1999, quilting, SelfTuning}, statistics based, estimating statistics of the original texture and transferring them to a new texture \cite{Heeger1995, Galerne2011, Galerne2012, Gilet2014, Galerne2017, Heitz2018}, and neural-network-based, reproducing texture \new{or material} appearance using machine learning \cite{Ulyanov2016, Bergmann2017, TexSyn18, InGAN, Henzler20, henzler21neuralmaterial, Niklasson21}. We use the second family of methods to reproduce the local variations of our decomposed atomic components \textit{sub-materials}. Alone, these methods can reproduce micro-structures well, but fail for larger scale patterns. They are also limiting in the scope of possible editability. But by making use of progressive decomposition, extending them in a multi-layer fashion and incorporating them with an optimizable post-processing step, our method can represent complex SVBRDFs as procedural models with arbitrary scale and resolution as well as high level editability.

\subsection{Procedural Modeling}
\begin{figure}
    \centering
    \includegraphics[width=0.48\textwidth] {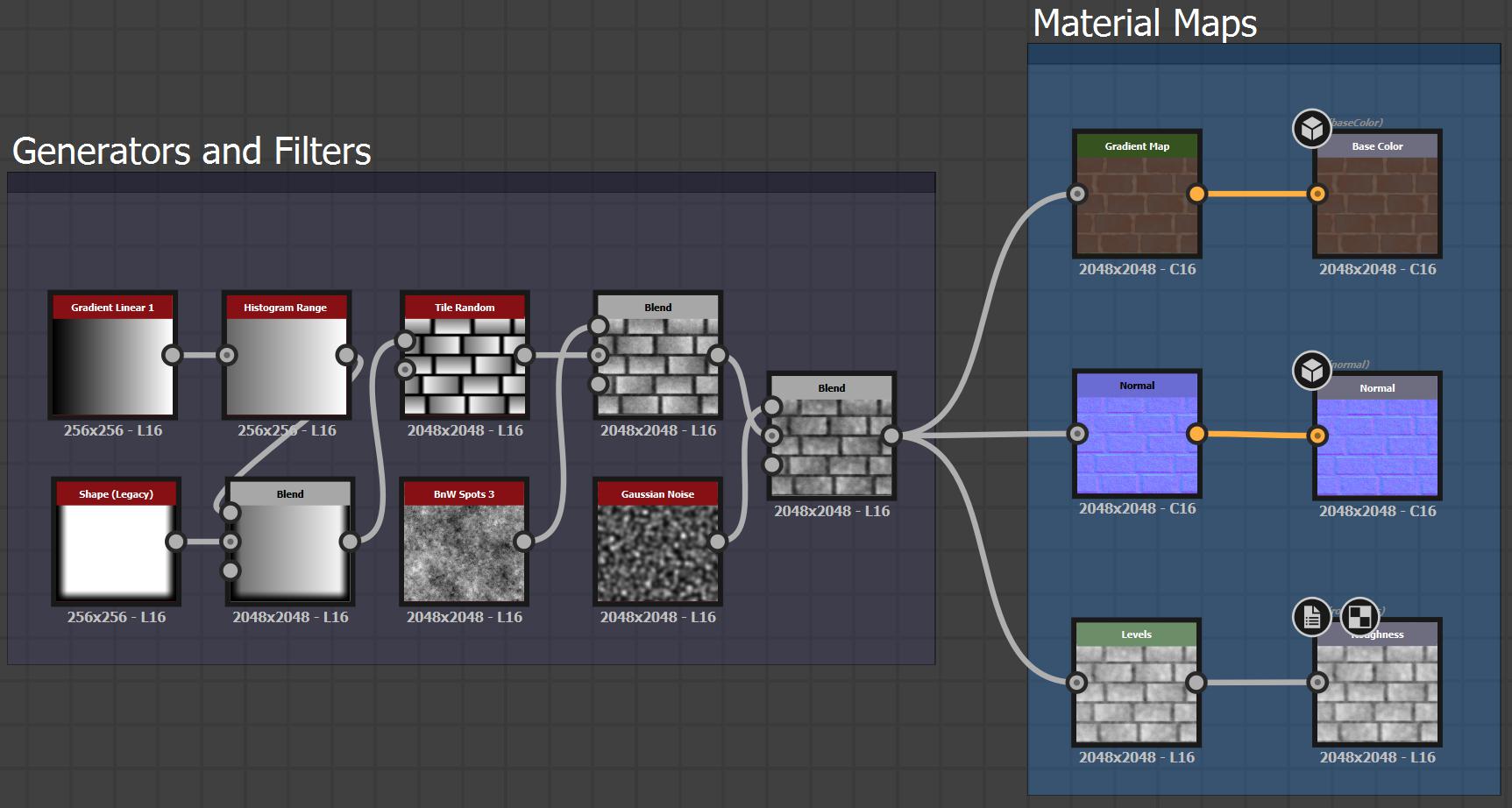}
    \caption{A simple example of a brick material created using node graphs in Substance Designer \cite{SubstanceDes}. This material graph is composed by a set of procedures to generate brick patterns and add details upon them. The designed material is fully procedural thus can be further edited or expanded to any resolution. }
    \label{fig:substance-example}
\end{figure}
Procedural models allow artists to retain editability and produce arbitrary scale and resolution. Specialized procedural models have been proposed but are limited to specific materials such as wood or leather and mostly stationary materials~\cite{GuoBayesian20}. A more generic option lies in the design software allowing artists to combine procedures to create a new material (see Fig.~\ref{fig:substance-example}). While this procedural representation provides great editability and versatility, it remains time consuming and challenging, even for expert artists, to reproduce an existing or imagined material.

Recent methods explore inverse procedural modeling of materials, aiming at the reproduction of a given material appearance using a procedural model. Hu et al.~\shortcite{hu2019} and Shi et al. \new{(MATch)}~\shortcite{Shi20} both propose different frameworks that leverage existing Substance graph procedural models and infer their parameters to match the appearance of input textures. To do so, Hu et al. relies on neural networks trained for each individual material graph, while Shi et al. \new{(MATch)} relies on gradient descent and a differentiable version of the Substance Engine. Both of these methods, however,  work on the premise of  known pre-existing procedural node graphs, manually created by artists\new{. This approach requires a large, expressive enough dataset of procedural graphs to be available and non-trivial search methods to find the closest ones among hundred of options}. \new{As opposed to these methods}, our proposed pipeline, does not require  predefined material graphs \new{and simply relies on a few user scribble, which typically only require 2 minutes of interaction}.

A key component of our inverse procedural modeling pipeline is the procedural representation of the structure of materials. That is, the global spatial distributions of decomposed sub-materials, which we represent using a set of binary mask maps. To preserve the procedural aspect of our results we need to represent these masks procedurally. Rosenberger et al.~\shortcite{Rosenberger2009} propose a shape synthesis method to generate layered control maps but is limited to unstructured shapes with fractal-like boundaries. Alternatively, we could use an L-system \cite{vst2010}, but that approach requires predefined grammars. Rather, we make procedural generation of structured mask maps possible by proposing a by-example inverse Point Process Texture Basis Functions (PPTBF)~\cite{Guehl20} modeling approach and leveraging random sampling.

\subsection{Image Segmentation and Matting}
Image segmentation aims at separating an image into different regions, based on some specific criteria. In our material proceduralization framework, segmentation is used for partitioning the given material into several sub-materials. Image segmentation has been extensively studied in the last decades~\cite{khan2014survey}, with recent methods leveraging deep learning~\cite{minaee20}. Close to our challenge are the recent work of Cimpoi et al. and Bell et al.~\shortcite{cimpoi2016, bell2015} which semantically split natural scenes into their different components (such as wood, plastic, etc.). However, these methods focus on the segmentation of complete scene images and use  context~\cite{bell2015} to better recognize materials, which is not available in our material exemplars.

Most existing segmentation methods do not perform well in our context as they result in significant error on the boundaries between two sub-materials.
We therefore use image matting,  originally developed to segment background and foreground with fuzzy %
boundaries. We found alpha matting to better represent the transition between sub-materials in a SVBRDF and to preserve good quality boundaries, even after thresholding.

Many methods for image matting suggest affinity-based solutions, solving a Laplacian clustering problem~\cite{levin2008, chen2013, aksoy2017}. Recent work \cite{cho2016, xu2017} attempts to learn alpha maps directly using deep neural networks, but is limited to two-layer (foreground and background) alpha matting due to available training data. Aksoy et al. \shortcite{aksoy2018} proposed an improved affinity-based method by combining deep semantic features with affinity Laplacian. While it supports multiple layers, it is geared toward natural scene images and requires user to provide the number of layers. %
\new{Our decomposition approach of materials is similar to the one adopted by Lawrence et al.~\shortcite{Lawrence06}: we aim to decompose the input spatially-varying material maps into regions of similar \emph{sub-materials}. The approach by Lawrence et al. was further developed by AppWand~\cite{Pellacini07}, AppProp~\cite{An08} and Material Matting~\cite{Lepage11}, which introduce segmenting measured or analytical material images to make them easy to edit. However, in practice, these approaches  seem to handle limited normal variations. More importantly, they do not target the creation of fully procedural representations. They treat the materials as SVBRDF images, limiting the editing of results to pixelwise (rather than parametric) modifications of the segmentation and/or uniform (rather than multi-scale) material editing. Beyond the editing limits, these methods do not provide for the generation of textures of arbitrary spatial extent. }

\new{We choose to build upon KNN matting \cite{chen2013} by proposing a spectrum-sensitive affinity-based image matting method. KNN matting supports multiple layers and enables user control to conveniently define  layers of interest, which are useful features for our material segmentation purpose. The aforementioned material segmentation approaches~\cite{Pellacini07, An08, Lepage11} could produce interesting segmentation results, but we chose KNN matting because of its  generality, the availability of its implementation and its efficient runtime.}

\section{Overview}

\begin{figure*}
\includegraphics[width=\textwidth]{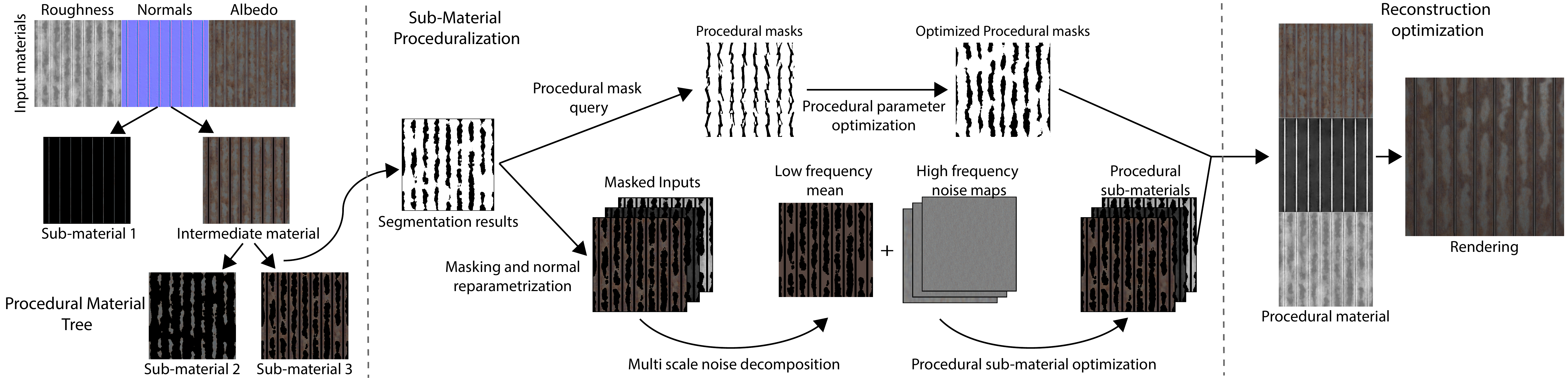}
\vspace{-5mm}
\caption{Our procedural modeling pipeline allows a user to guide the segmentation of an input material to automatically generate a matching fully procedural material model. On the left we show the procedural tree generated by our method and on the right we show the proceduralization steps taken for each sub material. We represent the spatial variation using segmentation masks and Point Process Texture Basis Functions to create a procedural version through a query-and-optimization method. We represent the sub-material parameters with a multi-scale noise matching operation after reparameterizing normals to height values. Finally we use differentiable rendering to compose all sub-materials into a complete procedural material, rendered here lit from the top. \new{Please see our supplemental material for a graph visualisation of a few results.}}
\label{fig:overview}
\end{figure*}

Given a set of SVBRDF maps, we want to generate a procedural representation of the material. Our inverse procedural modeling pipeline starts by hierarchically decomposing input SVBRDF maps into multiple sub-materials (Sec. \ref{sec:svbrdf-decomp}) organized as a tree structure. Each sub-material represents a statistically similar region and its local variation, while the associated segmentation masks encode the global spatial variations of these sub-materials. The tree structure provides a layered relationship between different sub-materials. We traverse the material tree to convert each component to their procedural counterparts. We represent procedural sub-materials with a multi-layer procedural noise model combining procedural noise maps (Sec. \ref{sec:multi-layer-noise-model}) and mask maps using Point Process Texture Basis Functions (PPTBF)~\cite{Guehl20} and random sampling (Sec. \ref{sec:mask-synthesis}). After proceduralization, we compose these procedural representations and use a differentiable rendering-based optimization to match the appearance of the input material (Sec. \ref{sec:recomposition}). We show an overview of our method in Fig.~\ref{fig:overview}.

\begin{figure}
	\centering
	\addtolength{\tabcolsep}{-3pt}
	\begin{tabular}{cccc}
		\includegraphics[width=0.15\textwidth]{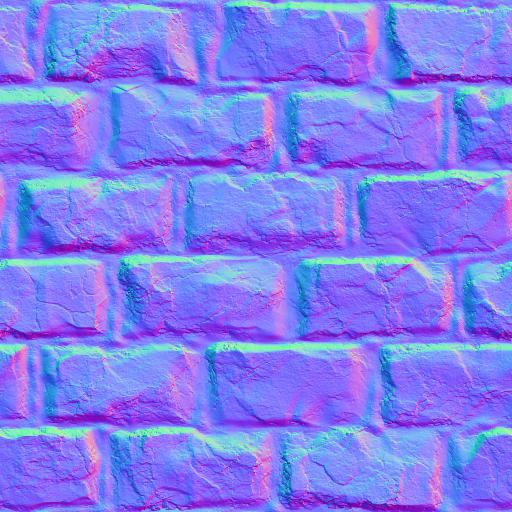} &
		\includegraphics[width=0.15\textwidth]{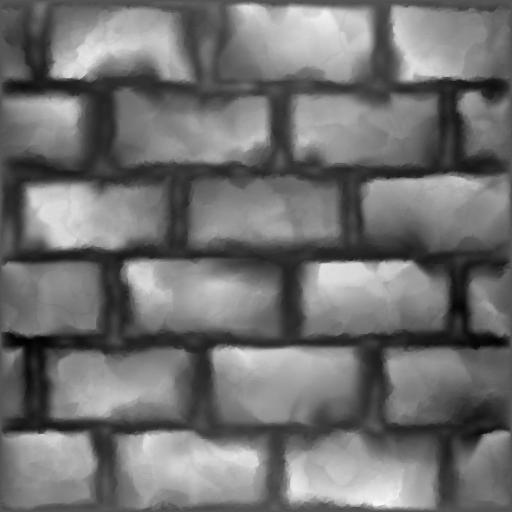} &
		\includegraphics[width=0.15\textwidth]{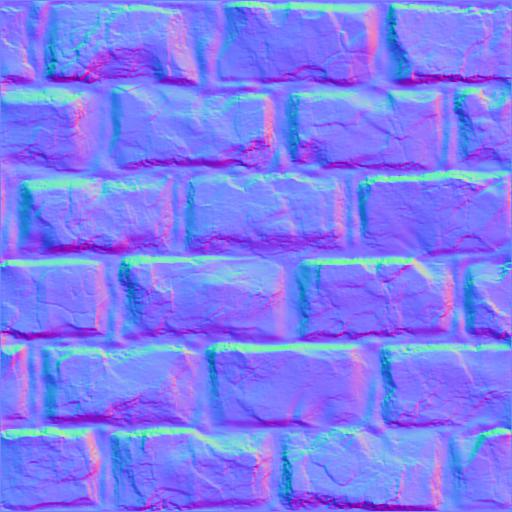} \\
		Normal map & Height map & Normal map* \\
	\end{tabular}
\vspace{-10pt}
\caption{Reconstruction of a height map from a normal map. Values on the height map are normalized between 0 and 1. The Normal map* is computed from reconstructed height map by a Sobel operator. While the recomputed Normal map* shows slightly less fine-grain detail, the general structures are well preserved.}
\label{fig:normal2height}
\end{figure}

In this paper we demonstrate our method on common physically-based material parameters, that can be easily acquired using recent methods~\cite{DDB20}: albedo maps, normal maps and roughness maps. As normal maps encode vectors, their channels \new{represent 3D directions which are} difficult to proceduralize. Instead of directly working in normal map space, we convert to a height map using Poisson reconstruction \cite{Perez2003} a produce a better proceduralization. Fig.~\ref{fig:normal2height} provides an example of this reconstruction. Our approach can easily handle additional gray-scale or color maps. %

\section{SVBRDF Decomposition} \label{sec:svbrdf-decomp}
The first step of our pipeline is a hierarchical decomposition of the input SVBRDF maps into multiple sub-materials in a semi-automatic way. Although it is possible to spatially decompose SVBRDF maps into multiple sub-parts in one step, we propose an iterative decomposition into a material hierarchy, allowing the encoding of a layered relationship between sub-materials for proceduralization. Specifically, given a set of SVBRDF maps, we decompose it into multiple sub-materials using a spectrum-aware matting algorithm (Sec. \ref{sec:spectrum-decomp}), which generates mask maps for each segmented sub-material. For each masked sub-material, we let the user decide whether to further decompose it using either our matting algorithm, or a lightweight instance-based decomposition algorithm (Sec. \ref{sec:instance-seg}). Using this process, users can define the elements they consider important in the texture and iterate with new sub-divisions until no salient sub-material is left, creating a tree structure of decomposed sub-materials.

\begin{figure}
	\centering
	\addtolength{\tabcolsep}{-3pt}
	\begin{tabular}{ccccc}
		\includegraphics[width=0.1\textwidth]{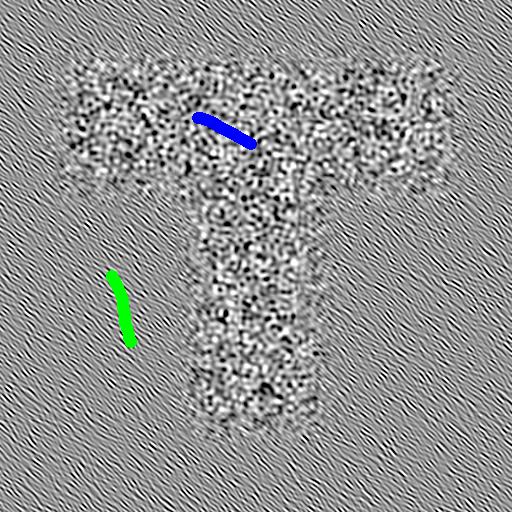} &
		\includegraphics[width=0.1\textwidth]{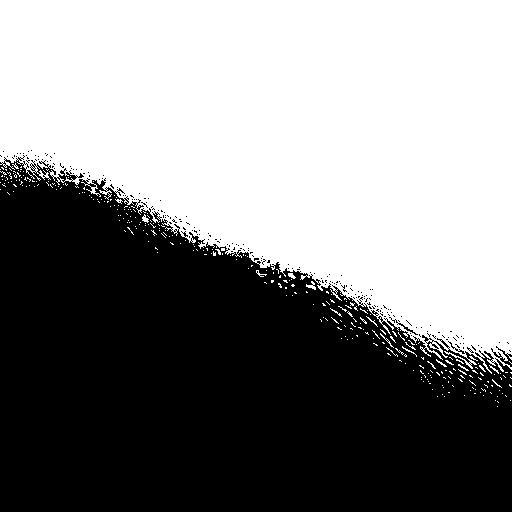} &
		\includegraphics[width=0.1\textwidth]{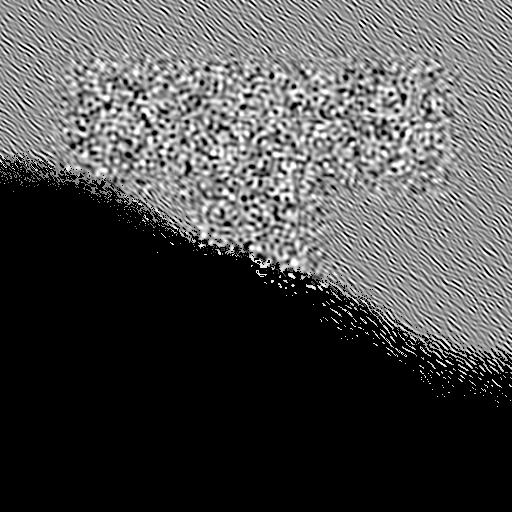} &
		\includegraphics[width=0.1\textwidth]{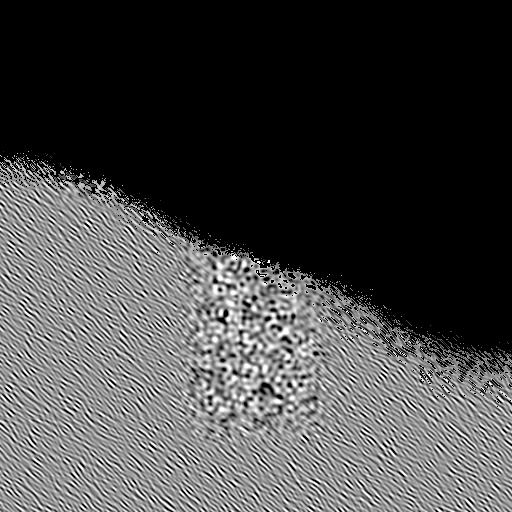} &
		\raisebox{25pt}{\rotatebox[origin=c]{-90}{No spectrum}} \\
		Input & Mask map & Layer 0 & Layer 1 \\
		&
		\includegraphics[width=0.1\textwidth]{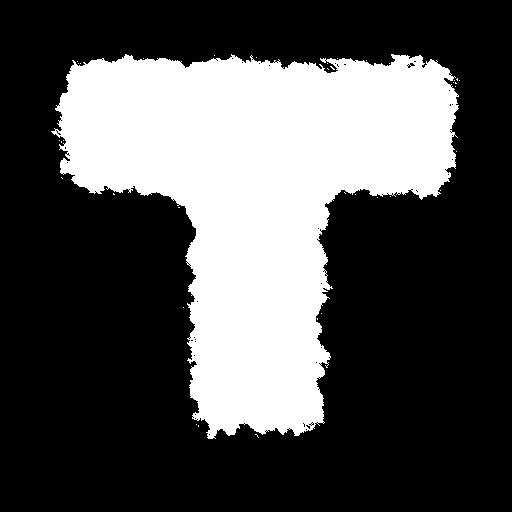} &
		\includegraphics[width=0.1\textwidth]{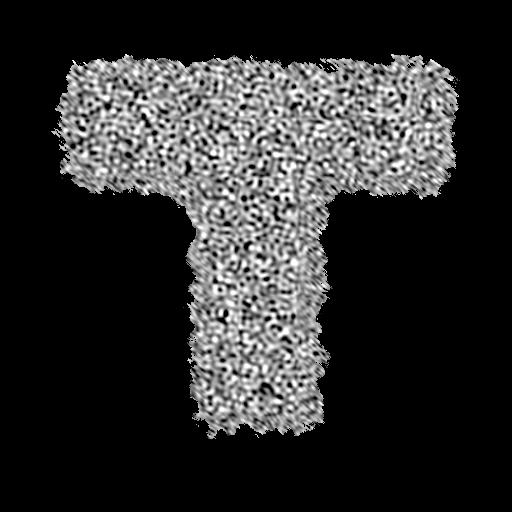} &
		\includegraphics[width=0.1\textwidth]{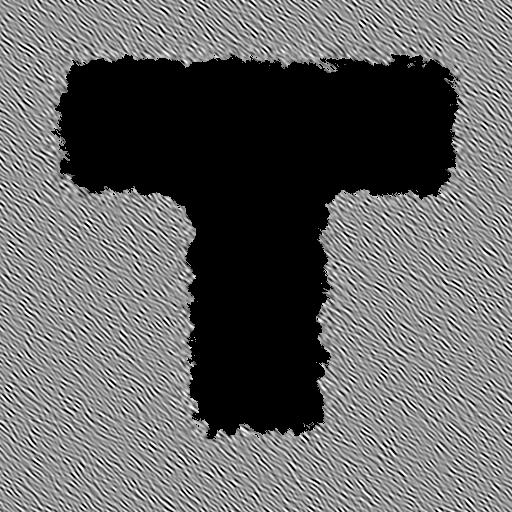} &
		\raisebox{21pt}{\rotatebox[origin=c]{-90}{With spectrum}} \\
		& Mask map & Layer 0 & Layer 1 \\
		\includegraphics[width=0.1\textwidth]{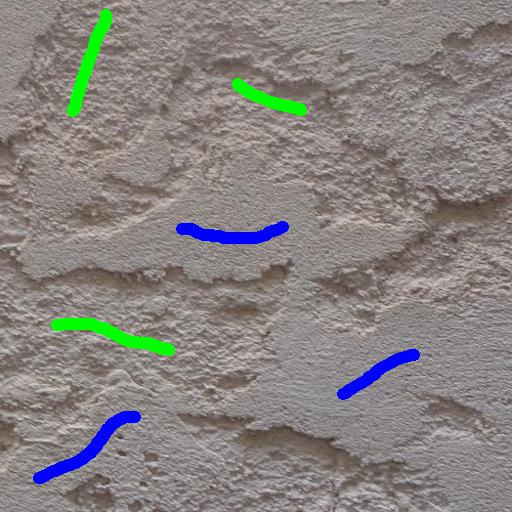} &
		\includegraphics[width=0.1\textwidth]{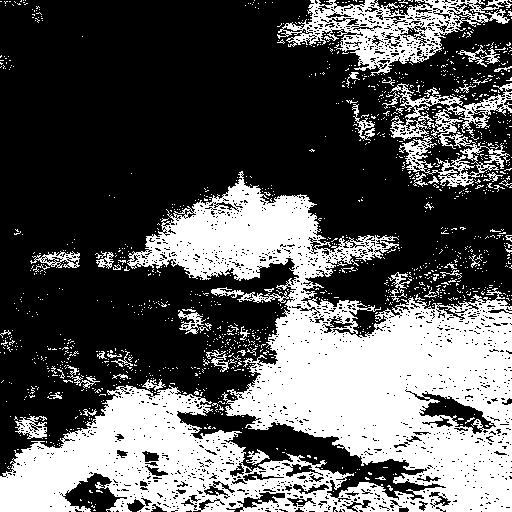} &
		\includegraphics[width=0.1\textwidth]{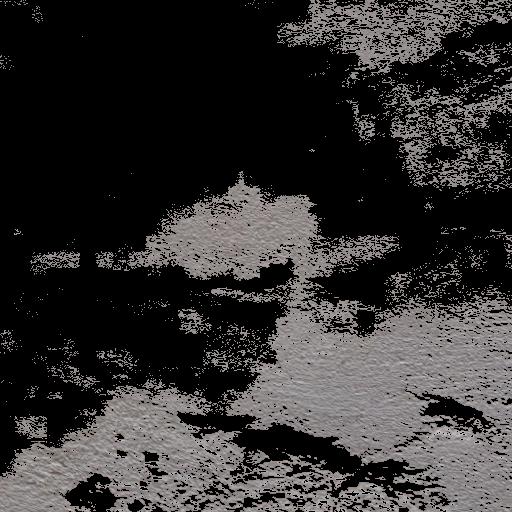} &
		\includegraphics[width=0.1\textwidth]{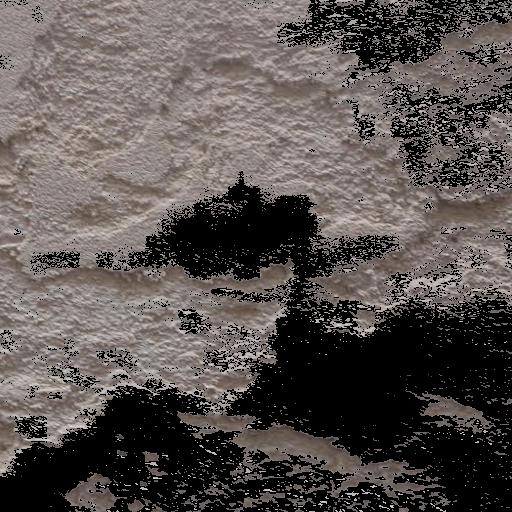} &
		\raisebox{25pt}{\rotatebox[origin=c]{-90}{No spectrum}} \\
		Input & Mask map & Layer 0 & Layer 1 \\
		&
		\includegraphics[width=0.1\textwidth]{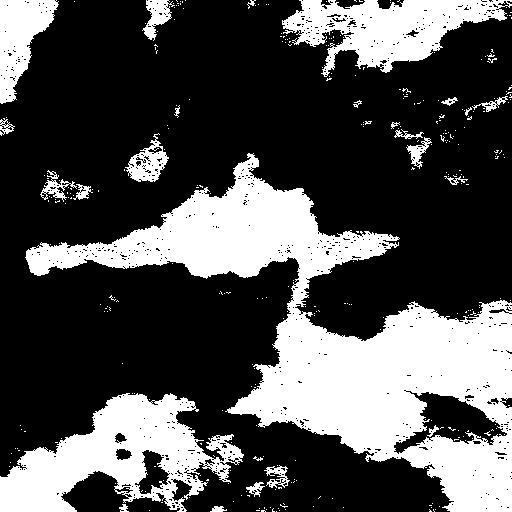} &
		\includegraphics[width=0.1\textwidth]{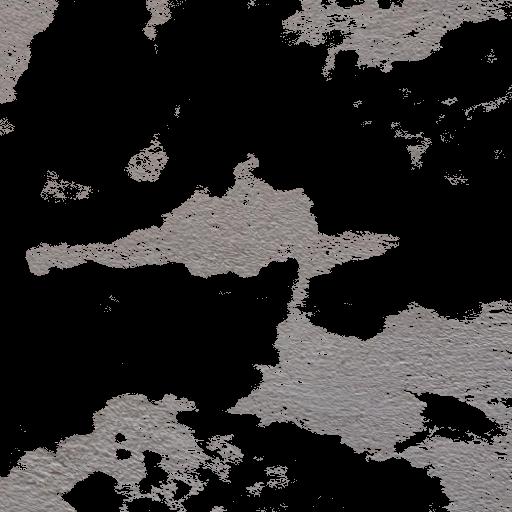} &
		\includegraphics[width=0.1\textwidth]{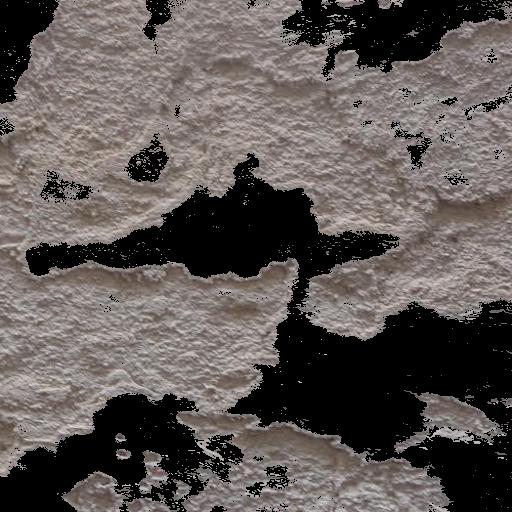} &
		\raisebox{21pt}{\rotatebox[origin=c]{-90}{With spectrum}} \\
		& Mask map & Layer 0 & Layer 1 \\
	\end{tabular}
\vspace{-10pt}
\caption{Image decomposition with and without spectrum features. The first image is a synthetic image while the second image is real-world texture. User scribbles are visualized as an additional layer superimposed over the input image, where blue and green scribbles indicate different matting layers. Matting results with spectrum features (bottom row) are better than the results without spectrum features (upper row) because color features here cannot provide sufficient information to separate two layers. Mask maps generated using spectrum features are also more integrated and less fragmented.}
\label{fig:matting}
\end{figure}
\subsection{Spectrum-aware SVBRDF Decomposition} \label{sec:spectrum-decomp}
We first consider how to decompose a set of SVBRDF maps, into sub-materials. As classical segmentation \cite{khan2014survey} does not allow for smooth boundaries, we use alpha matting \cite{chen2013}. Inspired by KNN matting, our algorithm allows users to draw a few strokes to conveniently indicate different regions of interest. %

As an affinity-based algorithm, KNN matting~\shortcite{chen2013} relies on a feature vector $X(i)$ for each pixel $i$ of the image to compute an affinity matrix. For SVBRDF maps, traditionally used features are albedo color (r, g, b), height (h), roughness ($\alpha$) and position (x,y): 
\begin{equation}
X(i)=(r(i), g(i), b(i), h(i), \alpha(i), x(i), y(i))
\end{equation}
We propose an additional feature, based on the noise spectrum, enforcing the statistical uniformity expected from decomposed sub-materials. To distinguish between the differences in local noise, we take the noise Fourier spectrum into account. \new{Notice that although position features $(x, y)$ are taken into account, we do \emph{not} enforce spatial continuity of the regions because we sample multiple non-local neighborhoods with different weights of $(x, y)$, to explore nonlocality (as done in the original KNN matting)}

We estimate the local spectrum at each pixel $i$ using Welch's method \cite{Welch1967} and reduce its dimensionality to 3 using Principal Component Analysis (PCA). We therefore compute the affinity matrix and matting using a feature vector $X(i)$ composed of albedo color, height, roughness, position and our spectrum estimation.

Alpha matting results in multiple alpha maps which we process to generate binary mask maps.  Each pixel in the image is assigned to the binary for which it has the highest alpha map value. The thresholded  alpha-matting  better represent the transition between sub-materials than direct segmentation methods.
Fig.~\ref{fig:matting} shows a comparison between the decomposition results with and without spectrum features on example textures. In these two images, color and position features do not provide enough hints to separate two layers because they are similar in color but differ in noise spectra. \new{More decomposition results on material maps with user scribbles can be found in our supplemental document.}

Once a material decomposed, we provide the option to further process each sub-material using the same matting algorithm with the newly generated mask map(s) as an additional constraint.

\begin{figure} 
	\centering
	\addtolength{\tabcolsep}{-3pt}
	\begin{tabular}{c|cc}
		\includegraphics[width=0.15\textwidth]{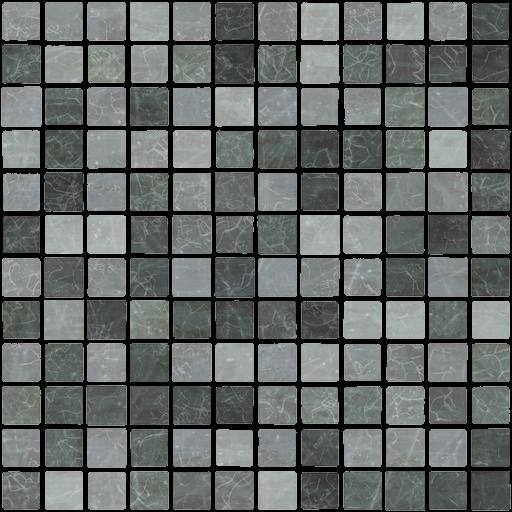} &
		\includegraphics[width=0.15\textwidth]{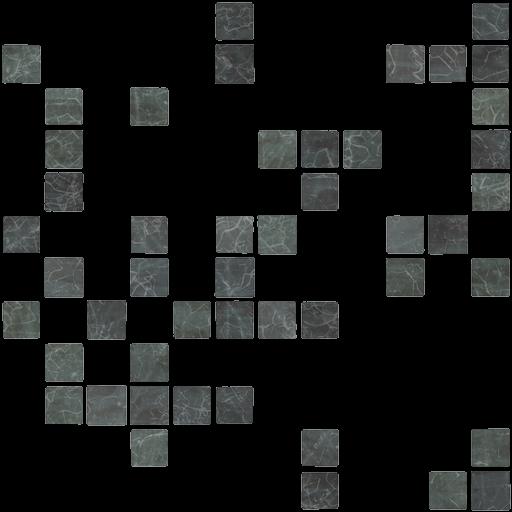} &
		\includegraphics[width=0.15\textwidth]{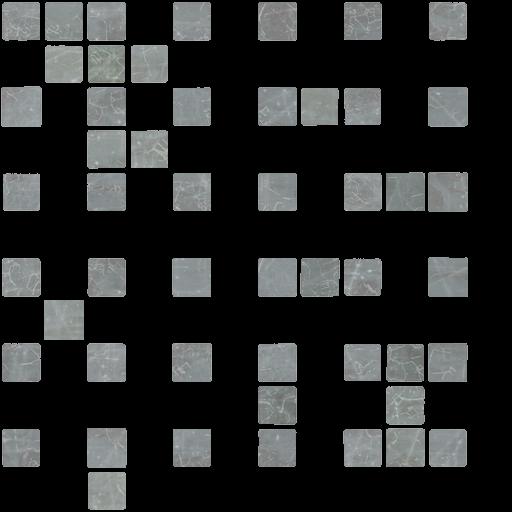} \\
		Input & Cluster 0 & Cluster 1 \\
		\includegraphics[width=0.15\textwidth]{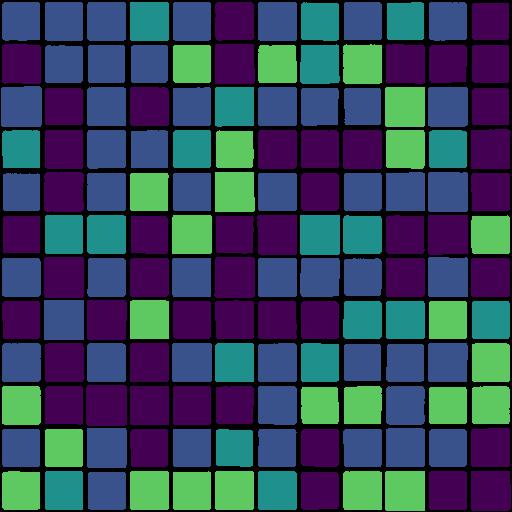} &
		\includegraphics[width=0.15\textwidth]{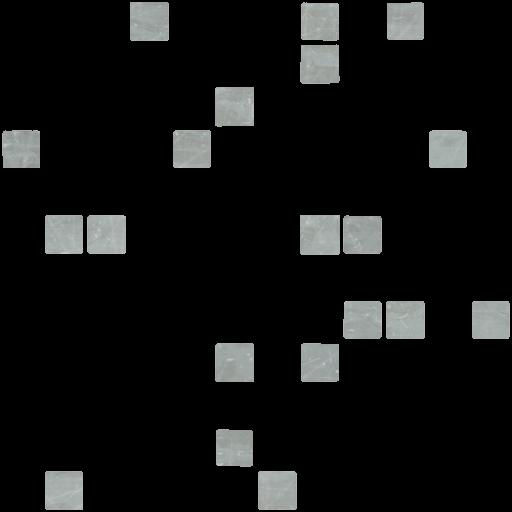} &
		\includegraphics[width=0.15\textwidth]{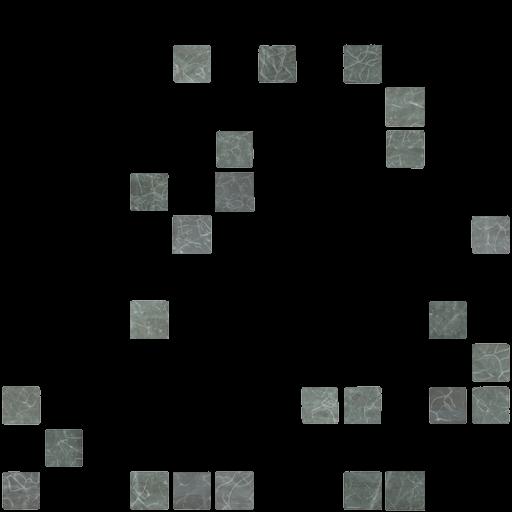} \\
		Label Map & Cluster 2 & Cluster 3 \\
	\end{tabular}
\caption{Instance-based decomposition for sub-materials. Instead of manually segmenting each sub-material, we extract instances in the input texture and build a feature matrix for each of them. Agglomerative clustering is then performed on these feature matrices, yielding multiple clusters with similar sub-materials. Their spatial distribution is visualized as a label map, where different colors indicate different clusters. Finally, we fit a different procedural model for each cluster.}
\label{fig:instace}
\end{figure}
\subsection{Instance-based Decomposition} \label{sec:instance-seg}
As an alternative to a progressive decomposition into sub-materials, we provide a lightweight solution for instance-based decomposition. This is particularly useful for repeating materials such as a tile wall composed of different types of tiles (the input shown in Fig.~\ref{fig:instace}). Rather than manually segmenting each different tile, we frame this as an instance detection and clustering problem, using -- in this example -- the mask map of segmented tiles to extract each instances of tile. \new{Different instances correspond to disconnected regions in the mask map segmented from the previous layer in the decomposed material tree.} We then scale each instance to the same size \new{based on its bounding box. For each instance, we estimate its color histogram and local spectrum as features and build a feature matrix for agglomerative clustering as shown in Fig.~\ref{fig:instace}. The clustering result} allows us to extract the different types of tiles, their frequency and their sub-material to assign them with a similar frequency in the procedural model.

\section{Material Proceduralization and Recomposition} \label{sec: mat-proc}
After decomposing the input SVRBDF maps into a tree of sub-materials and mask maps, we traverse the tree and convert each component into a procedural version layer by layer. Finally, we introduce optimizable parameters during the final recomposition phase to best match the input SVBRDF.

\subsection{Multi-layer Procedural Noise Model} \label{sec:multi-layer-noise-model}
In this section we describe the conversion of the segmented sub-material into procedural models. Each leaf node in our tree is a sub-material while the intermediate nodes store the mask maps. Each sub-material is represented by a set of masked SVBRDF maps.
We use a multi-layer procedural noise model to proceduralize their texture appearance.
As our images are masked and incomplete, the spectrum of the entire image is unavailable, preventing 
the direct fitting of procedural noise models using noise synthesis methods e.g. \cite{Galerne2011, Galerne2012, Gilet2014, Heitz2018}. While Guingo et al.~\shortcite{Guingo2017} propose estimating local spectra in the valid regions only, their approach cannot capture the global variation of the noise textures, e.g. when the scale of the randomness is comparable to the scale of the full image. \new{We leverage a similar sliding window approach but propose a multi-layer procedural model --shown in Algorithm~\ref{algo:noise}-- to deal with incomplete images}. Given an input masked image, we decompose it into several noise layers using a progressive filtering strategy. For each layer, we filter our input image with a Gaussian kernel. The kernel size of the filter becomes larger as the number of layers grows, aiming at capturing larger scale spatial variation. For each filtered layer, we estimate its local spectrum and convert it to a procedural noise model similar to \cite{Guingo2017}. Finally, our algorithm uses the mean value of filtered image $I$ as the base color $C$, and extracts the final noise layer $N \leftarrow I-C$.

\begin{figure} 
	\centering
	\addtolength{\tabcolsep}{-3pt}
	\begin{tabular}{ccccc}
		\multicolumn{5}{c}{
		\begin{tabular}{ccc}
		\includegraphics[width=0.15\textwidth]{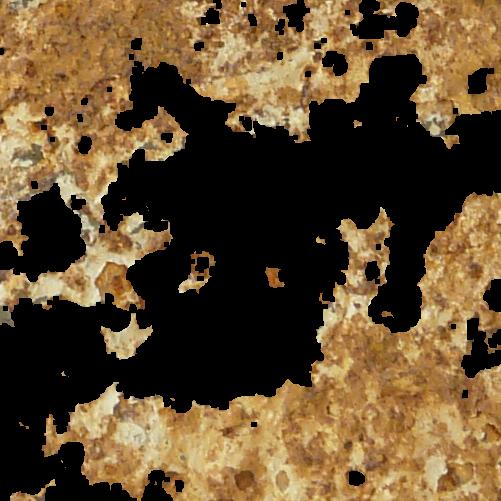} & 
		\includegraphics[width=0.15\textwidth]{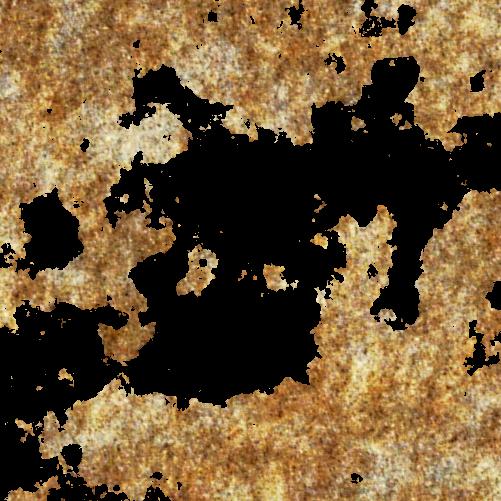} & 
		\includegraphics[width=0.15\textwidth]{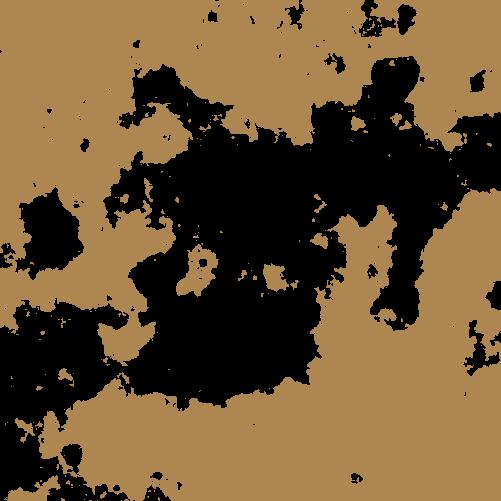}\\
		Input texture & Synthetic texture & Base color
		\end{tabular}
		} \\
		\raisebox{22pt}{\rotatebox[origin=c]{90}{Filtered noises}} &
		\includegraphics[width=0.105\textwidth]{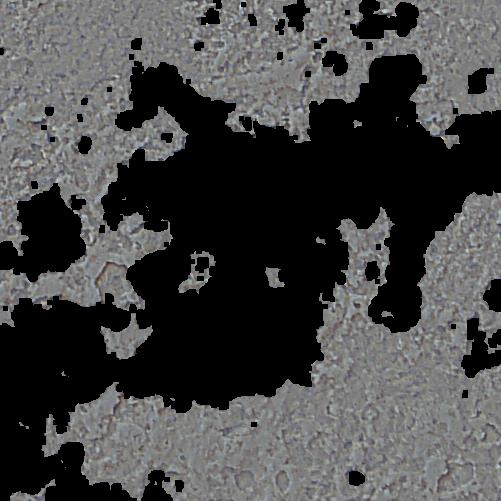} &
		\includegraphics[width=0.105\textwidth]{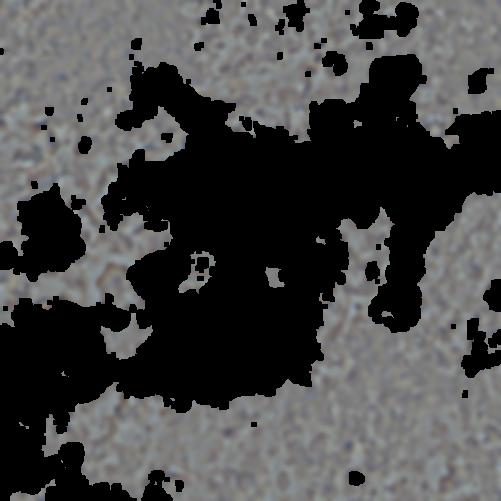} &
		\includegraphics[width=0.105\textwidth]{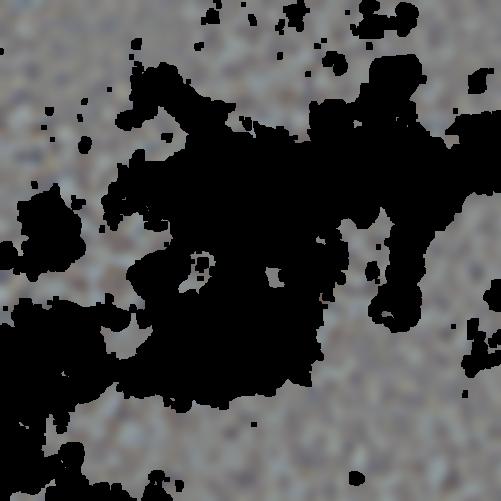} &
		\includegraphics[width=0.105\textwidth]{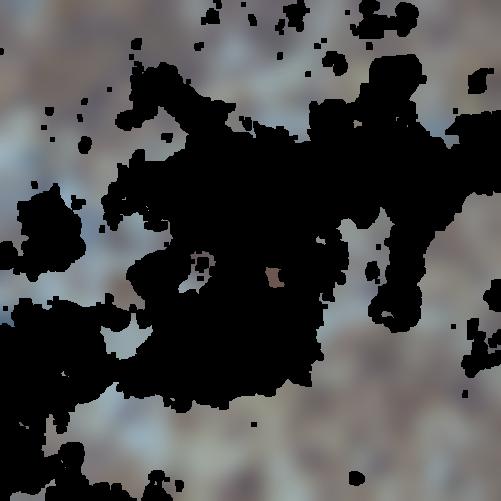} \\
		& Layer 0 & Layer 1 & Layer 2 & Layer 3 \\
		\raisebox{22pt}{\rotatebox[origin=c]{90}{Synthetic noises}} &
		\includegraphics[width=0.105\textwidth]{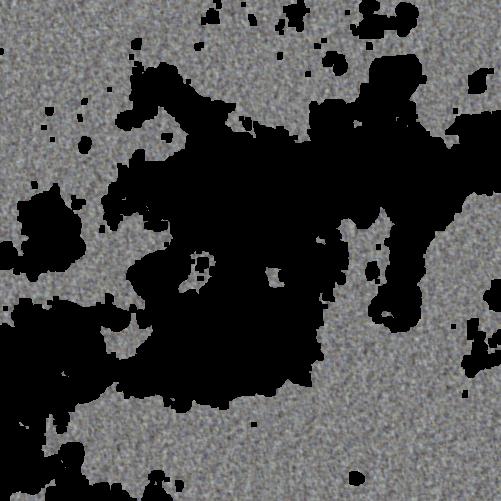} &
		\includegraphics[width=0.105\textwidth]{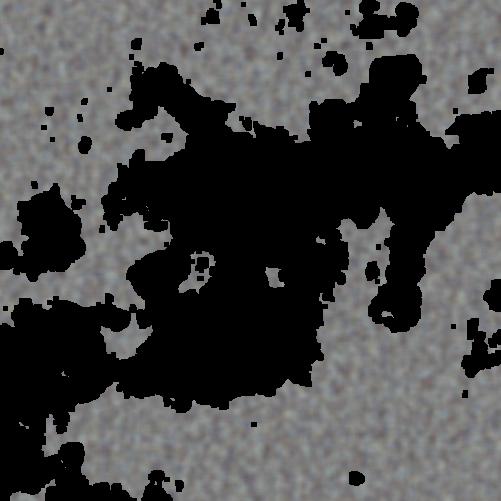} &
		\includegraphics[width=0.105\textwidth]{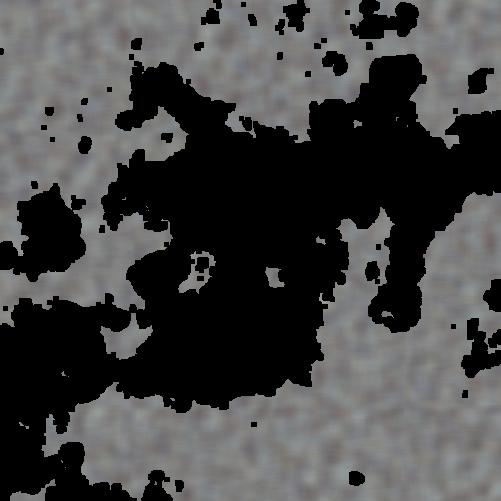} &
		\includegraphics[width=0.105\textwidth]{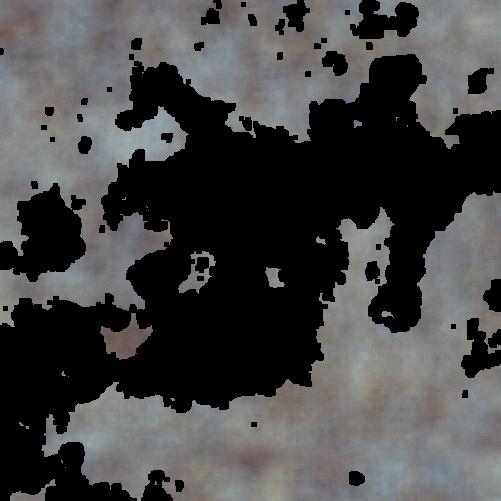} \\
		& Layer 0 & Layer 1 & Layer 2 & Layer 3 \\
	\end{tabular}
\vspace{-10pt}
\caption{Our Multi-layer procedural noise model on a colored texture image. Black regions shows masked unavailable regions. The color of noise images are more contrasted for visualization. The synthetic noise (bottom row) is procedural noise estimated from filtered noise (middle row), and the synthetic texture (middle image in the top row) is computed by adding all the synthetic procedural noise together with the base color.}
\label{fig:noise}
\end{figure}

\begin{algorithm}
 \SetKwInOut{Input}{input}
 \SetKwInOut{Output}{output}
 \Input{image $I$, maximum number of layers $n_{max}$, image variance threshold $\epsilon$, a set of filter kernel sizes $K$, a set of local window sizes $T$, a set of step sizes $S$}
 \Output{a set of procedural models $p$, base color $C$}
 $p \leftarrow \varnothing$\;
 $i \leftarrow 0$\;
 \While{$i < n_{max}$}{
  $\sigma$ $\leftarrow$ Variance($I$)\;
  \If{$\sigma \leq \epsilon$}{
  	break\;
  }
  $I'$ $\leftarrow \textrm{Gaussian filter}(I, K_i)$\;
  $N_i \leftarrow I-I'$\;
  $p \leftarrow p \cup \textrm{Random Phase Noise matching}(N_i, T_i, S_i)$ \;
  $I \leftarrow I'$\;
  $i \leftarrow i+1$\;
 }
 $C \leftarrow \textrm{mean}(I)$\;
 $N \leftarrow I-C$\;
 $p \leftarrow p \cup \textrm{Random Phase Noise matching}(N, T_{n_{max}}, S_{n_{max}})$\;
 \caption{Multi-layer procedural noise model}
 \label{algo:noise}
\end{algorithm}
This last layer, however, represents the lowest frequency of the input image, preventing the use of sliding window spectrum estimation. As each sub-material is masked, the full image spectrum is unavailable. %
We therefore inpaint the missing data \cite{telea2004, PatchMatch}. To reduce artifacts introduced by this step, we apply a set of Gabor noises as a basis to approximate the power spectrum of the inpainted image \cite{Galerne2012}. This last step and our use of multi-layer noise allows our method to produce a fully procedural representation of the sub-material's properties. Fig.~\ref{fig:noise} shows an example of our multi-layer decomposition on a masked colored texture image, where each layer captures different levels of details from an input image and the last layer shows the global variation of the noise image.

\begin{figure} 
	\centering
	\addtolength{\tabcolsep}{-3pt}
	\begin{tabular}{cccc}
		\includegraphics[width=0.15\textwidth]{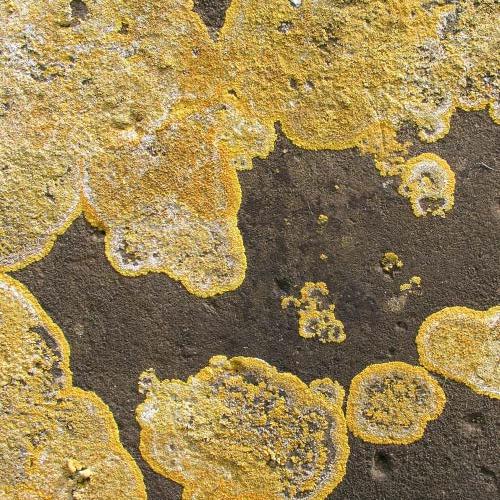} &
		\includegraphics[width=0.15\textwidth]{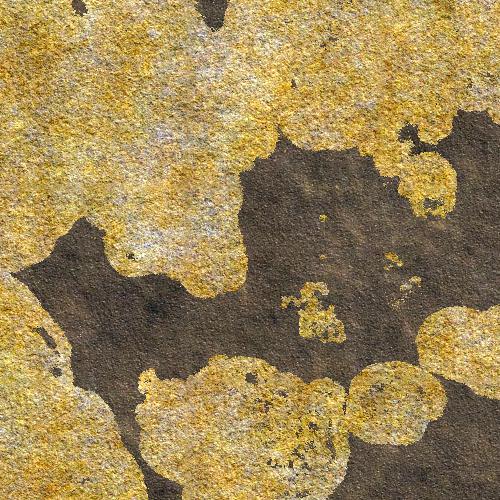} &
		\includegraphics[width=0.15\textwidth]{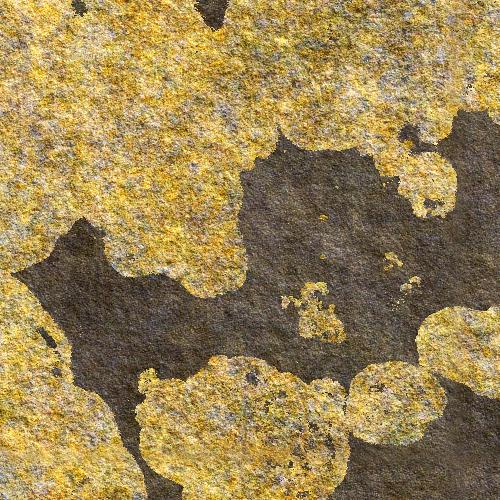} \\
		\includegraphics[width=0.15\textwidth]{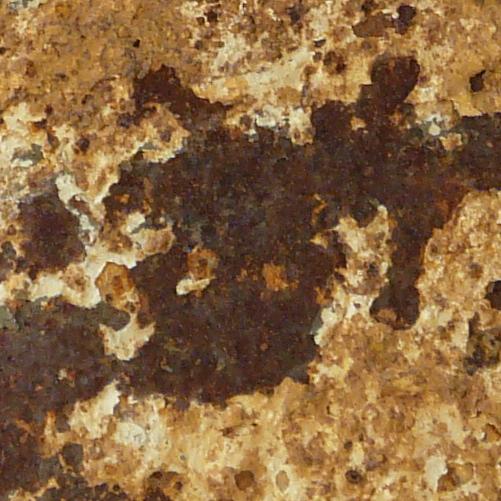} &
		\includegraphics[width=0.15\textwidth]{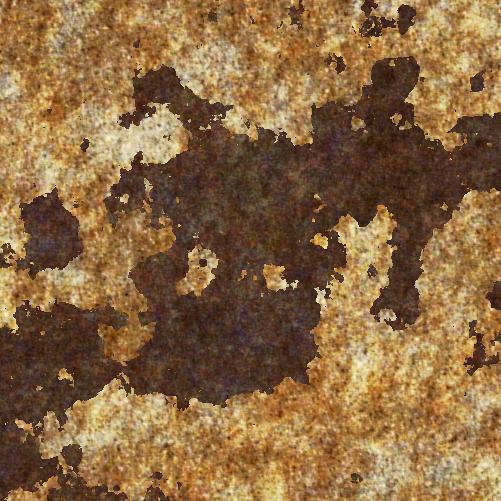} &
		\includegraphics[width=0.15\textwidth]{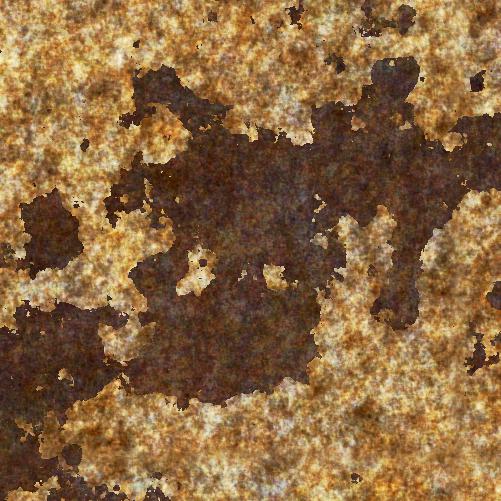} \\
		\includegraphics[width=0.15\textwidth]{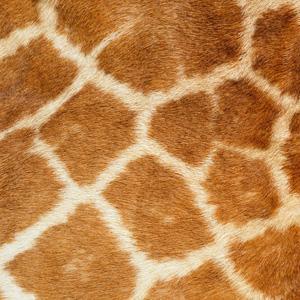} &
		\includegraphics[width=0.15\textwidth]{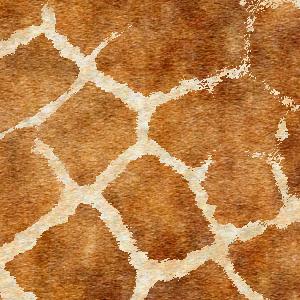} &
		\includegraphics[width=0.15\textwidth]{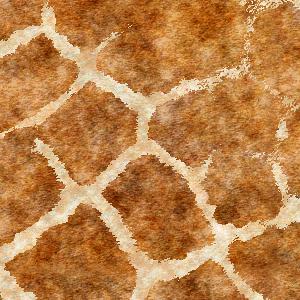} \\
		Input & Multi-layer & Single-layer \\
	\end{tabular}
\vspace{-10pt}
\caption{Comparison between multi-layer and single-layer noise on colored texture images. Compared to a single-layer approach, our multi-layer method can reconstruct a smoother texture image and capture more fine-grained details, while the single-layer generates some apparent repetitive and "dirty" patterns. Textures are segmented by our spectrum-aware matting method, fitted and composed to generate the results.}
\label{fig:multi-single}
\end{figure}

Using the procedural noise $p_i$ fitted to each layer $i$ and the base color $C$, we reconstruct the original texture as $\sum p_i + C, i=1,2,...,n$ where $n$ is the number of noise layers. We further improve the results of this reconstruction for SVBRDF maps by introducing a differentiable rendering-based optimization scheme, described in Sec.~\ref{sec:recomposition}. Our approach provides a fully procedural representation and can reproduce fine-grained details and yield a smoother noise texture than single layer methods (see Fig.~\ref{fig:multi-single}). Furthermore, our multi-layer approach gives users  more control over level of detail.

\subsubsection{Images with Multiple Channels} \label{Sec:tech-detail}

Synthesizing colored images, such as the albedo maps could result in poor color mixing during synthesis. Similar to previous approaches \cite{Heeger1995, Galerne2011, Guingo2017}, given a multi channel noise texture, we synthesize it in a PCA color space. We project the image from the original RGB space to the PCA space, allowing us to synthesize each channel independently before projecting it back to the original RGB space. We then match the histogram of the synthesized image to the input image as a post-processing step to ensure a matching color distribution.

\begin{figure} 
	\centering
	\addtolength{\tabcolsep}{-3pt}
	\begin{tabular}{ccc}
		\includegraphics[width=0.15\textwidth]{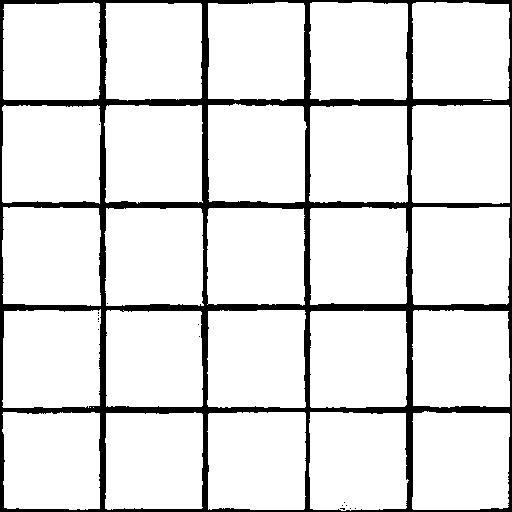} &
		\includegraphics[width=0.15\textwidth]{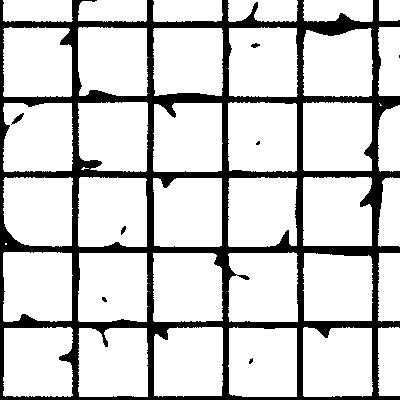} &
		\includegraphics[width=0.15\textwidth]{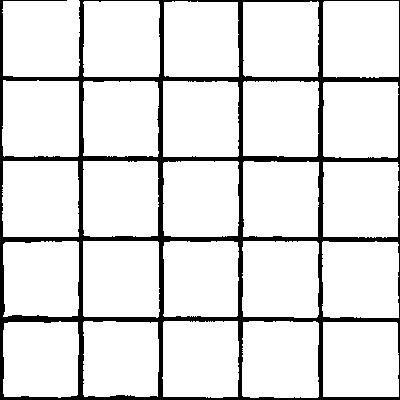} \\
		\includegraphics[width=0.15\textwidth]{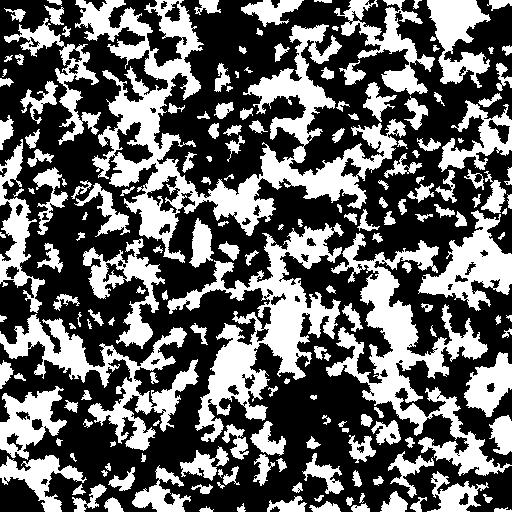} &
		\includegraphics[width=0.15\textwidth]{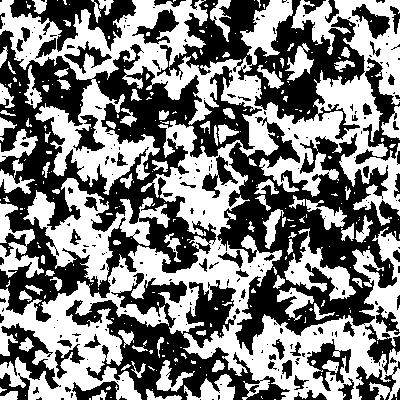} &
		\includegraphics[width=0.15\textwidth]{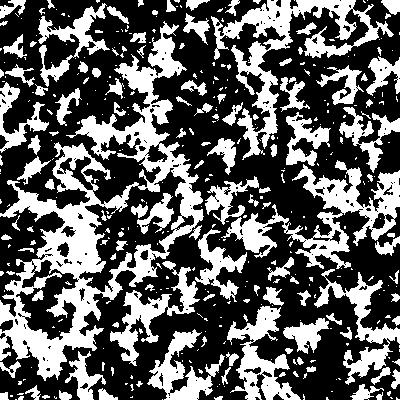} \\
		\includegraphics[width=0.15\textwidth]{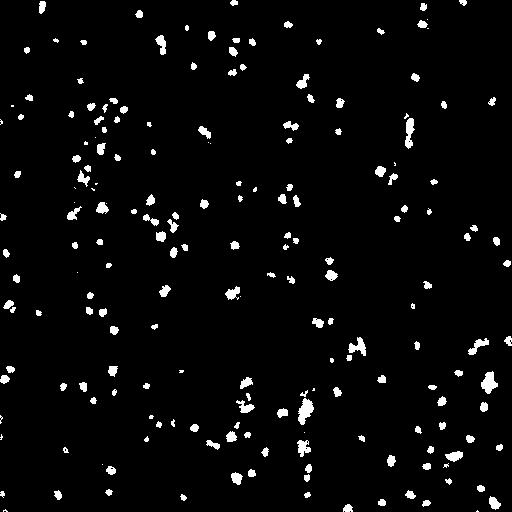} &
		\includegraphics[width=0.15\textwidth]{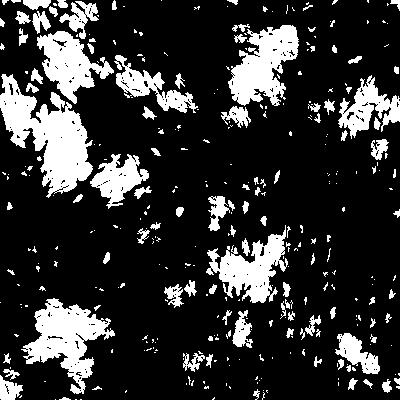} &
		\includegraphics[width=0.15\textwidth]{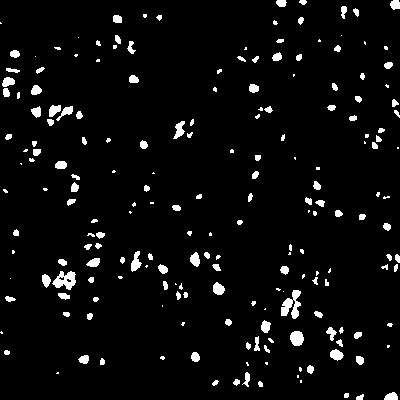} \\
		Input & Query & Optimized \\
	\end{tabular}
\vspace{-10pt}
\caption{Examples of our procedural mask synthesis method. Given an input binary mask map (left) segmented by our spectrum-aware matting method, we first query a similar mask map (middle) from a database, and use it as an initialization of our alternating optimization algorithm to correct and match its structure to the input (right).}
\label{fig:mask-synthesis}
\end{figure}
\subsection{Procedural Mask Synthesis} \label{sec:mask-synthesis}
While we represent the sub-material local variations using our multi-layer noise model, we use mask maps to represent the global spatial distributions of the sub-materials. Proceduralizing these mask maps allows us to reach a full procedural representation of the material. With it, we can easily edit, resample and extend the resolution of the spatial distribution. Similar to sub-material proceduralization, mask proceduralization is done recursively during traversal of the material tree. We model mask maps by two methods 1) Point Process Texture Basis Functions (PPTBF) \cite{Guehl20} to model decomposed binary mask maps by matting algorithm (Sec. \ref{sec:spectrum-decomp}); 2) Random Sampling to model decomposed instances (Sec. \ref{sec:instance-seg}).
\subsubsection{Inverse Mask Fitting by PPTBF}
\new{As introduced in \cite{Guehl20}, Point Process Texture Basis Functions (PPTBF) are defined by the sparse convolution of randomly-sampled 2D points $\mathbf{x_i}$ with a kernel function being the product between a visual feature $f$ and a blending window $w$:
\begin{equation}
\label{eq:pptbf}
PPTBF_k(\mathbf{x})=\sum_{\mathbf{x_i} \in \mathcal{N}_k(\mathbf{x})} f(\mathbf{x}-\mathbf{x_i}) w(\mathbf{x}-\mathbf{x_i}),
\end{equation}
where $\mathcal{N}_k(\mathbf{x})=\{\mathbf{x_1},...,\mathbf{x_k}\}$ describes the $k$ closest sample points around $\mathbf{x}$.
According to Eq. (\ref{eq:pptbf}), the behavior of PPTBF depends on 2D spatial point distribution of $\mathbf{x_i}$, visual feature function $f$, and blending window function $w$, which are in turn controlled by intuitive parameters such as kernel size, degree of smoothing etc. to produce a continuous scalar field. After applying a threshold, we use PPTBF to model binary mask maps segmented by our matting algorithm. In this section, We first describe the fitting of mask maps on the top level of the material tree where the mask covers the entire image, and then discuss fitting maps in the lower hierarchy where mask contains missing regions.}

As PPTBF is a forward generation process, %
\cite{Guehl20} did not focus on by-example modeling. The authors suggested an approach that queries a precomputed PPTBF database and then optimizes parameters, but did not provide details. We therefore experimented with different feature representations and optimization routines and here discuss the techniques we chose. Upon acceptance we will publish the implementation of our \textit{query-and-optimization} method for by-example mask fitting with our entire pipeline.

\textbf{Query:} As PPTBF are not fully differentiable and partly rely on discrete parameters, starting from a good initialization is particularly important for efficient optimization. We first query a database, uniformly covering the variations allowed by PPTBF, to retrieve the nearest neighbor parameters of our input mask. \new{We use the mask image database provided in ~\cite{Guehl20}. The database contains 450K images and was generated by non-regular sampling with three different thresholds for binarization.} To measure the similarity between the input mask and pre-sampled mask maps in the database, we choose Local Binary Pattern (LBP) to encode local statistics, Gram Matrix of pre-trained VGG19~\cite{vgg19} deep features for global statistics, and Fourier power spectrum for regularity. We weight and concatenate these three features to build a high dimensional vector acting as a descriptor for binary mask maps. We precompute such a vector for each mask map in the database, and reduce their dimensionality by PCA to 512 dimensions for a more robust representation. The storage of the processed database is 7.81 GB in total. We build an acceleration structure for fast nearest neighbor search based on $L_2$ distance of two feature vectors. The precomputation takes around 36 hours, but each query then requires less than 1 second. %

\textbf{Optimization:} After we retrieve a pre-sampled binary mask map from the database, we use its parameters to initialize our optimization algorithm to better match the input mask. Since PPTBF contains both continuous parameters as well as discrete parameters, we apply coordinate descent to optimize continuous and discrete ones alternatively. As evaluation of PPTBF is expensive and not easily differentiable, we apply gradient-free approaches to avoid costly finite differentiation. \new{Our implementation for PPTBF optimization is CPU-based. For continuous parameters, we apply the Powell method (SciPy \shortcite{scipy2020}) and for discrete parameters, we adopt a Bayesian optimization method  with Gaussian Process (GPyOpt \shortcite{gpyopt2016}).} %

As Fig.~\ref{fig:mask-synthesis} shows, our query results provide an initial guess about the general structure but it can be hard to find a perfect match from the database due to sparse sampling. Our optimization step can correct the parameters and match the structure of the input image.

\begin{figure} 
	\centering
	\addtolength{\tabcolsep}{-5pt}
	\begin{tabular}{ccc}
		\includegraphics[width=0.15\textwidth]{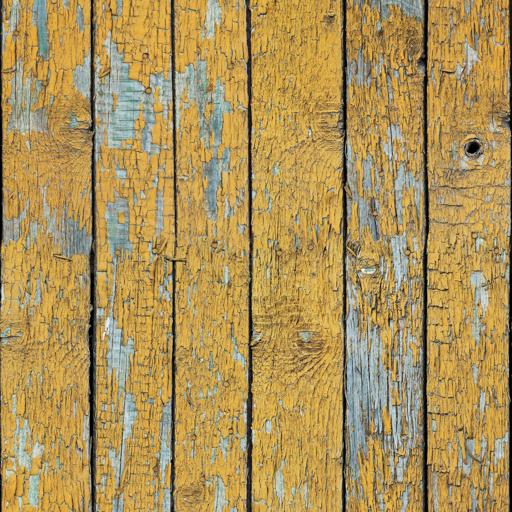} &
		\includegraphics[width=0.15\textwidth]{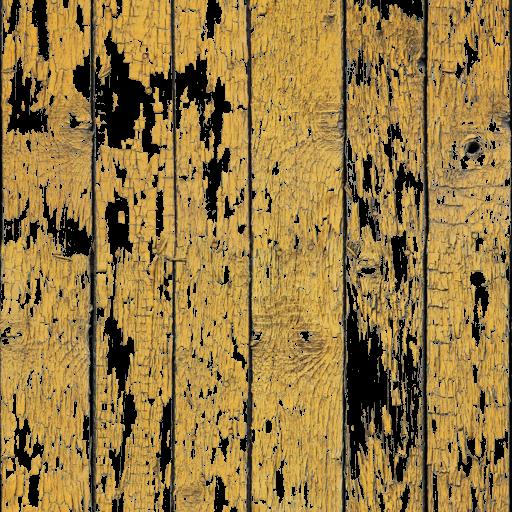} &
		\includegraphics[width=0.15\textwidth]{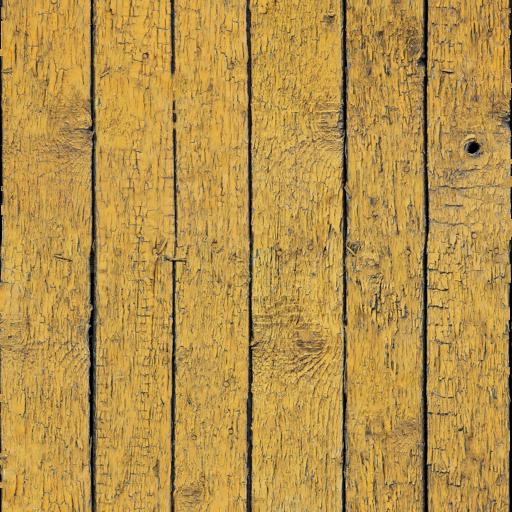} \\
		Original Input & Masked Input & Inpainted Input \\
		\includegraphics[width=0.15\textwidth]{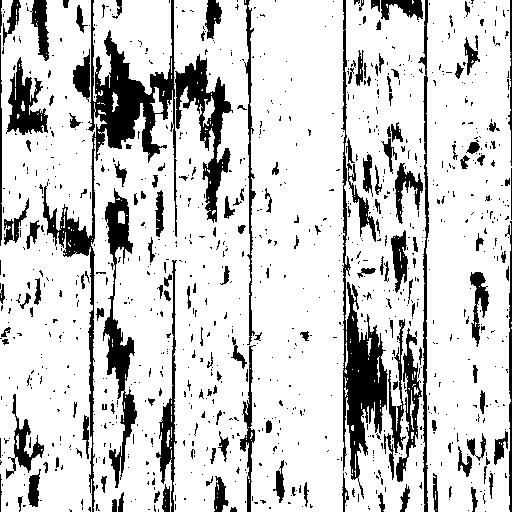} &
		\includegraphics[width=0.15\textwidth]{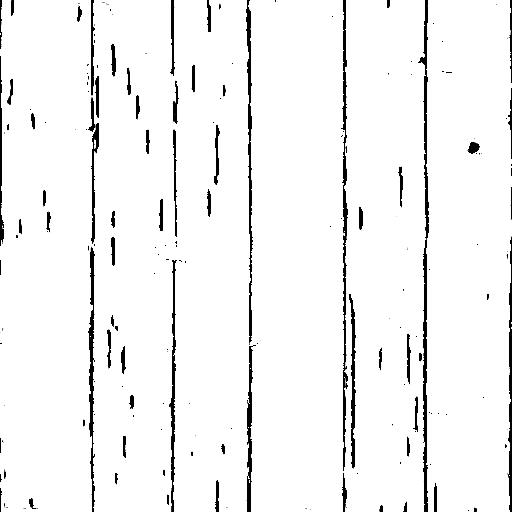} &
		\includegraphics[width=0.15\textwidth]{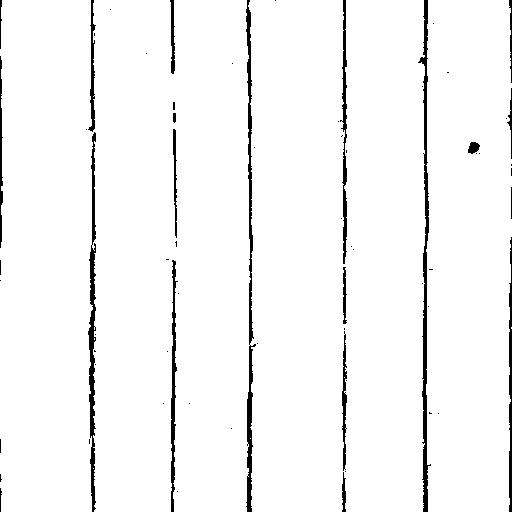} \\
		 Incomplete Mask & Directly Inpainted Mask & Segmented Mask \\
	\end{tabular}
\caption{Mask inpainting. We apply a hierarchical segmentation approach. We first segment the chipped area, leaving the yellow paint and spaces between the planks (Masked Input). If we directly segmented the masked input, the result (Incomplete Mask) is difficult to fit procedurally. We need to separate the influence of the parent mask (chipped area) from the sub parts to separate (planks from space in between). To do so we use in-painting. Directly inpainting the incomplete mask fails as binary
masks cannot provide sufficient hints to guide the inpainting (Directly Inpainted Mask). Instead, we inpaint the input image using PatchMatch (Inpainted Input) on which we compute the segmentation, producing a better mask (Segmented Mask)}

\label{fig:mask-inpaint}
\end{figure}
\subsubsection{Incomplete mask maps}
Different from masks at the top layer of our tree, masks for sub-materials in the lower hierarchy are masked by higher level masks. If naively processed, this leads to poor procedural reconstruction of the sub-material distribution. \\
In order to fit incomplete mask maps, we propose two solutions. First, we adjust our optimization by computing losses only for unmasked pixels between input mask and PPTBF output. Second, we propose proceduralizing an inpainted version -- using PatchMatch~\cite{PatchMatch} -- of incomplete mask maps. However, features in the mask map are not sufficient to guide inpainting. Instead, we inpaint the sub-material and compute the mask map on it as shown in Fig.~\ref{fig:mask-inpaint}. Because PatchMatch inpainting is a sample-based method, this second approach works best for stochastic distributions.

\subsubsection{Random Sampling}
When mask maps are generated using our instance-based decomposition solution, we can model their distributions by random sampling. Suppose we have $n$ mask maps, each representing one type of segmented instance. We count the number of instances in each mask map, and estimate their probability of occurrence. During mask synthesis, we randomly -- following the estimated distribution -- assign a label between 1 and $n$ to instances in the procedural version of the mask map. The corresponding sub-materials are then synthesized in the labeled regions.

\begin{figure} 
	\centering
	\addtolength{\tabcolsep}{-4pt}
	\begin{tabular}{cccccc}
	    \raisebox{22pt}{\scalebox{1.0}{\rotatebox[origin=c]{90}{Inputs}}} &
		\includegraphics[width=0.11\textwidth]{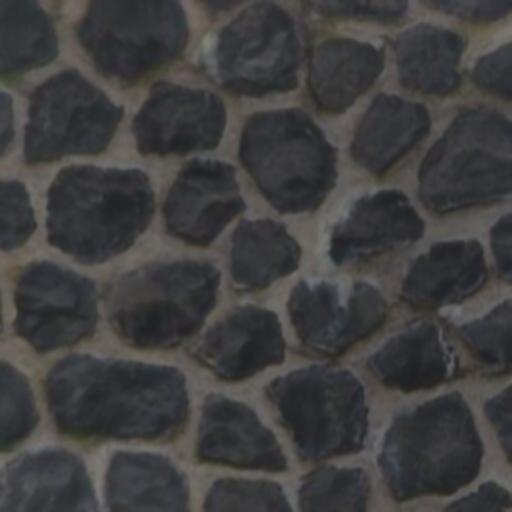} &
		\includegraphics[width=0.11\textwidth]{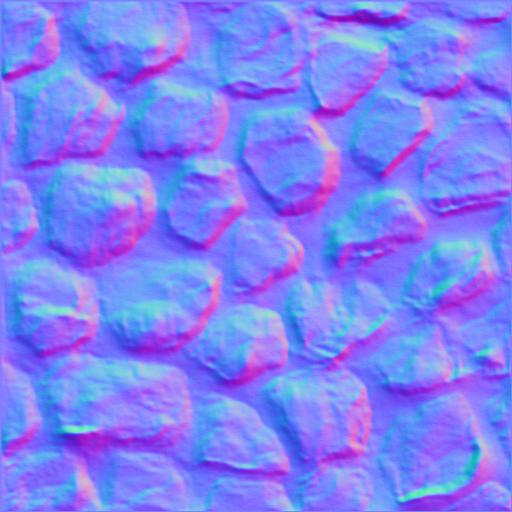} &
		\includegraphics[width=0.11\textwidth]{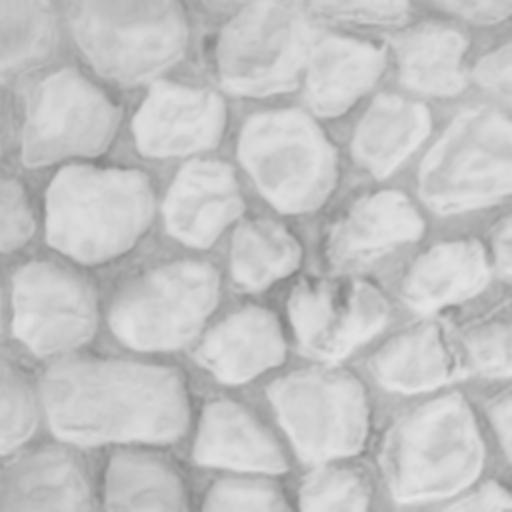} &
		\includegraphics[width=0.11\textwidth]{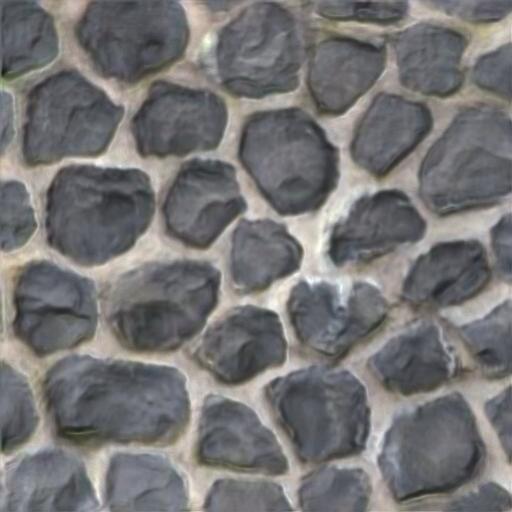} \\
		
		\raisebox{22pt}{\scalebox{1.0}{\rotatebox[origin=c]{90}{Optimized}}} &
		\includegraphics[width=0.11\textwidth]{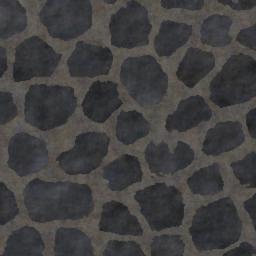} &
		\includegraphics[width=0.11\textwidth]{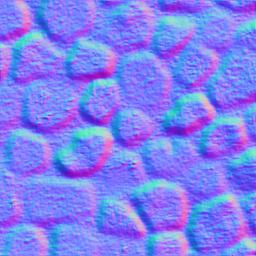} &
		\includegraphics[width=0.11\textwidth]{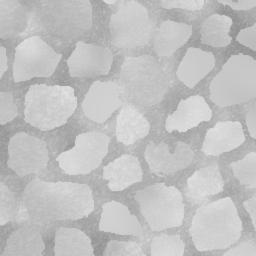} &
		\includegraphics[width=0.11\textwidth]{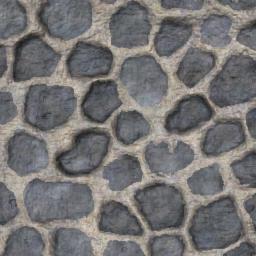} \\
		
		\raisebox{22pt}{\scalebox{1.0}{\rotatebox[origin=c]{90}{W/o optim.}}} &
		\includegraphics[width=0.11\textwidth]{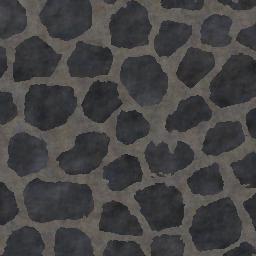} &
		\includegraphics[width=0.11\textwidth]{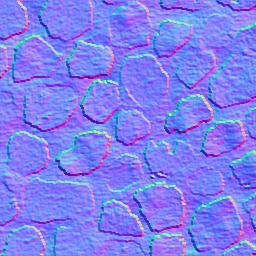} &
		\includegraphics[width=0.11\textwidth]{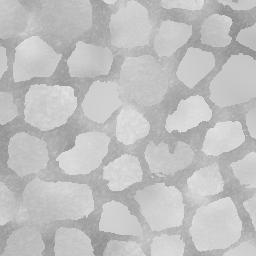} &
		\includegraphics[width=0.11\textwidth]{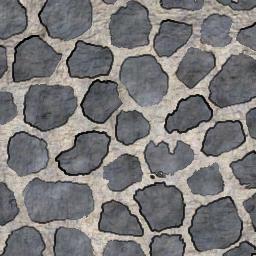} \\
		
		&
		\scalebox{1}{Albedo} & 
		\scalebox{1}{Normal} & 
		\scalebox{1}{Roughness} &
		\scalebox{1}{Rendered}
	\end{tabular}
\vspace{-10pt}
\caption{Validation of our optimization-based recomposition. In this case, the "input" material maps are generated using \cite{DDB20} from a natural image data where lighting is not completely removed from material maps. Additionally, local variations of the different material maps cannot be well represented with simple Gaussian noise especially in the normals. Directly fitting these local textures using our multi-layer noise model leads to large fitting errors and visual artifacts (3rd row). Using our differentiable optimization approach, our method automatically refines the procedural parameters of our output during the recomposition step (2nd row). Thanks to this optimization, we improve the stability of our method for arbitrary inputs. For better visualization we directly use the KNN segmented mask rather than a procedural counterpart.}
\label{fig:optimizationAblation}
\end{figure}
\subsection{Recomposition} \label{sec:recomposition}
Finally, using our generated procedural noise maps and masks tree, we compose them into an output SVBRDF to reproduce the appearance of the original SVBRDF inputs. To relate this approach to a classical Substance Designer~\shortcite{SubstanceDes} pipeline, our procedural noise maps and binary masks function as \emph{generators nodes}, which do not rely on existing artist designed graphs.

To better match the original SVBRDF inputs, we add optimizable operators to control the appearance of procedural noise and binary maps. Given a noise map $I$, we modify its appearance by $G(I*\alpha+\delta, \sigma)$, where $G$ is a Gaussian filter, $\alpha$ controls the intensity, $\delta$ biases the noise, and $\sigma$ is the standard deviation of the Gaussian filter. For a binary mask map $M$ which represents the distribution of sub-materials, we model the smooth transition between the boundaries of sub-materials by $G(M, \sigma)$. The parameters $\alpha$, $\delta$ and $\sigma$ are optimizable  for each Gaussian filter per noise or mask map. In each leaf node of our tree, the noise maps are multiplied by their corresponding mask maps and linearly combined to reconstruct the SVBRDF maps for a procedural sub-material. Procedural Sub-materials from different layers are recursively computed and aggregated in a bottom-up fashion to build the final output SVBRDF maps.

These optimizable operators, together with our reconstructed noise maps and binary maps, build a small optimizable material graph. We optimize this material graph, using a differentiable rendering-based optimization routine. The reconstructed SVBRDF maps and the input SVBRDF maps are rendered using a Cook-Torrance, GGX shading model under randomly sampled lighting configurations. Considering that the structure of the procedural SVBRDF maps might not align perfectly with the original inputs, we define the loss as a combination of Style loss and SSIM loss. Style loss is defined as the $L_1$ difference between Gram matrices computed over VGG~\shortcite{vgg19} features of the renderings, similar to the style transfer literature~\cite{gatys2016}. SSIM loss is computed by the $L_1$ difference between the structural similarity (SSIM) indices \cite{ssim} of the renderings. The full loss is written as
\begin{equation}
    L =\sum \|GM(I_i)-GM(I^*_i)\|_1 + \beta \|SSIM(I_i) - SSIM(I^*_i)\|_1
\end{equation}
where $GM$ is the Gram matrix and SSIM is an operator to compute structure similarity index; $I_i$ and $I^*_i$ are input/procedural albedo map, normal map (computed from height map), roughness map and their renderings; $\beta$ balances the weight between the style term and the SSIM term. In Fig.  ~\ref{fig:optimizationAblation} we show an example with and without this optimization process. Often in real data local texture appearance cannot be well represented by Gaussian noise models, resulting in fitting errors and artifacts when generating procedural models (3rd row in Fig.~\ref{fig:optimizationAblation}). This is particularly problematic for normal/height map modeling. Our optimization helps refine the parameters of our procedural graph, significantly improving the results, even when the exemplar violates our assumption of locally uniform appearance such as in Fig.~\ref{fig:optimizationAblation}.

\subsection{Final procedural representation} \label{sec:procedural_representation}
Our final procedural representation of material maps is the equivalent of small material graphs with (i) procedural noise maps (ii) procedural mask maps (iii) optimizable operators, similar to  Fig. \ref{fig:substance-example}. 
In \cite{Shi20}'s terminology, each procedural noise and mask model is a generator while optimizable operators serve as filters. As a fully procedural model, editable parameters of our small material graph are (i) parameters of our generators e.g. noise models and mask models; (ii) parameters in the filter nodes. Editing can be done on individual maps or jointly on all of them. The procedural model can then generate the material at arbitrary resolution and scale.

\begin{figure*} %
	\centering
	\addtolength{\tabcolsep}{-3pt}
	\begin{tabular}{ccccccc}
			& Albedo & Normal & Roughness & Label Map & Render (TL) & Render (SL) \\
		
		\raisebox{33pt}{\scalebox{1.0}{\rotatebox[origin=c]{90}{Input}}} &
		\includegraphics[width=0.135\textwidth]{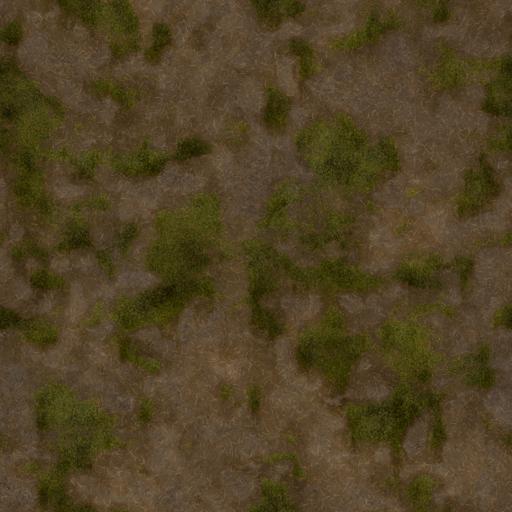} &
		\includegraphics[width=0.135\textwidth]{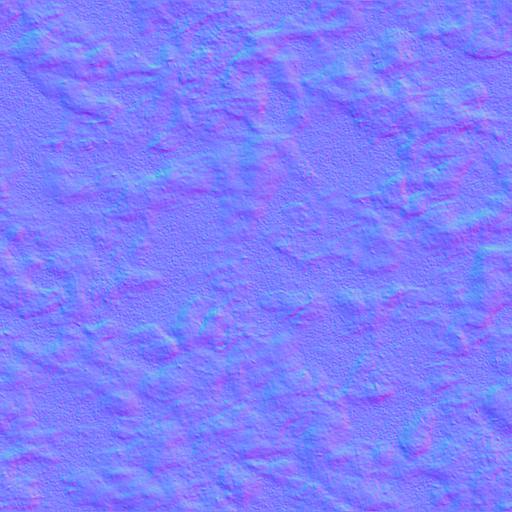} &
		\includegraphics[width=0.135\textwidth]{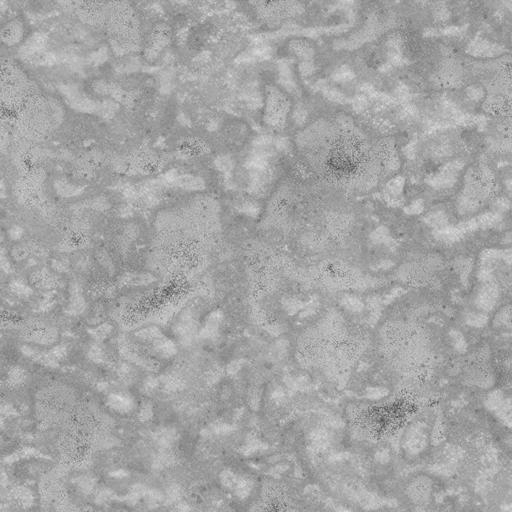} & 
		\includegraphics[width=0.135\textwidth]{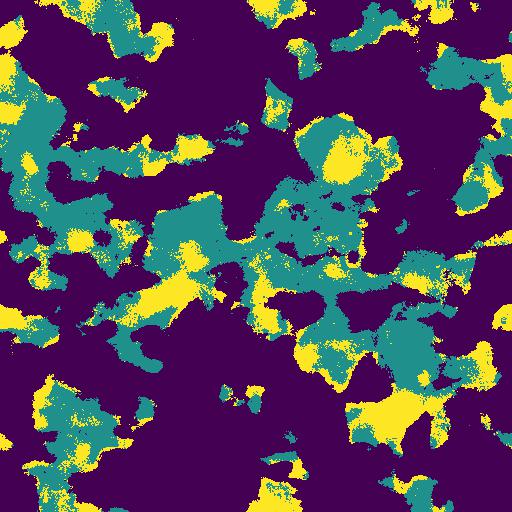} & 
		\includegraphics[width=0.135\textwidth]{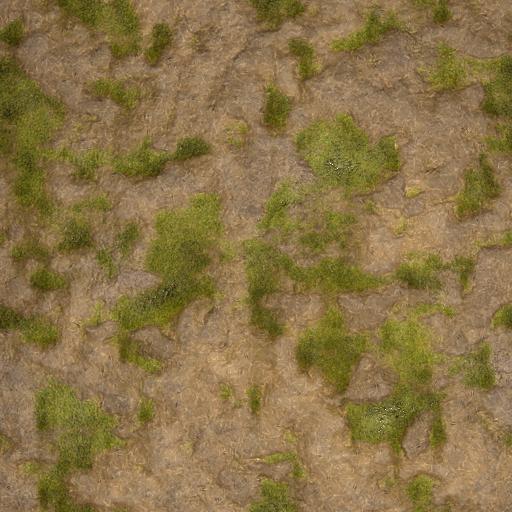} & 
		\includegraphics[width=0.135\textwidth]{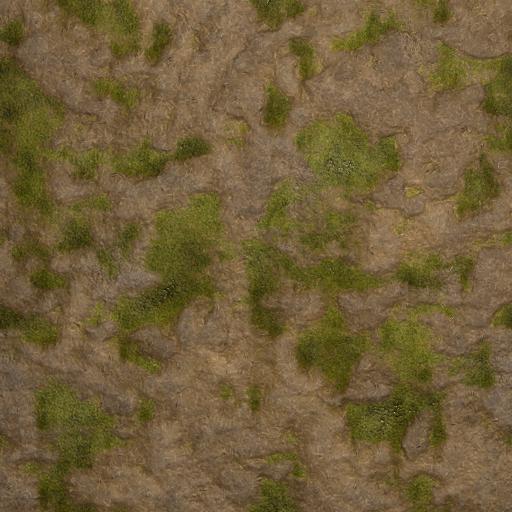} \\

		\raisebox{33pt}{\scalebox{1.0}{\rotatebox[origin=c]{90}{Ours (Proc.)}}} &
		\includegraphics[width=0.135\textwidth]{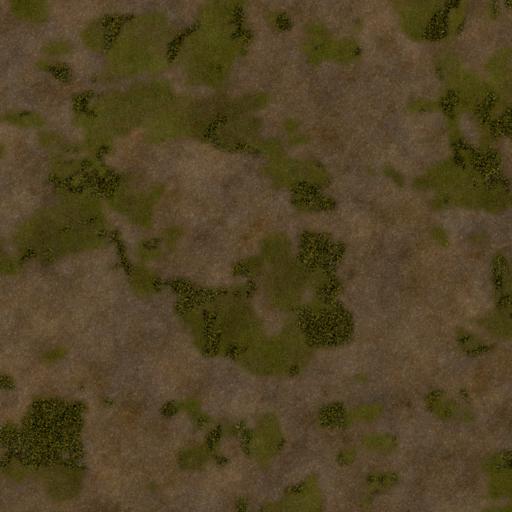} &
		\includegraphics[width=0.135\textwidth]{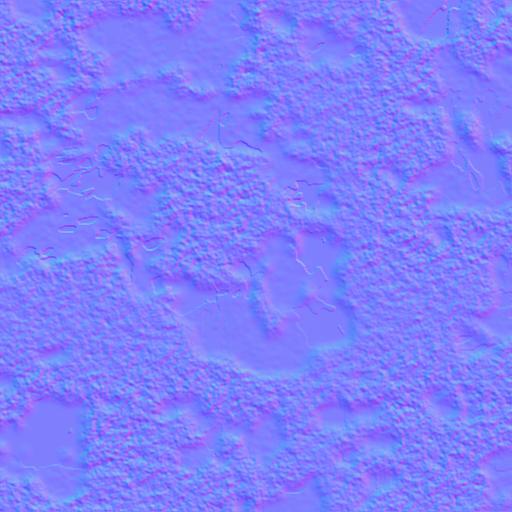} &
		\includegraphics[width=0.135\textwidth]{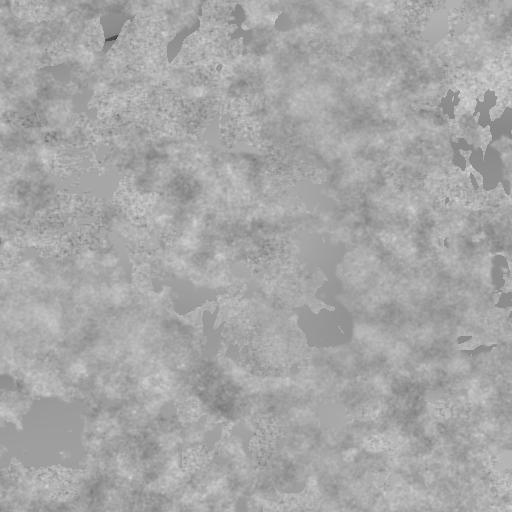} &
		\includegraphics[width=0.135\textwidth]{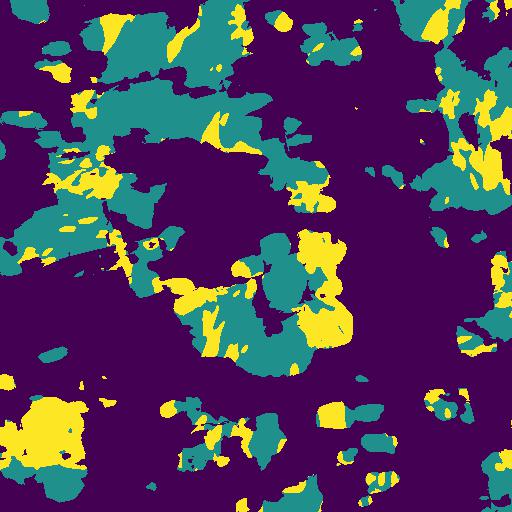} & 
		\includegraphics[width=0.135\textwidth]{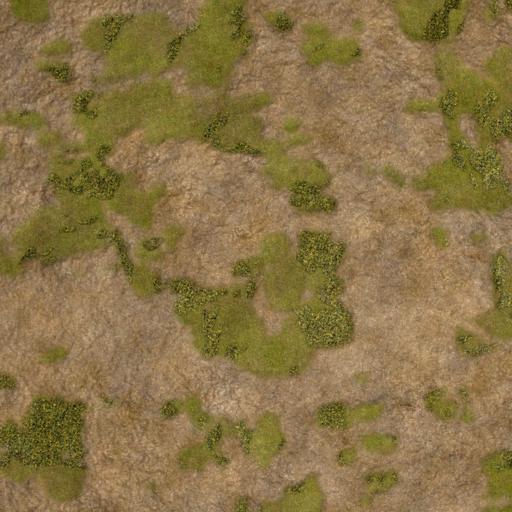} &
		\includegraphics[width=0.135\textwidth]{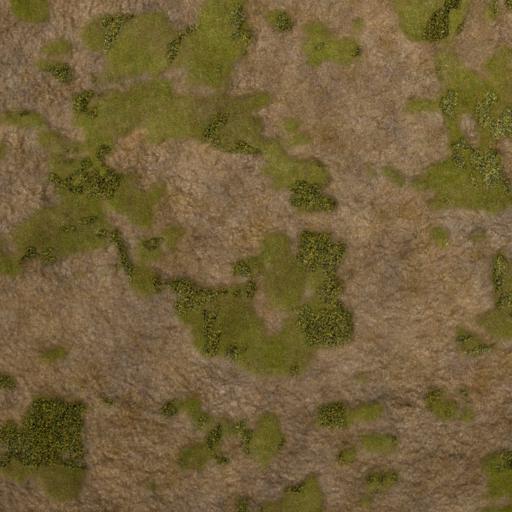} \\
		
		\raisebox{33pt}{\scalebox{1.0}{\rotatebox[origin=c]{90}{Input}}} &
		\includegraphics[width=0.135\textwidth]{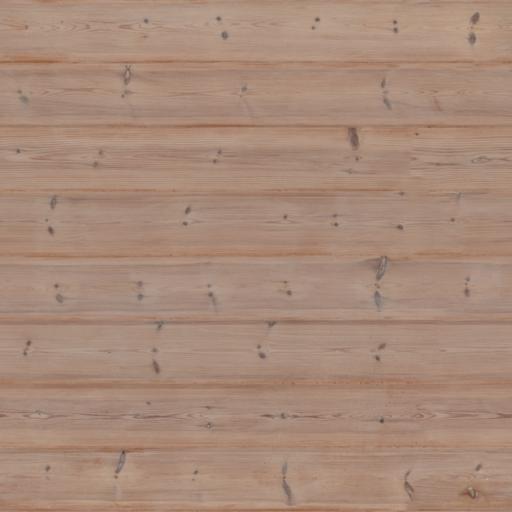} &
		\includegraphics[width=0.135\textwidth]{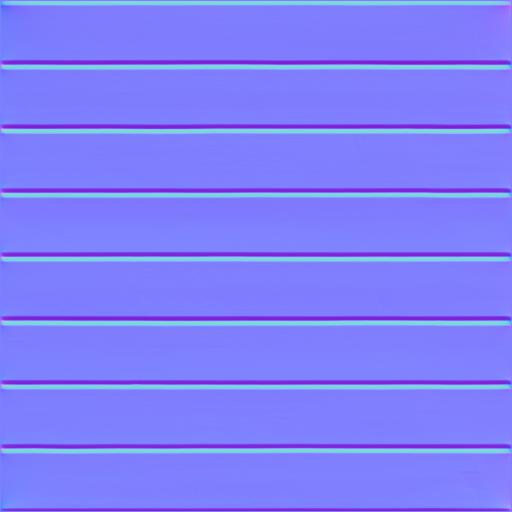} &
		\includegraphics[width=0.135\textwidth]{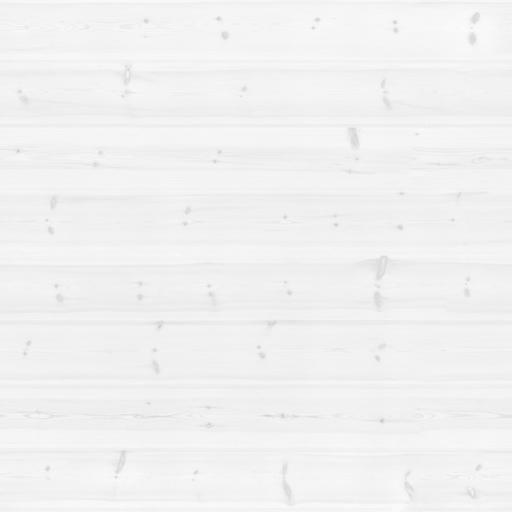} &
		\includegraphics[width=0.135\textwidth]{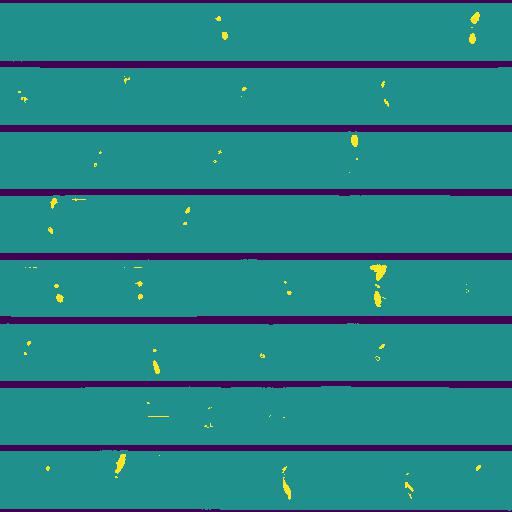} & 
		\includegraphics[width=0.135\textwidth]{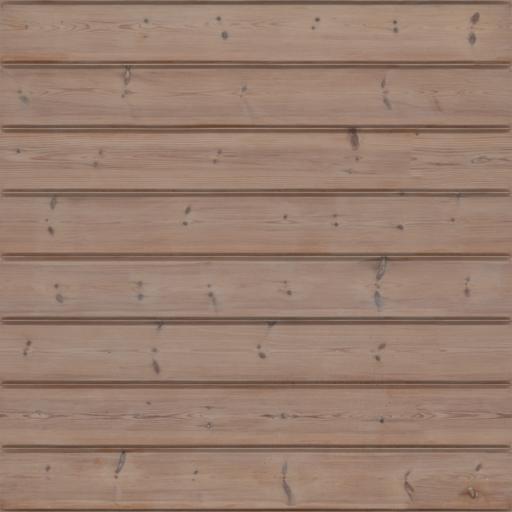} &
		\includegraphics[width=0.135\textwidth]{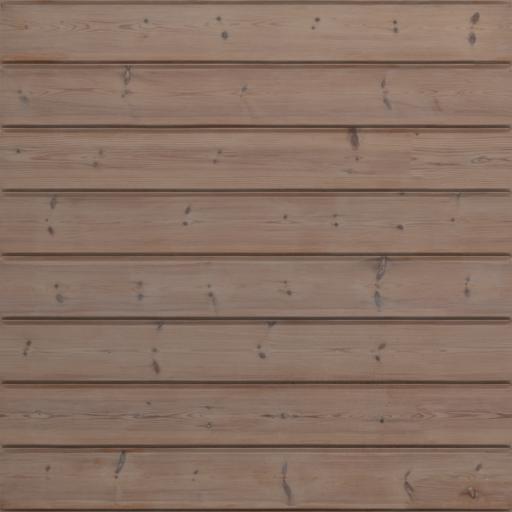} \\
		
		\raisebox{33pt}{\scalebox{1.0}{\rotatebox[origin=c]{90}{Ours (Proc.)}}} &
		\includegraphics[width=0.135\textwidth]{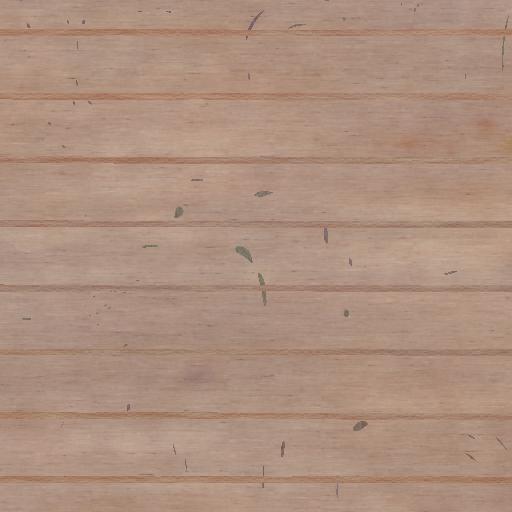} &
		\includegraphics[width=0.135\textwidth]{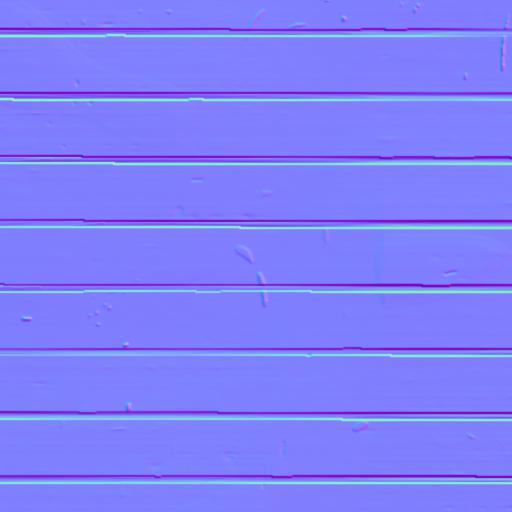} &
		\includegraphics[width=0.135\textwidth]{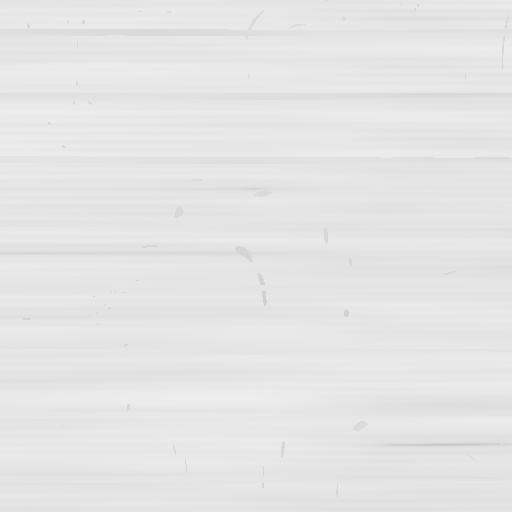} &
		\includegraphics[width=0.135\textwidth]{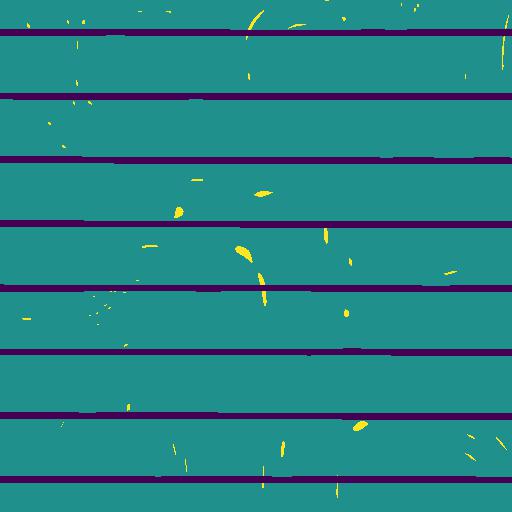} & 
		\includegraphics[width=0.135\textwidth]{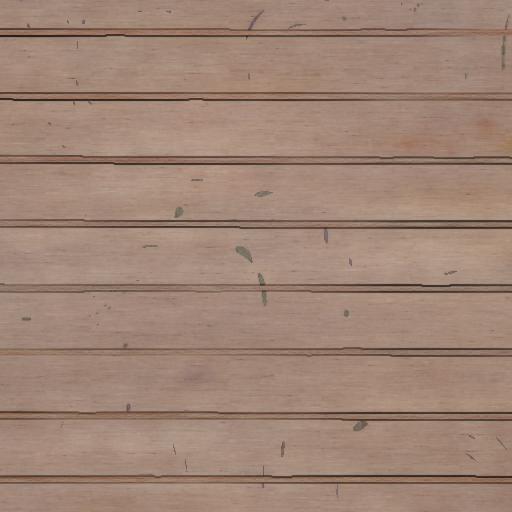} &
		\includegraphics[width=0.135\textwidth]{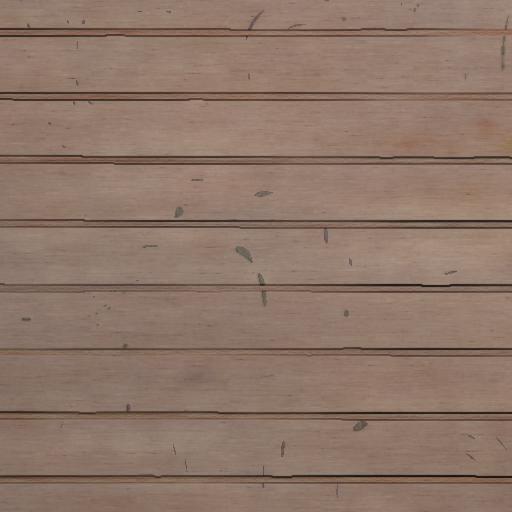} \\

	    \raisebox{33pt}{\scalebox{1.0}{\rotatebox[origin=c]{90}{Input}}} &
		\includegraphics[width=0.135\textwidth]{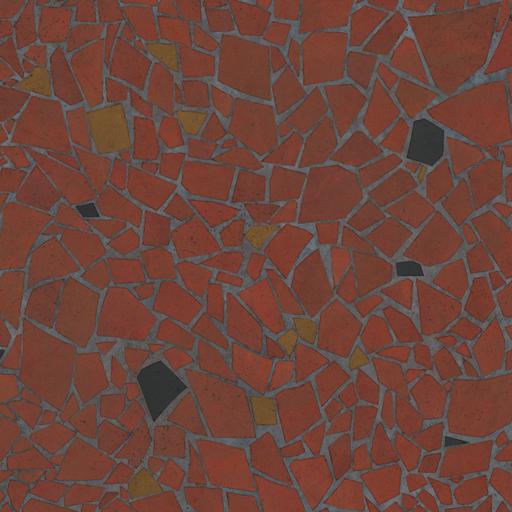} &
		\includegraphics[width=0.135\textwidth]{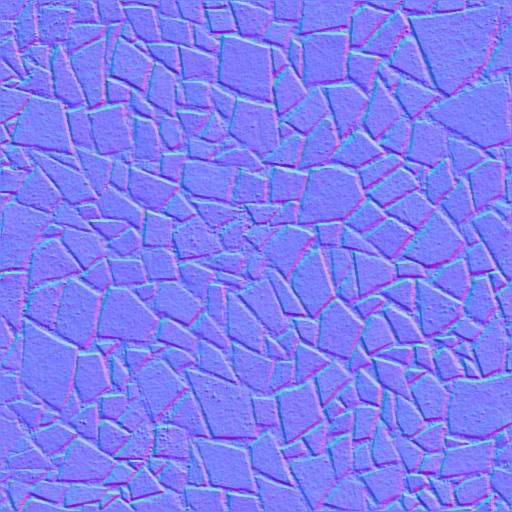} &
		\includegraphics[width=0.135\textwidth]{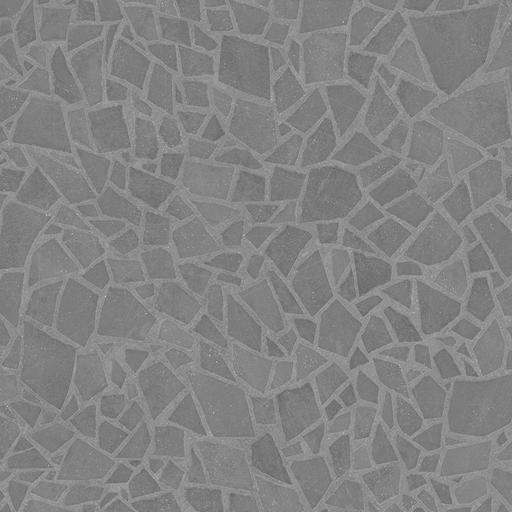} & 
		\includegraphics[width=0.135\textwidth]{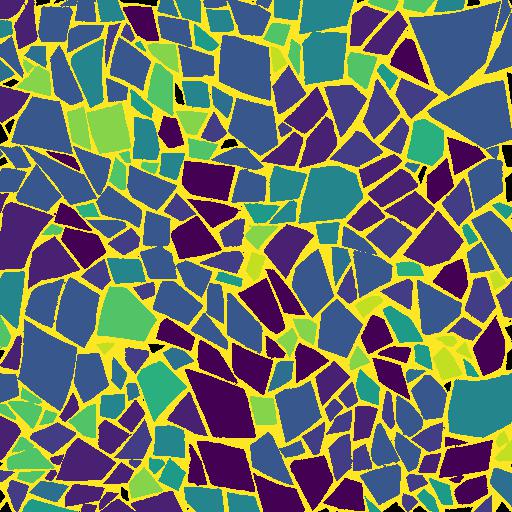} & 
		\includegraphics[width=0.135\textwidth]{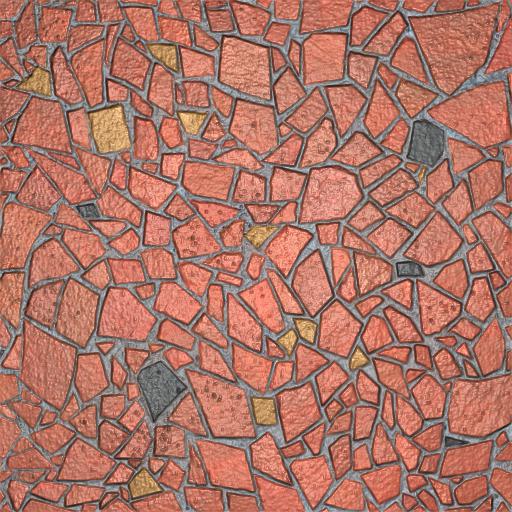} & 
		\includegraphics[width=0.135\textwidth]{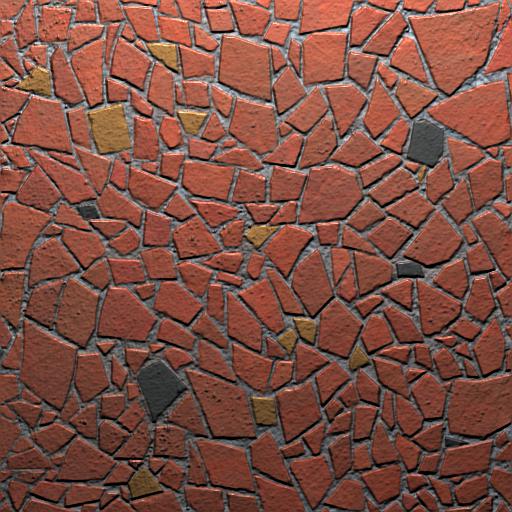} \\
		
		\raisebox{33pt}{\scalebox{1.0}{\rotatebox[origin=c]{90}{Ours (Proc.)}}} &
		\includegraphics[width=0.135\textwidth]{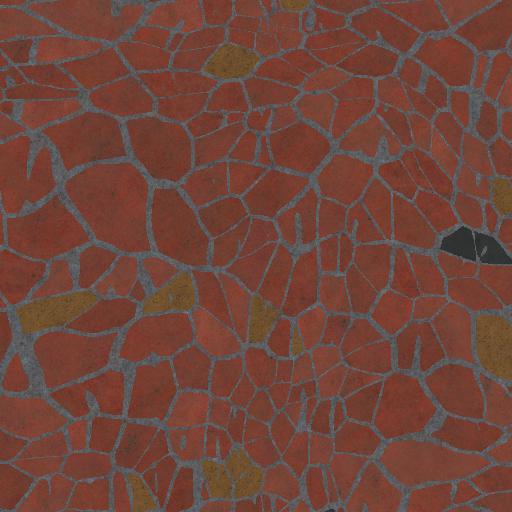} &
		\includegraphics[width=0.135\textwidth]{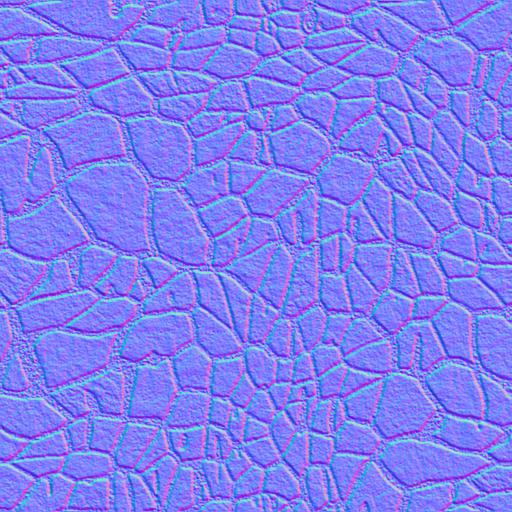} &
		\includegraphics[width=0.135\textwidth]{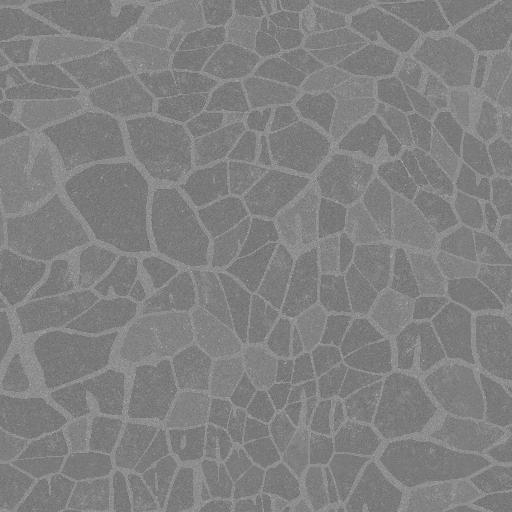} &
		\includegraphics[width=0.135\textwidth]{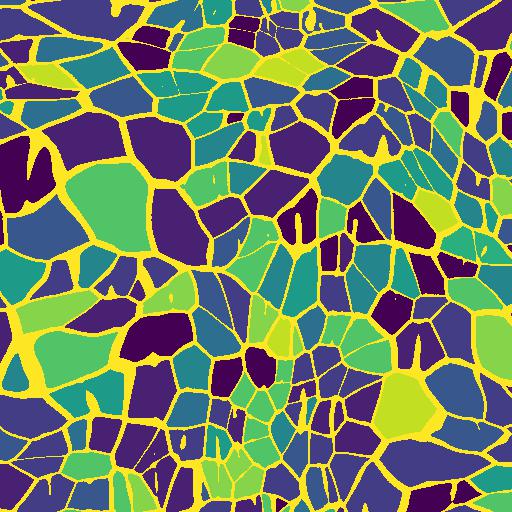} & 
		\includegraphics[width=0.135\textwidth]{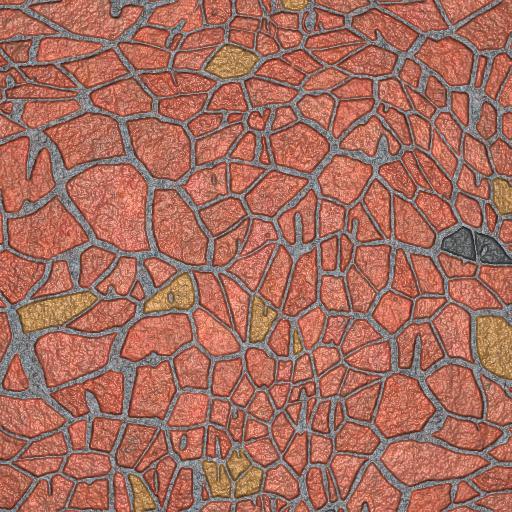} &
		\includegraphics[width=0.135\textwidth]{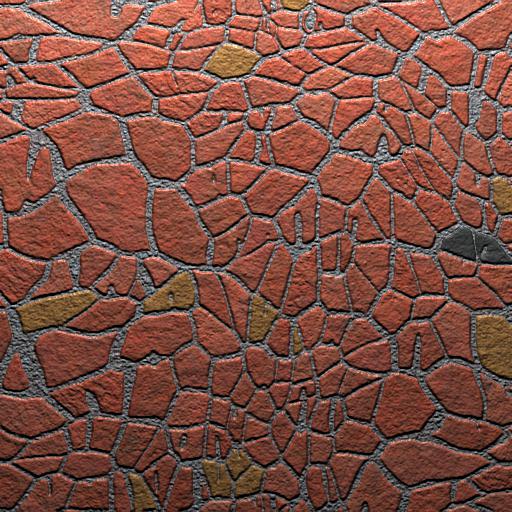} \\
		
		\raisebox{33pt}{\scalebox{1.0}{\rotatebox[origin=c]{90}{Input}}} &
		\includegraphics[width=0.135\textwidth]{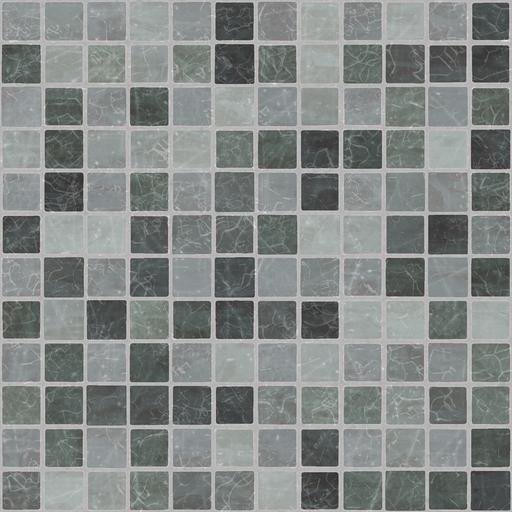} &
		\includegraphics[width=0.135\textwidth]{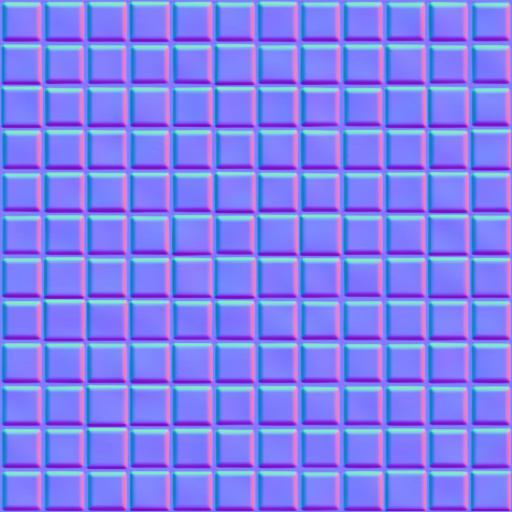} &
		\includegraphics[width=0.135\textwidth]{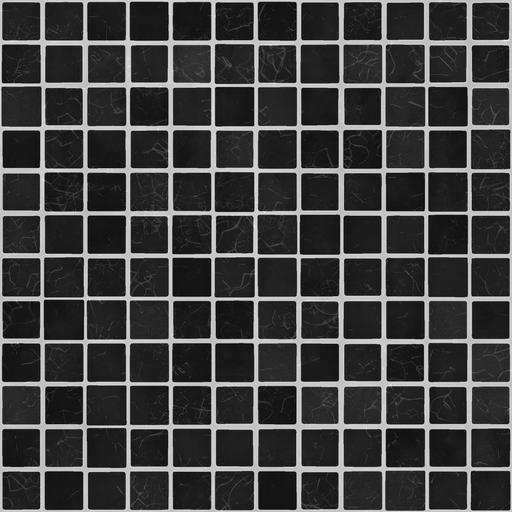} &
		\includegraphics[width=0.135\textwidth]{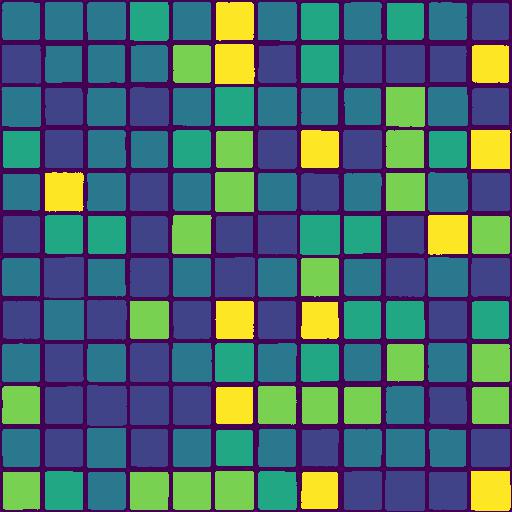} & 
		\includegraphics[width=0.135\textwidth]{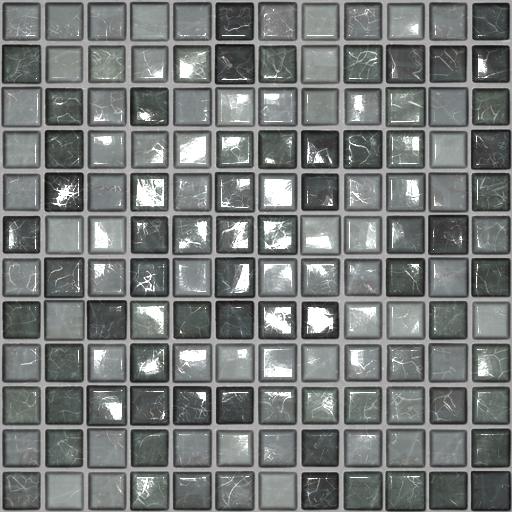} &
		\includegraphics[width=0.135\textwidth]{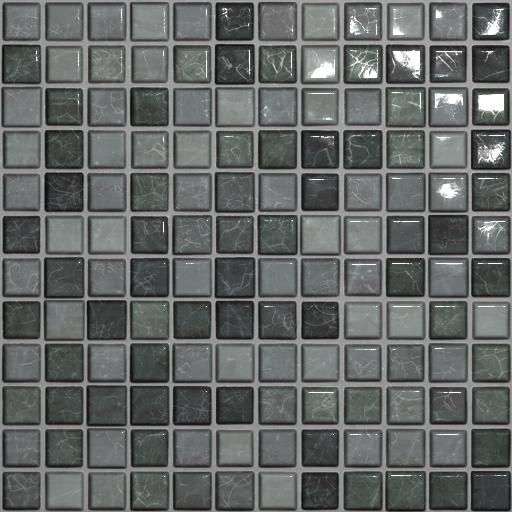} \\
		
		\raisebox{33pt}{\scalebox{1.0}{\rotatebox[origin=c]{90}{Ours (Proc.)}}} &
		\includegraphics[width=0.135\textwidth]{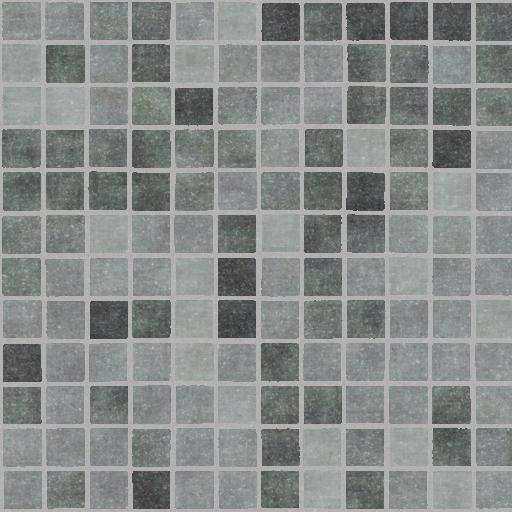} &
		\includegraphics[width=0.135\textwidth]{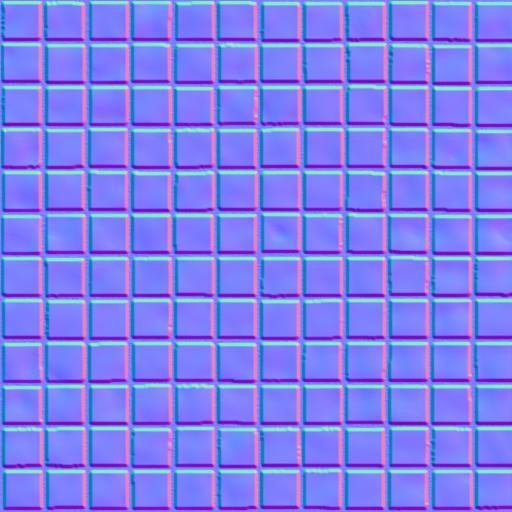} &
		\includegraphics[width=0.135\textwidth]{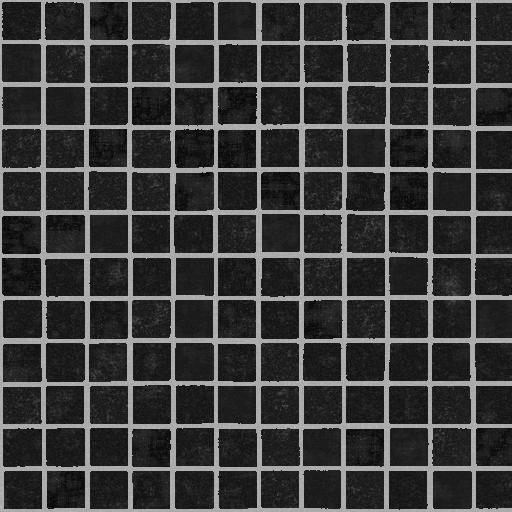} &
		\includegraphics[width=0.135\textwidth]{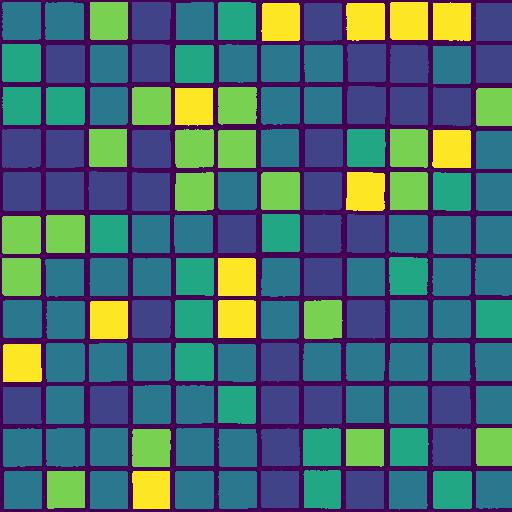} & 
		\includegraphics[width=0.135\textwidth]{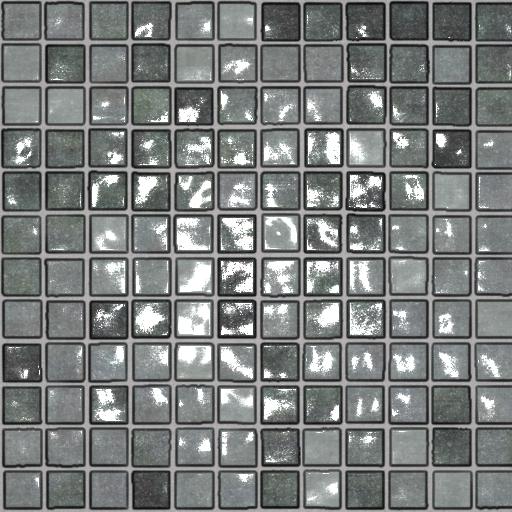} &
		\includegraphics[width=0.135\textwidth]{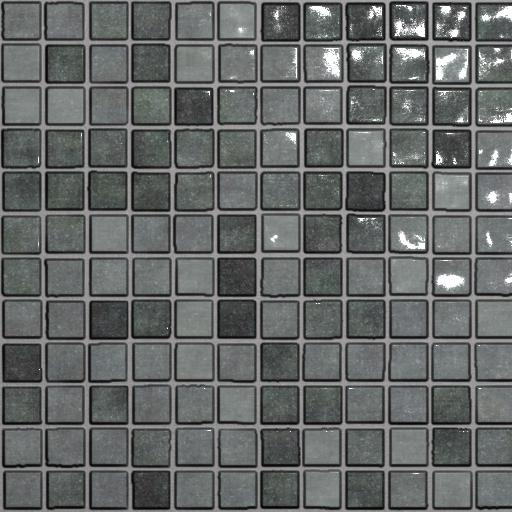} \\
	\end{tabular}
\vspace{-10pt}
\caption{Results of our method for different materials. We show that our pipeline can proceduralize a variety of materials, and  reproduce their global structures as well as local texture appearance. All of our results are entirely procedural. TL: Top lighting; SL: Side lighting. Please see our supplemental material for more results.}
\label{fig:main-results}
\vspace{-1mm}
\end{figure*}
\begin{figure*} %
	\centering
	\addtolength{\tabcolsep}{-3pt}
	\begin{tabular}{ccc @{} c @{} c @{} c @{} c @{} c @{} c}
		& Input & Albedo & Normal & Roughness & Label Map & Render (CTL) & Render (TSL) \\
			
		\raisebox{33pt}{\scalebox{0.7}{\rotatebox[origin=c]{90}{\cite{Guo20}}}} &
	    \includegraphics[width=0.135\textwidth]{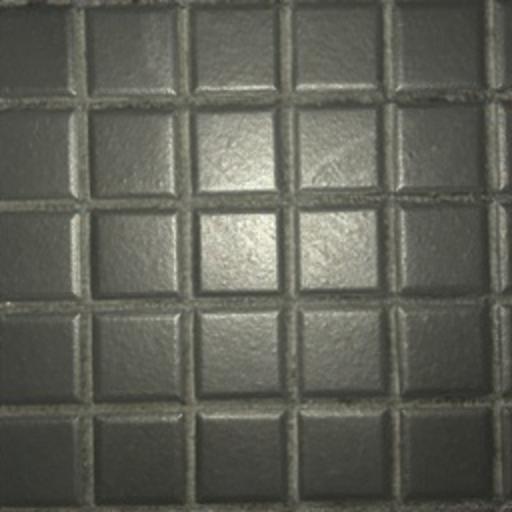} &
		\includegraphics[width=0.135\textwidth]{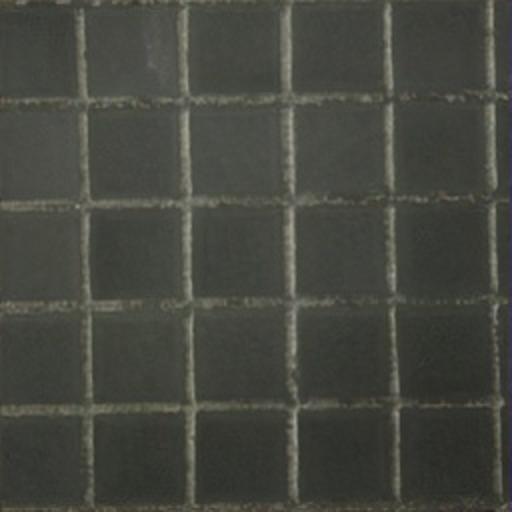} &
		\includegraphics[width=0.135\textwidth]{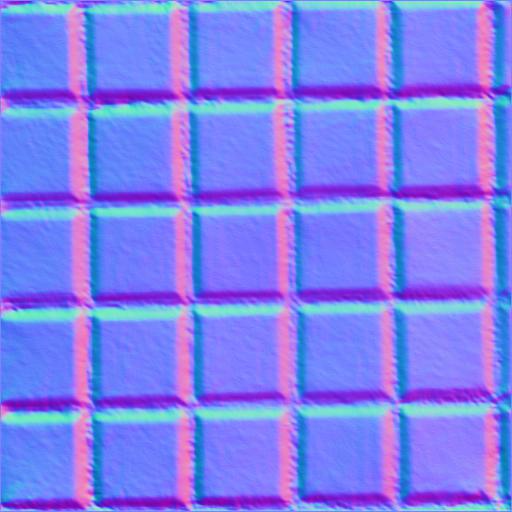} &
		\includegraphics[width=0.135\textwidth]{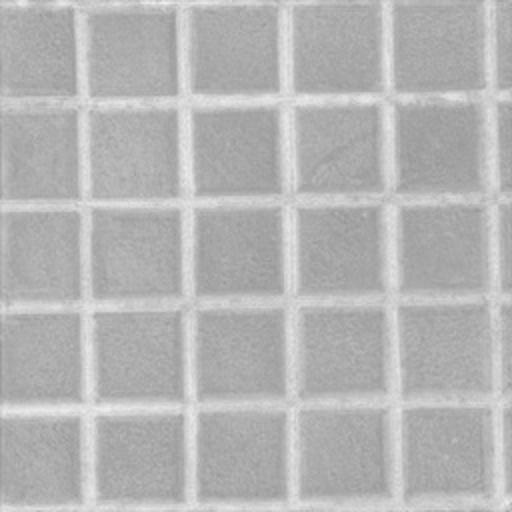} & 
		\includegraphics[width=0.135\textwidth]{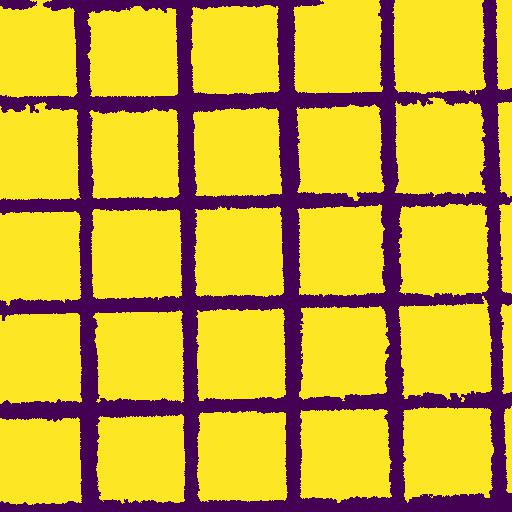} & 
		\includegraphics[width=0.135\textwidth]{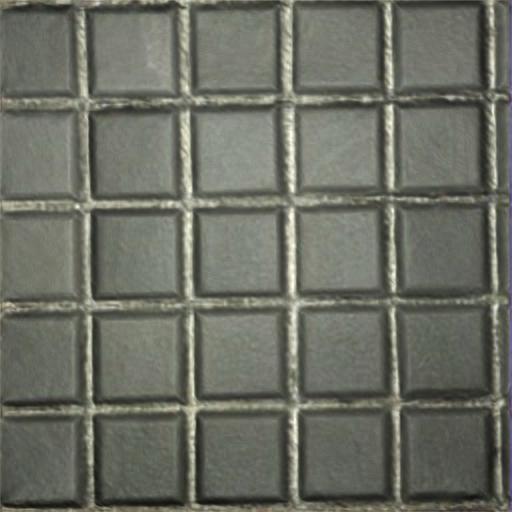} & 
		\includegraphics[width=0.135\textwidth]{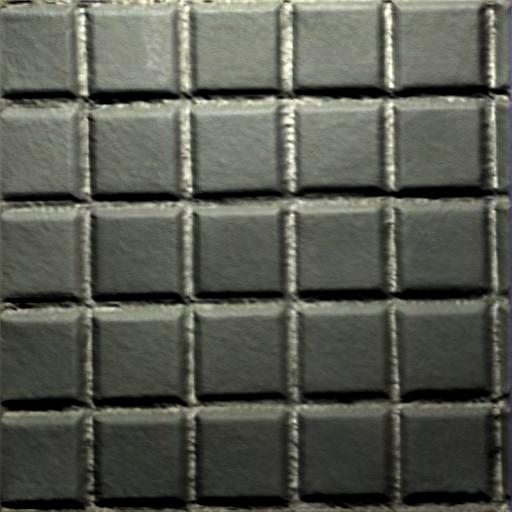} \\
		
		\raisebox{33pt}{\scalebox{0.7}{\rotatebox[origin=c]{90}{Ours}}} &
		&
		\includegraphics[width=0.135\textwidth]{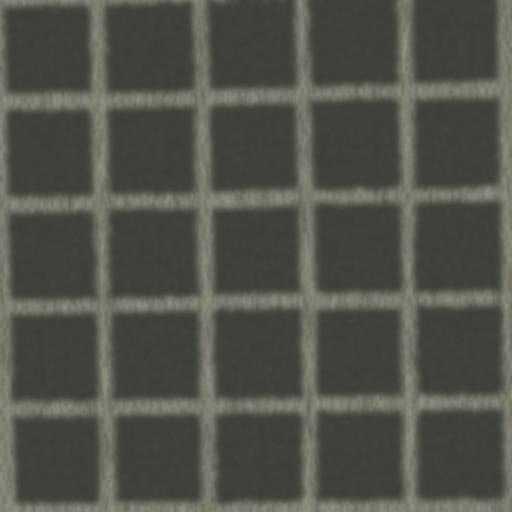} &
		\includegraphics[width=0.135\textwidth]{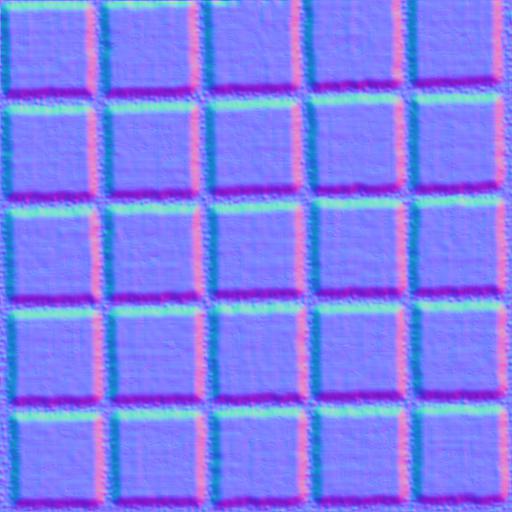} &
		\includegraphics[width=0.135\textwidth]{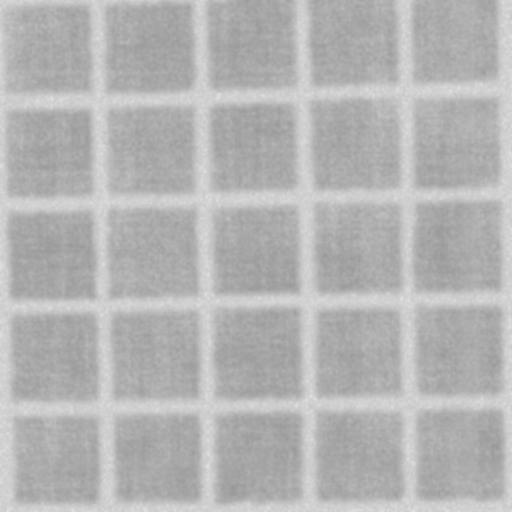} &
		\includegraphics[width=0.135\textwidth]{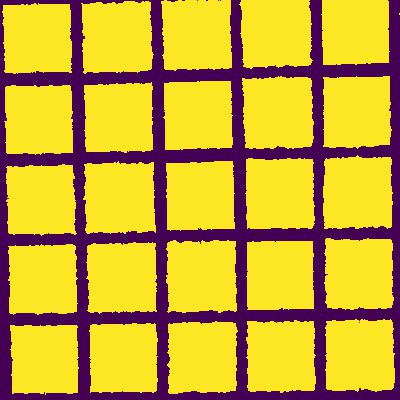} & 
		\includegraphics[width=0.135\textwidth]{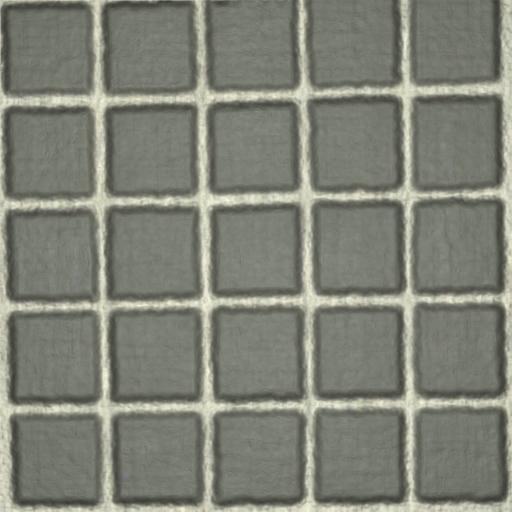} &
		\includegraphics[width=0.135\textwidth]{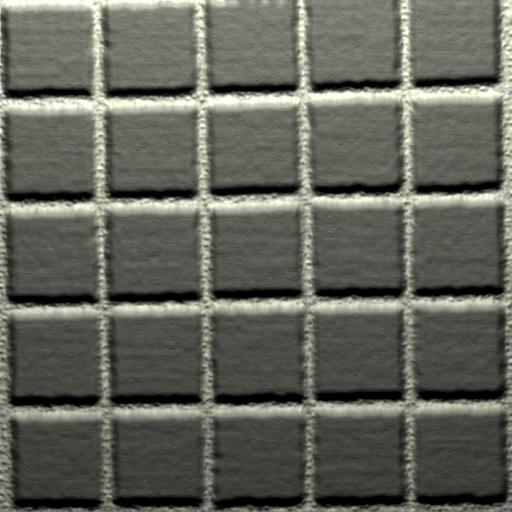} \\
		
		\raisebox{33pt}{\scalebox{0.7}{\rotatebox[origin=c]{90}{\cite{Deschaintre19}}}} &
	    \includegraphics[width=0.135\textwidth]{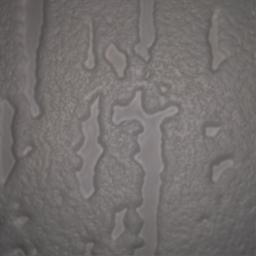} &
		\includegraphics[width=0.135\textwidth]{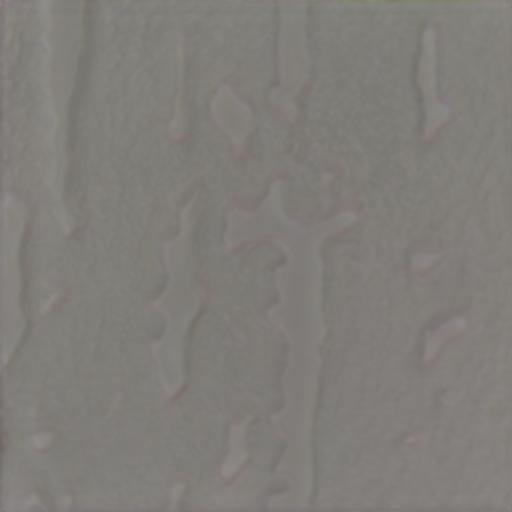} &
		\includegraphics[width=0.135\textwidth]{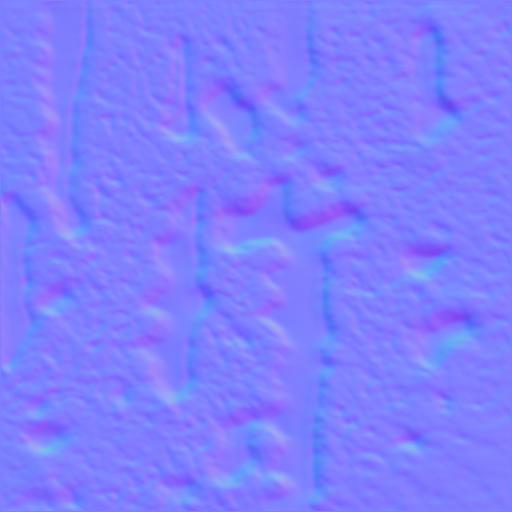} &
		\includegraphics[width=0.135\textwidth]{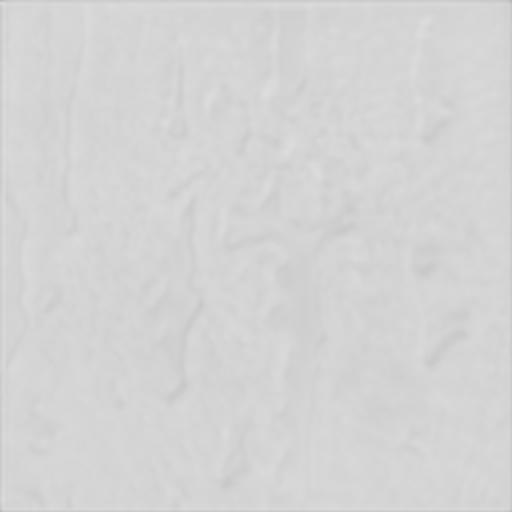} & 
		\includegraphics[width=0.135\textwidth]{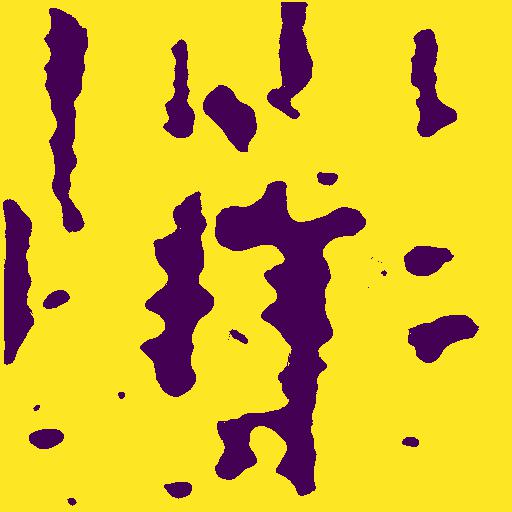} & 
		\includegraphics[width=0.135\textwidth]{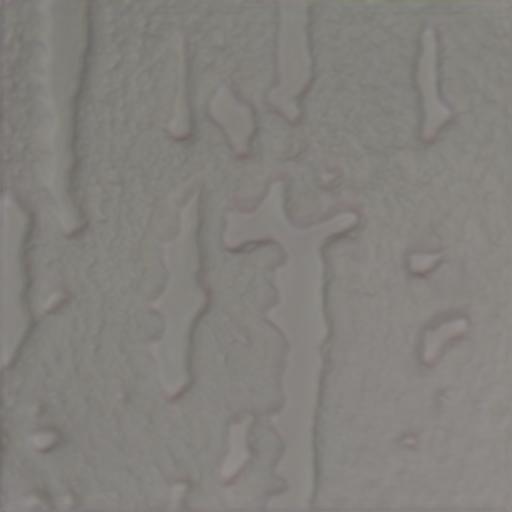} & 
		\includegraphics[width=0.135\textwidth]{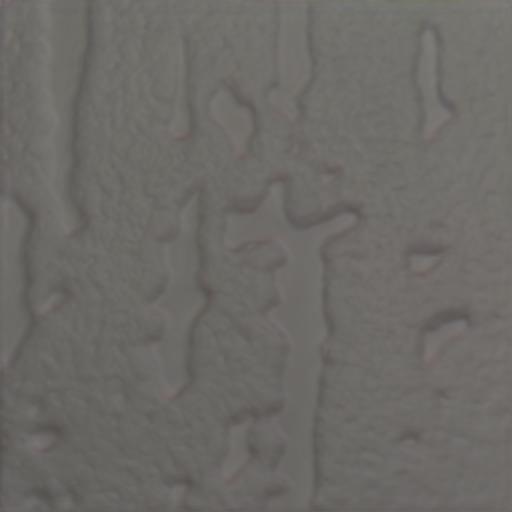} \\
		
		\raisebox{33pt}{\scalebox{0.7}{\rotatebox[origin=c]{90}{Ours}}} &
		&
		\includegraphics[width=0.135\textwidth]{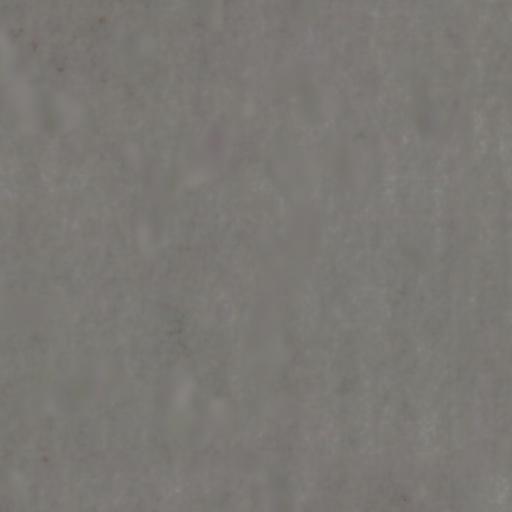} &
		\includegraphics[width=0.135\textwidth]{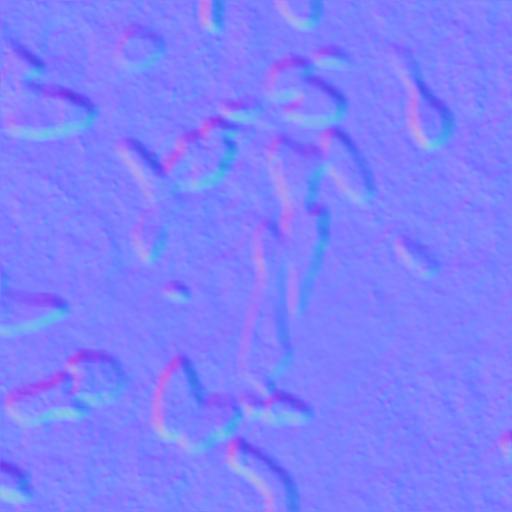} &
		\includegraphics[width=0.135\textwidth]{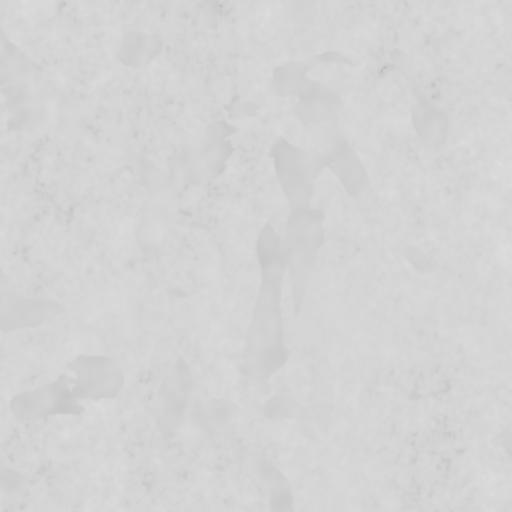} &
		\includegraphics[width=0.135\textwidth]{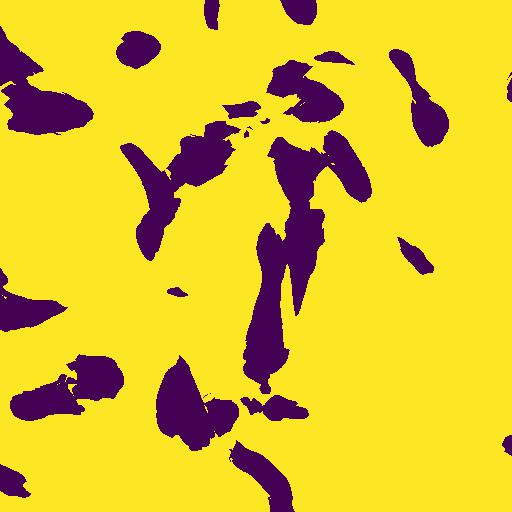} & 
		\includegraphics[width=0.135\textwidth]{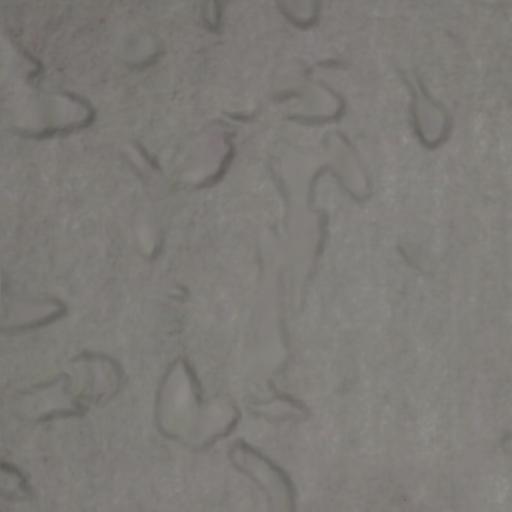} &
		\includegraphics[width=0.135\textwidth]{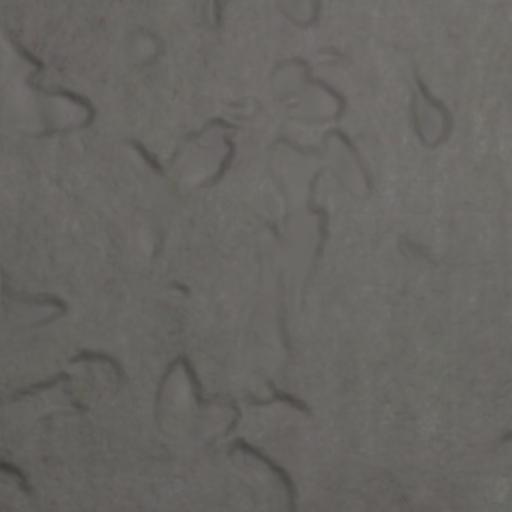} \\
		
	    \raisebox{33pt}{\scalebox{0.7}{\rotatebox[origin=c]{90}{\cite{DDB20}}}} &
	    \includegraphics[width=0.135\textwidth]{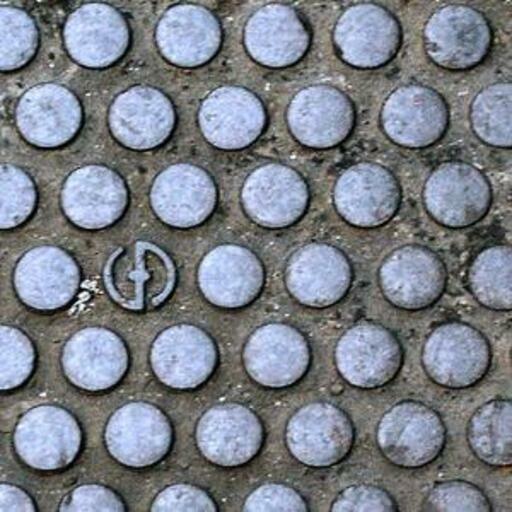} &
		\includegraphics[width=0.135\textwidth]{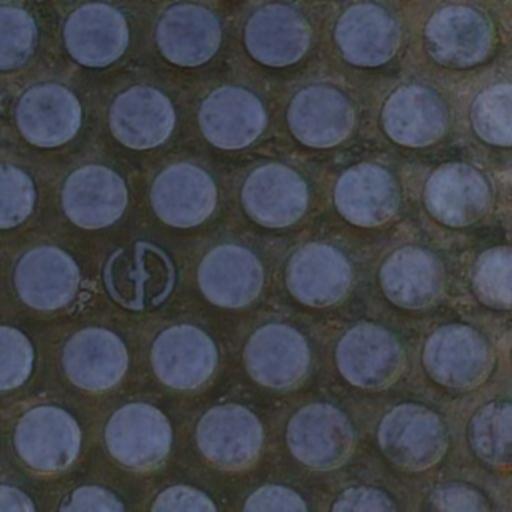} &
		\includegraphics[width=0.135\textwidth]{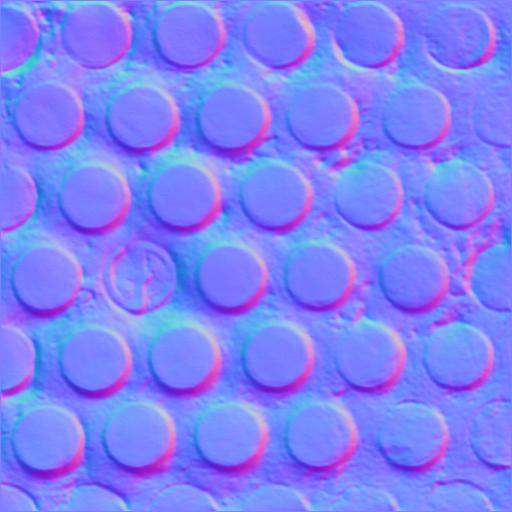} &
		\includegraphics[width=0.135\textwidth]{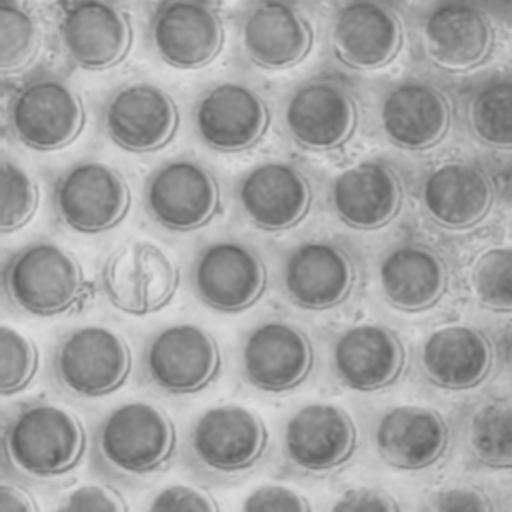} & 
		\includegraphics[width=0.135\textwidth]{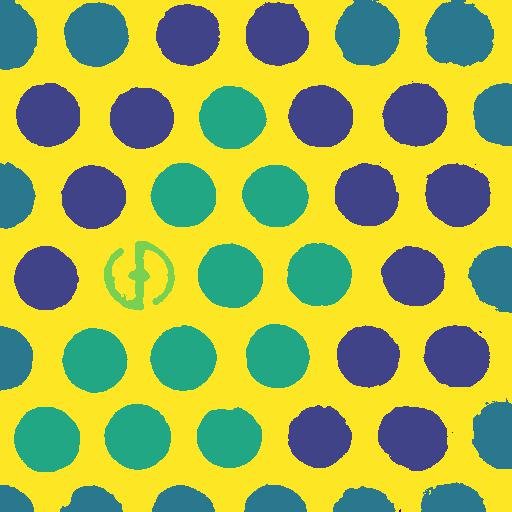} & 
		\includegraphics[width=0.135\textwidth]{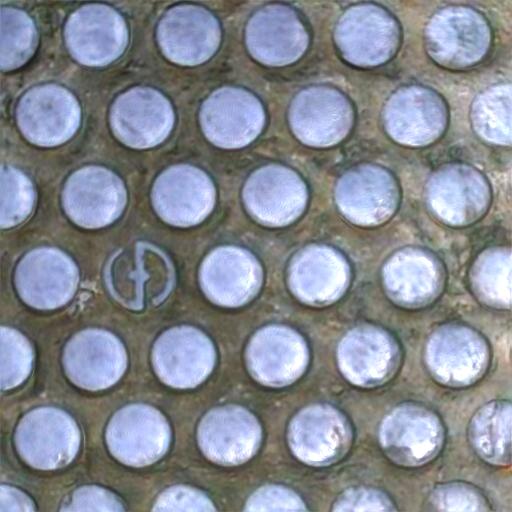} & 
		\includegraphics[width=0.135\textwidth]{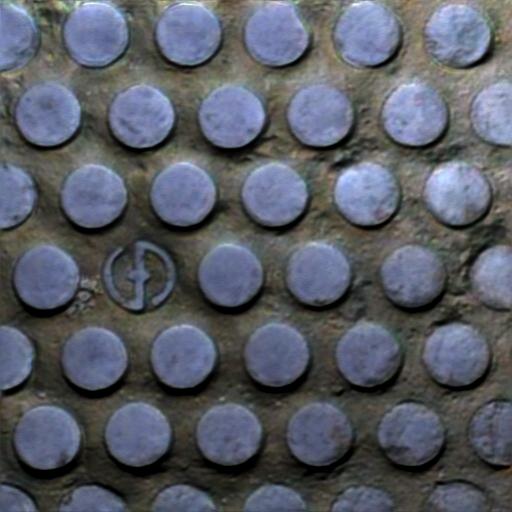} \\
		
		\raisebox{33pt}{\scalebox{0.7}{\rotatebox[origin=c]{90}{Ours}}} &
		&
		\includegraphics[width=0.135\textwidth]{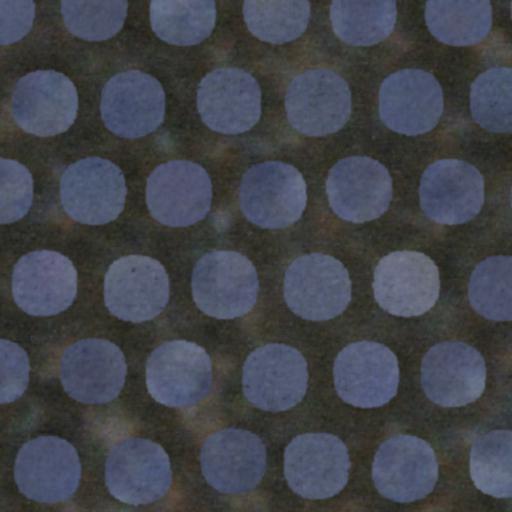} &
		\includegraphics[width=0.135\textwidth]{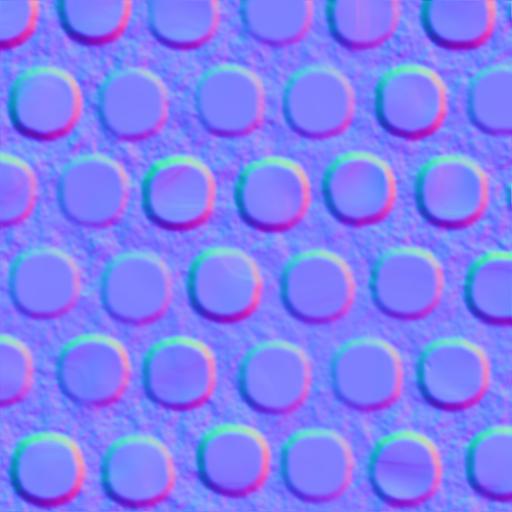} &
		\includegraphics[width=0.135\textwidth]{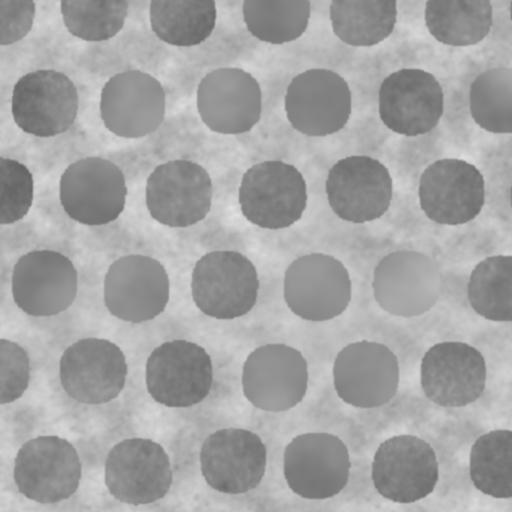} &
		\includegraphics[width=0.135\textwidth]{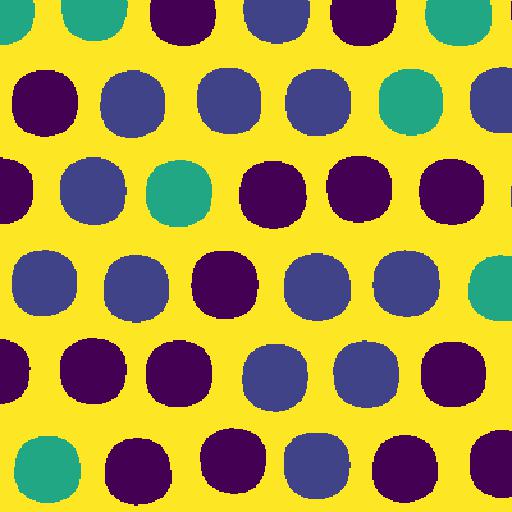} & 
		\includegraphics[width=0.135\textwidth]{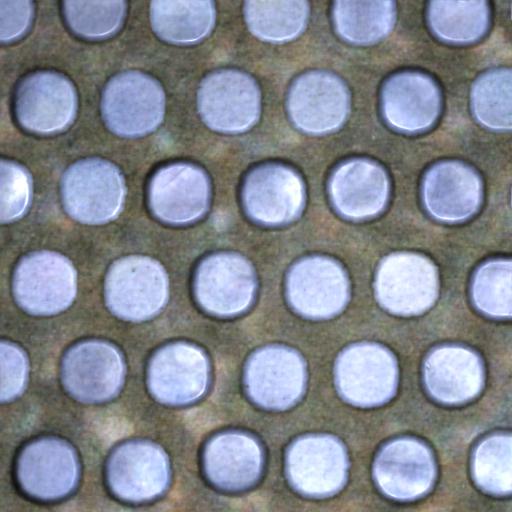} &
		\includegraphics[width=0.135\textwidth]{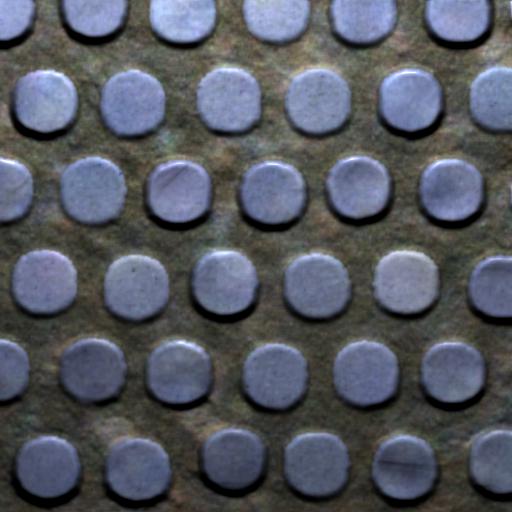} \\
		
		\raisebox{33pt}{\scalebox{0.7}{\rotatebox[origin=c]{90}{\cite{DDB20}}}} &
	    \includegraphics[width=0.135\textwidth]{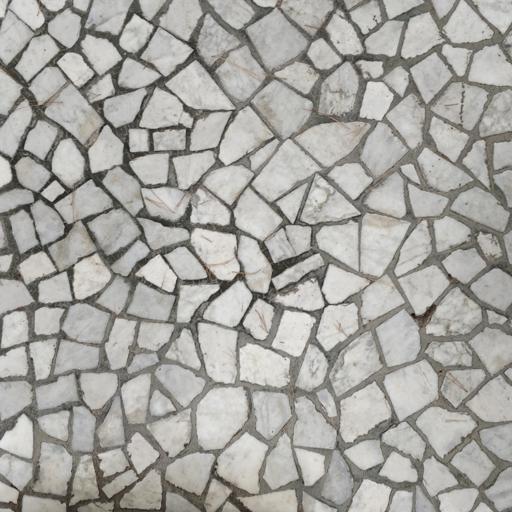} &
		\includegraphics[width=0.135\textwidth]{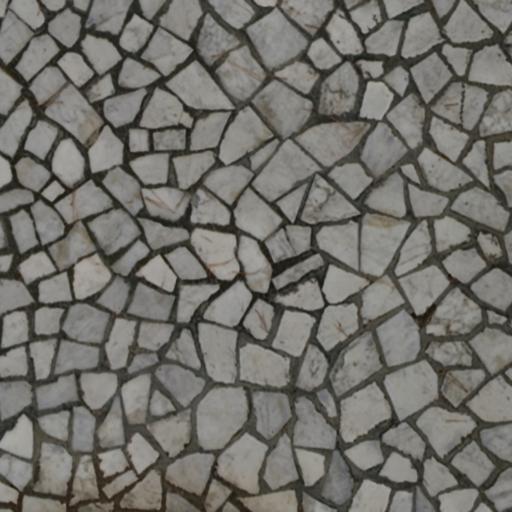} &
		\includegraphics[width=0.135\textwidth]{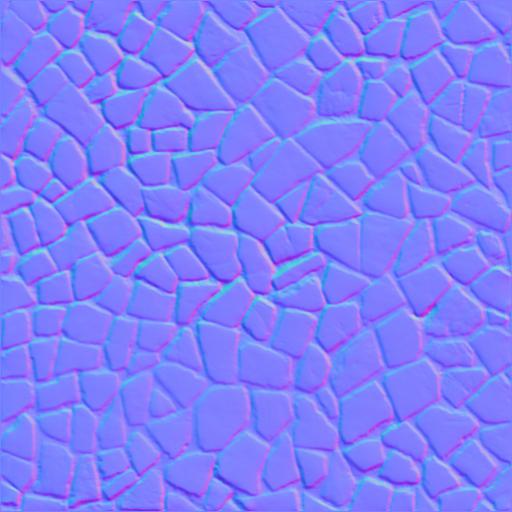} &
		\includegraphics[width=0.135\textwidth]{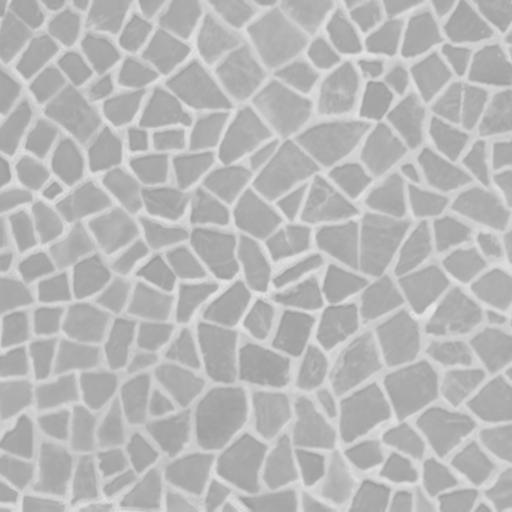} & 
		\includegraphics[width=0.135\textwidth]{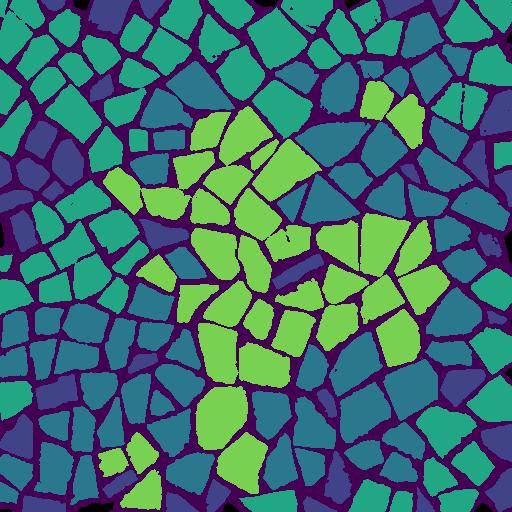} & 
		\includegraphics[width=0.135\textwidth]{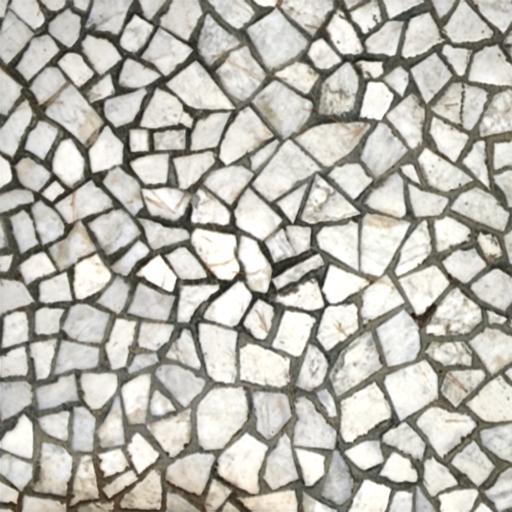} & 
		\includegraphics[width=0.135\textwidth]{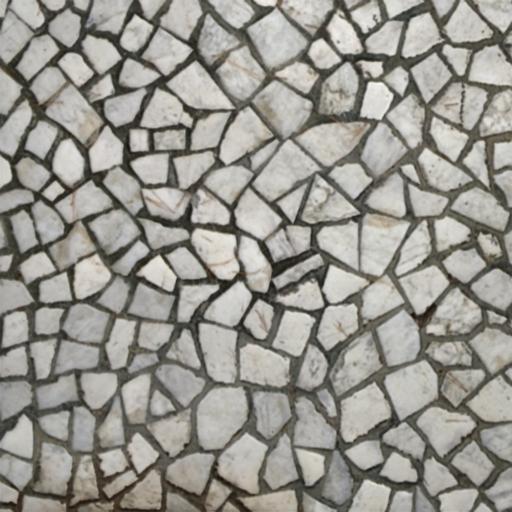} \\
		
		\raisebox{33pt}{\scalebox{0.7}{\rotatebox[origin=c]{90}{Ours}}} &
		&
		\includegraphics[width=0.135\textwidth]{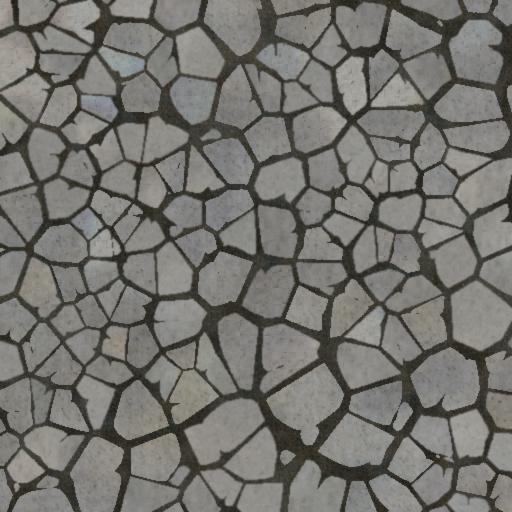} &
		\includegraphics[width=0.135\textwidth]{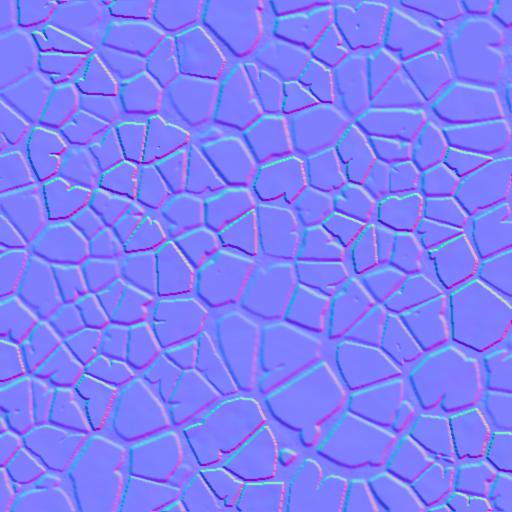} &
		\includegraphics[width=0.135\textwidth]{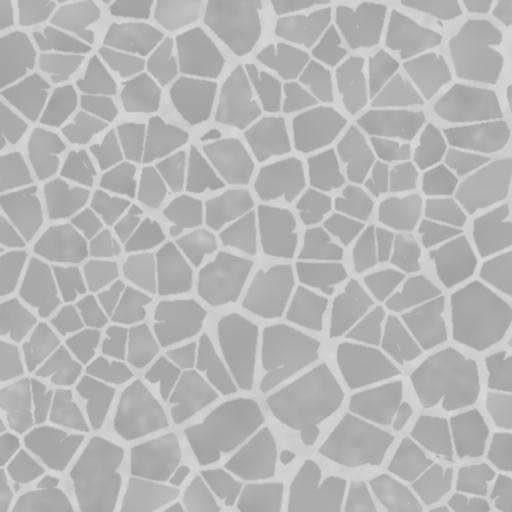} &
		\includegraphics[width=0.135\textwidth]{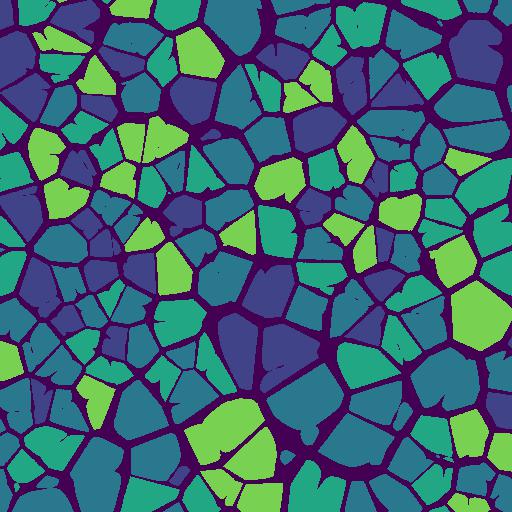} & 
		\includegraphics[width=0.135\textwidth]{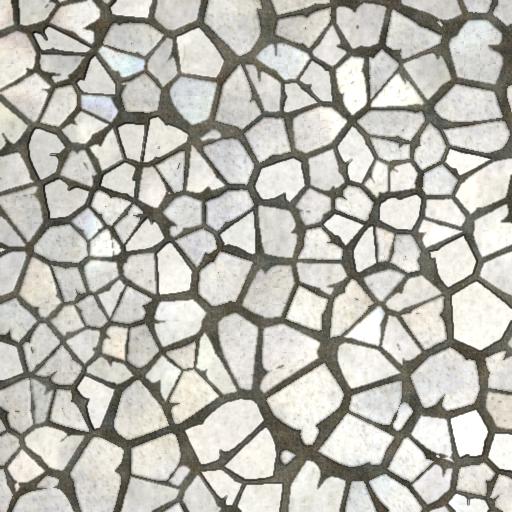} &
		\includegraphics[width=0.135\textwidth]{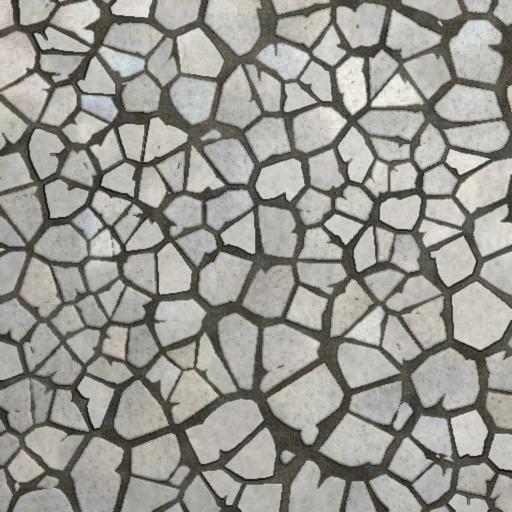} \\
	\end{tabular}
\vspace{-10pt}
\caption{Picture(s) as input. We apply our method on decomposed SVBRDF maps captured with a single or multiple image(s) with flash lighting~\cite{Deschaintre19, DDB20, Guo20}, showing the diversity of material source we can handle. Our procedural results enable further creation and provide more regularity and remove the baked in lighting and irregularities due to the few image(s) capture method. \new{The bottom two examples were segmented using the instance segmentation approach showing that it can also handle irregular patterns.} CTL: Central Top light; TSL: Top Side light. Please see our supplemental material for more results.}
\label{fig:image-input-results}
\end{figure*}
\begin{figure*} %
	\centering
	\addtolength{\tabcolsep}{-4pt}
	\begin{tabular}{ccc}
		\includegraphics[width=0.5\textwidth]{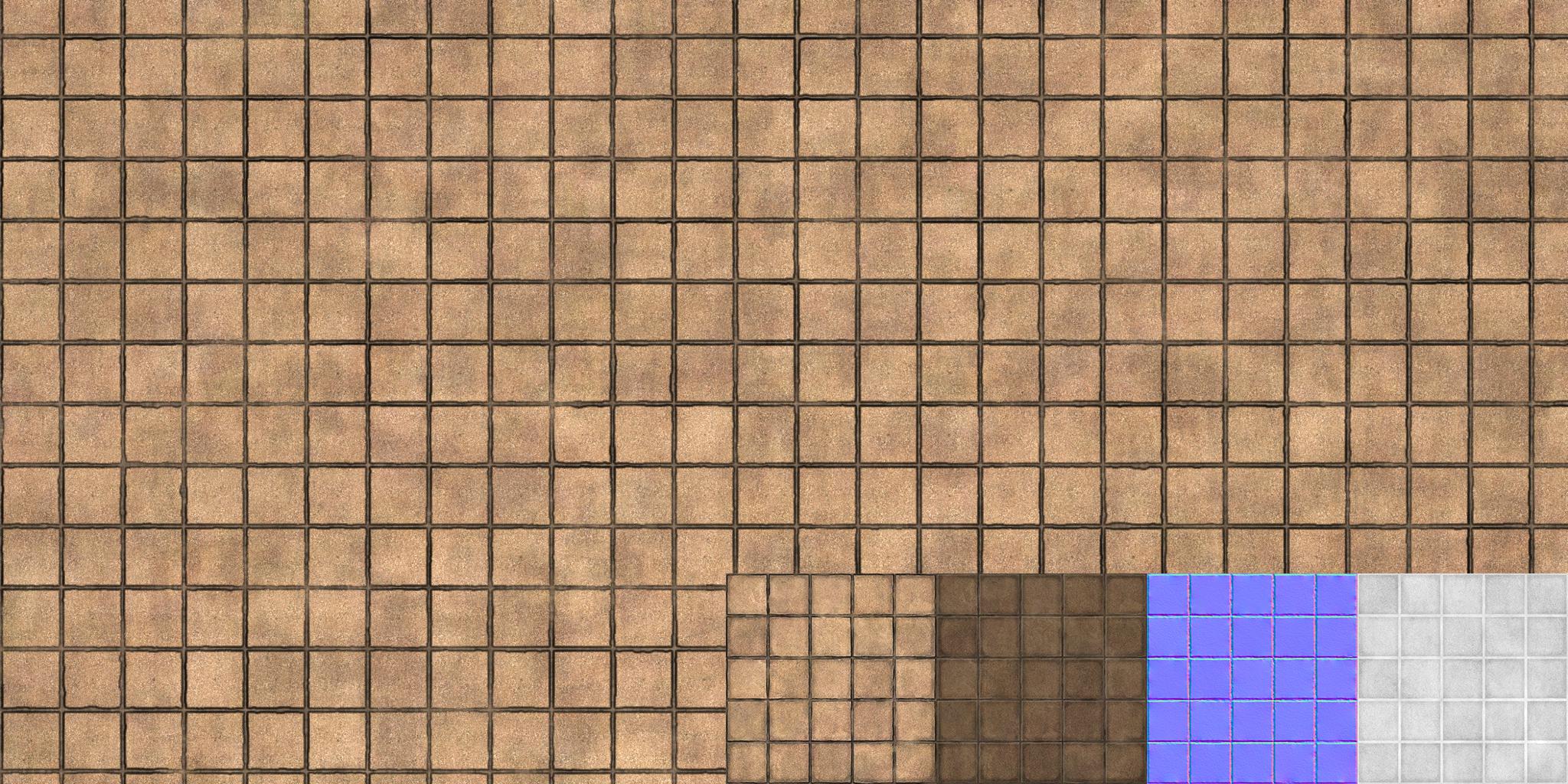} &
		\includegraphics[width=0.5\textwidth]{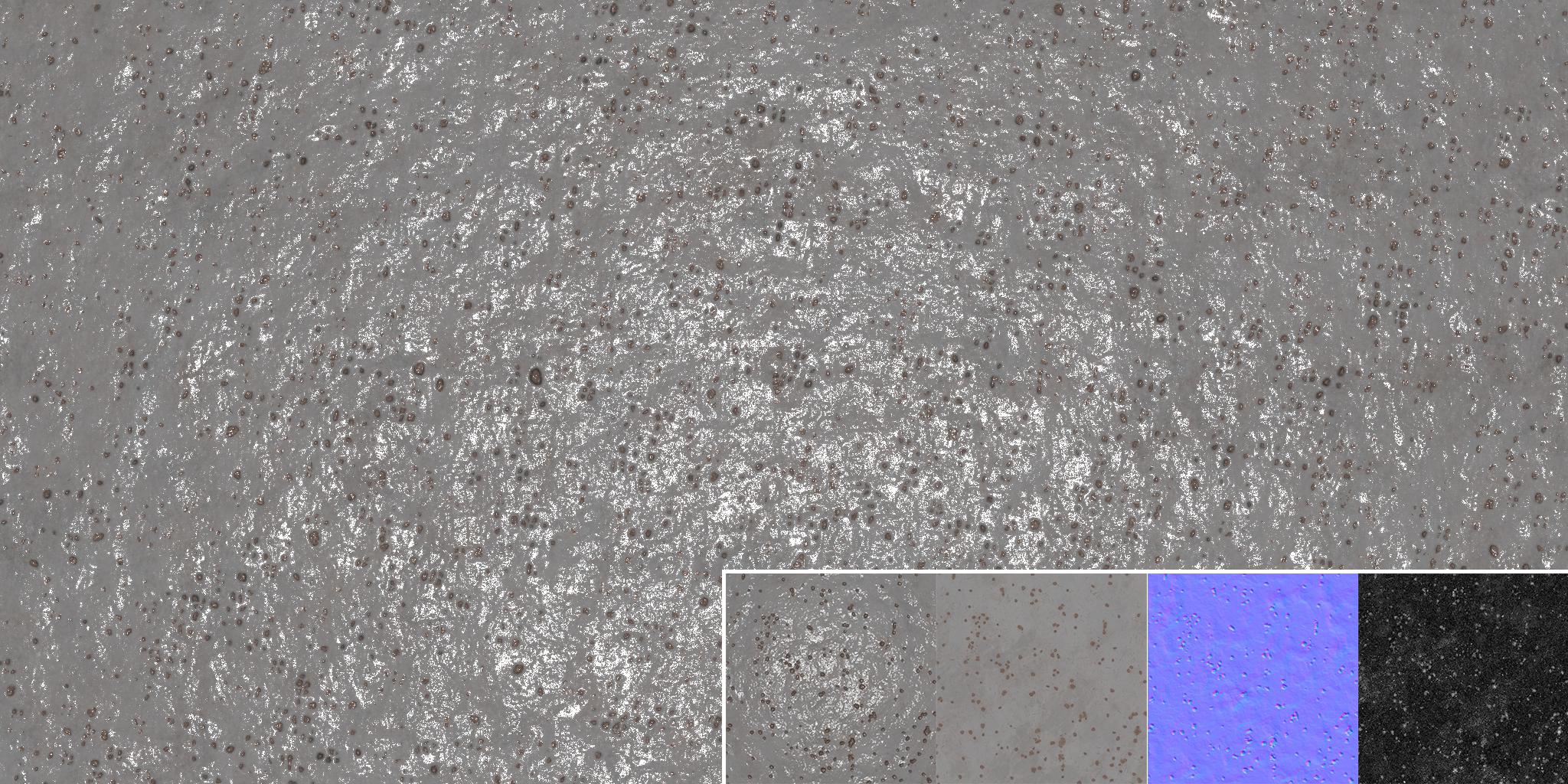}
	\end{tabular}
\vspace{-10pt}
\caption{High-resolution material synthesis. We show the input material in the bottom right corner and render a high resolution ($2048 \times 1024$) version generated with our method's results with lighting from the top. By applying our pipeline, we convert these pixel-based material maps to procedural representation thus enabling arbitrary scale and resolution expansion without repetition.}
\label{fig:high-res}
\end{figure*}
\section{Results}
We demonstrate our pipeline with a variety of materials as inputs. Each material is defined by a set of SVBRDF maps including albedo map, normal map and roughness map -- these are the maps our input materials use, but our method can adapt to any additional or different gray-scale or color maps.

Our pipeline is implemented in Python and partly relies on Matlab. Differentiable optimization of our material graphs (Sec.~\ref{sec:recomposition}) is implemented in PyTorch. We adopt an L-BFGS-B optimizer to optimize our material graph with a learning rate of 0.005. It takes around 200 steps -- 2 minutes on a Nvidia RTX 2070 Super GPU with a CPU of Intel Core i7-9700 -- to converge. A complete inverse material modeling process takes less than 5 minutes for user interaction and about 20 minutes for computing, depending on the complexity of the input material. The typical computation time is: 2 minutes for spectrum-aware matting; 13 minutes for procedural mask query (1s) and optimization (depending on initialization quality); 3 minutes for multi-layer noise modeling and 2 minutes for optimization-based recomposition. We provide this time for reference but highlight that our current implementation is not optimized. 

\begin{figure} 
	\centering
	\addtolength{\tabcolsep}{-4pt}
	\begin{tabular}{cccccc}
		\includegraphics[width=0.09\textwidth]{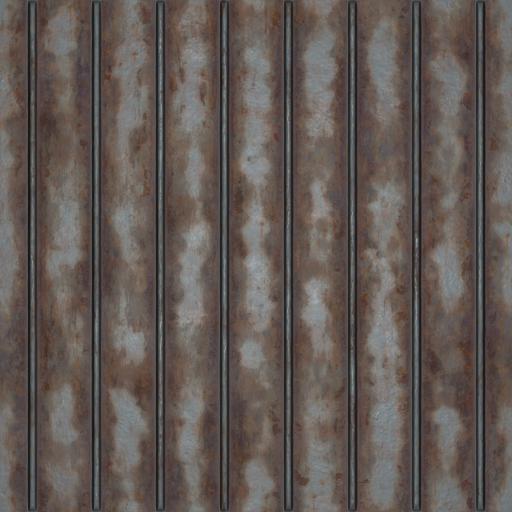} &
		\includegraphics[width=0.09\textwidth]{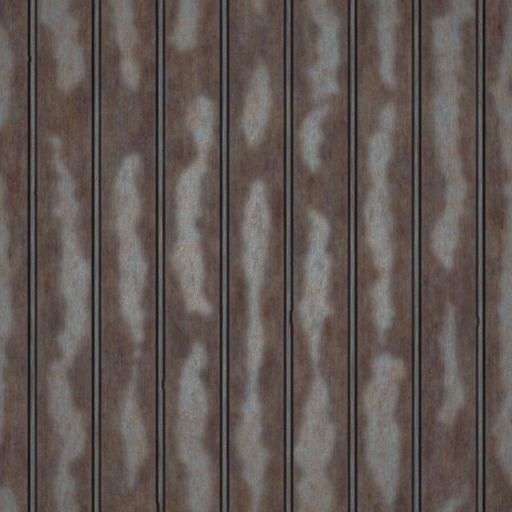} &
		\includegraphics[width=0.09\textwidth]{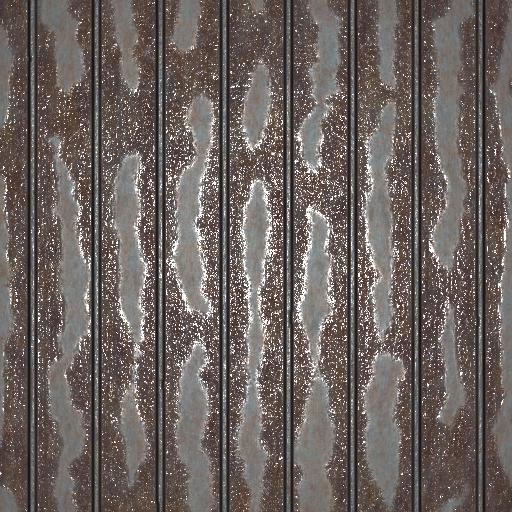} &
		\includegraphics[width=0.09\textwidth]{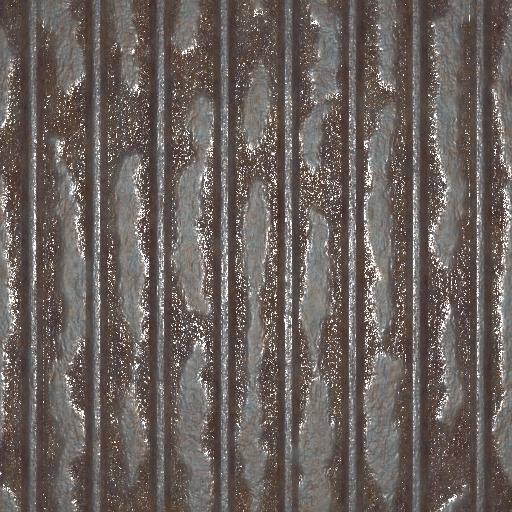} &
		\includegraphics[width=0.09\textwidth]{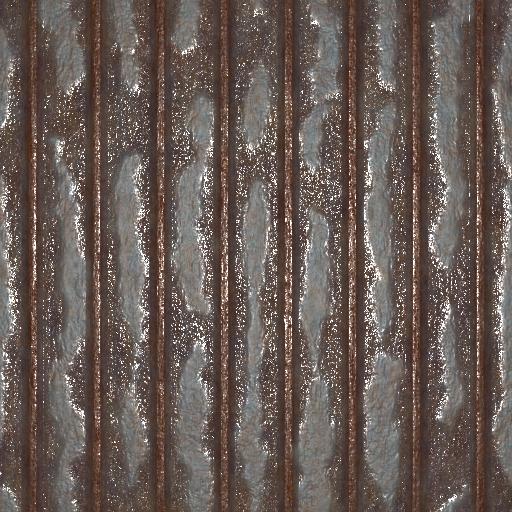} \\
		\includegraphics[width=0.09\textwidth]{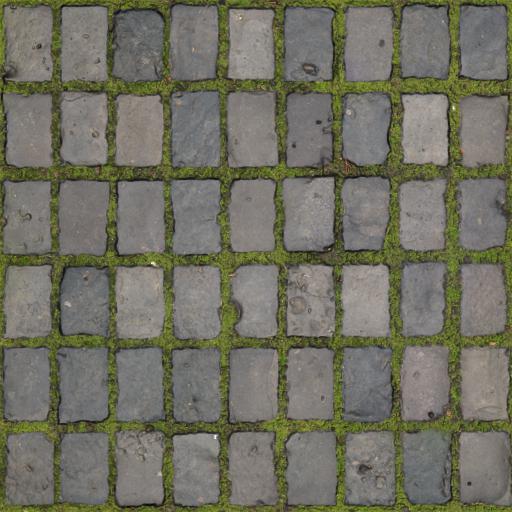} &
		\includegraphics[width=0.09\textwidth]{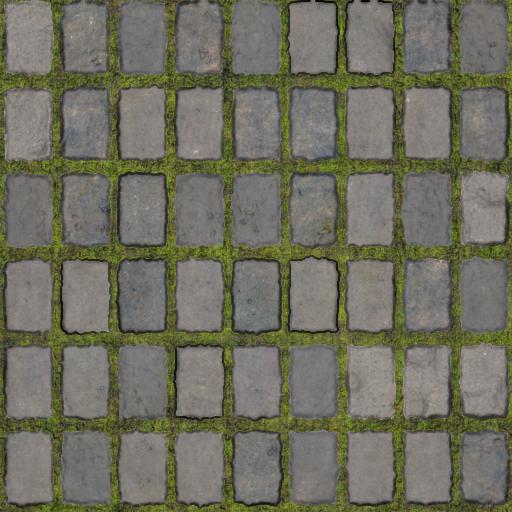} &
		\includegraphics[width=0.09\textwidth]{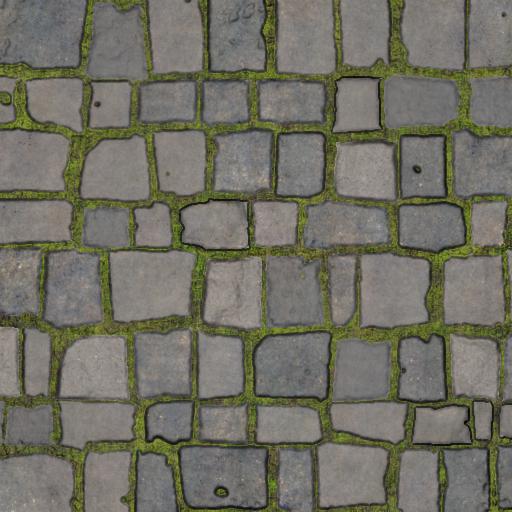} &
		\includegraphics[width=0.09\textwidth]{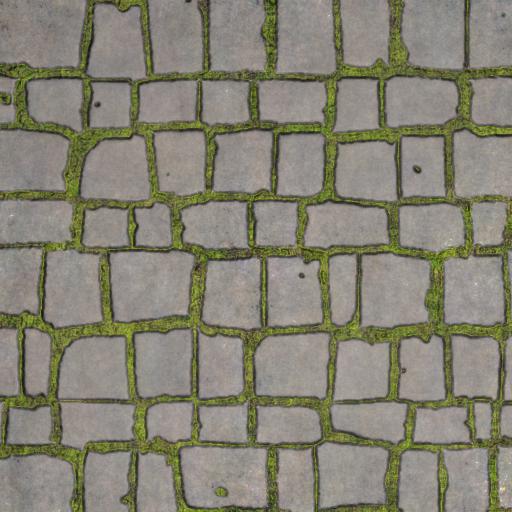} &
		\includegraphics[width=0.09\textwidth]{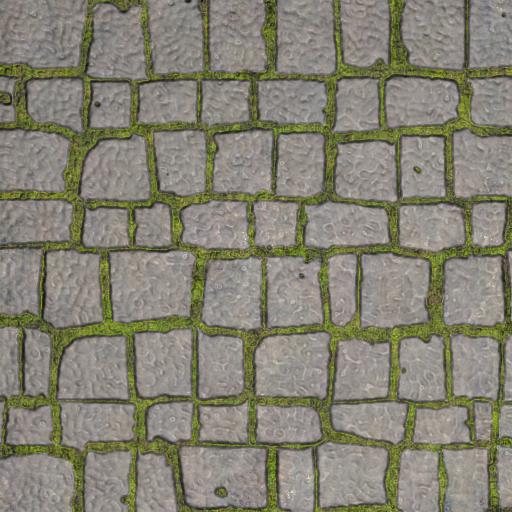} \\
		\includegraphics[width=0.09\textwidth]{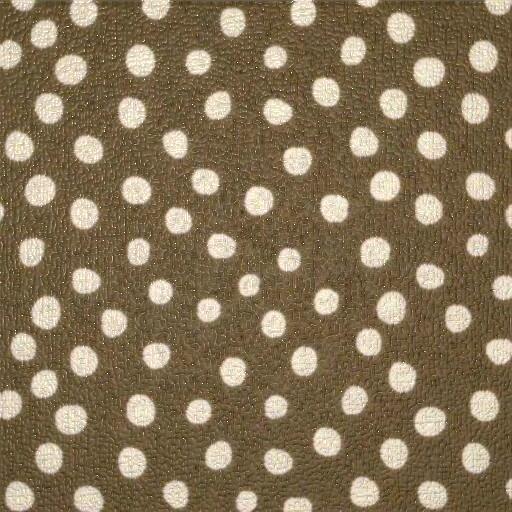} &
		\includegraphics[width=0.09\textwidth]{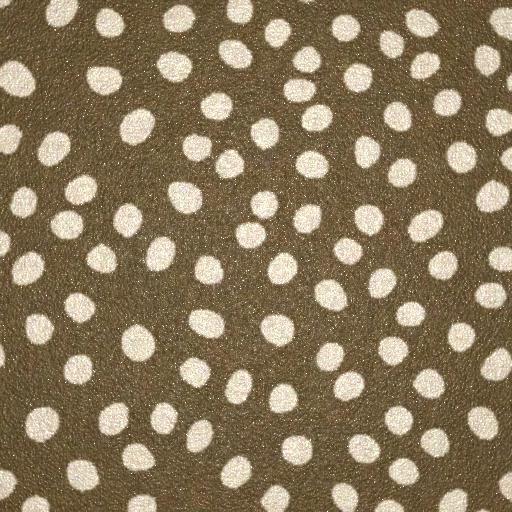} &
		\includegraphics[width=0.09\textwidth]{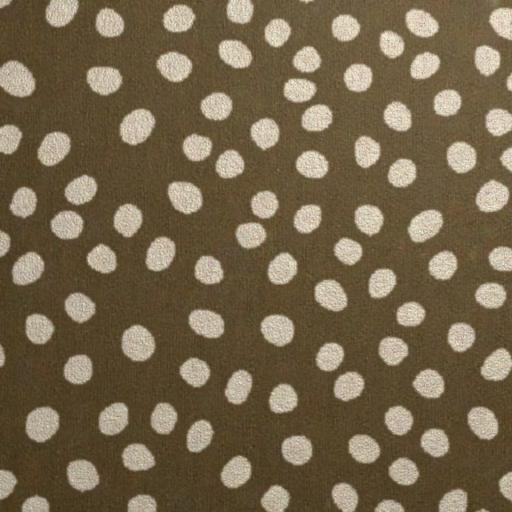} &
		\includegraphics[width=0.09\textwidth]{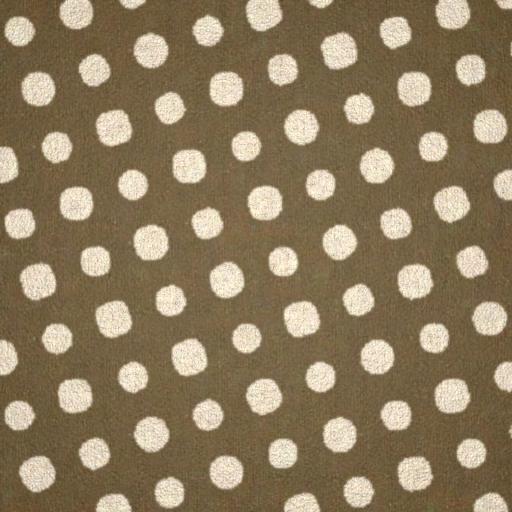} &
		\includegraphics[width=0.09\textwidth]{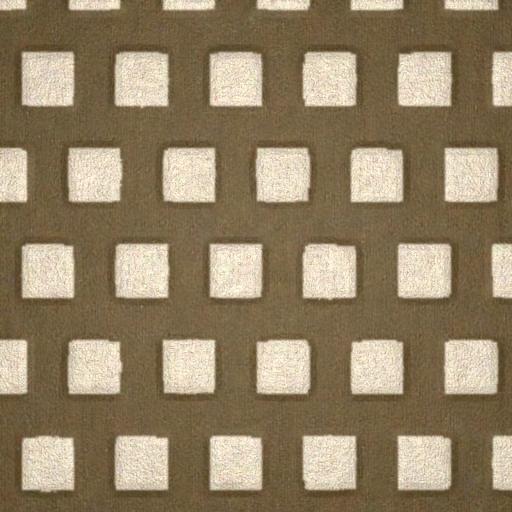} \\
		Original & 
		Ours Proc. & 
		\multicolumn{3}{c}{Edit Sequence}
	\end{tabular}
\vspace{-10pt}
\caption{Procedural material editing. Our procedural material model provides editability to users. In the top example, we edit the optimizable filter parameters to change the (1) roughness (2) normal and (3) color of the material sequentially; for the second example, we edit parameters in the (1) mask generators (first two edits) to change spatial arrangement and distribution of sub-materials; and (2) noise generators (last edit) to change the fine-scale normals on each tile; we edit all parameters for the third example where (1) high-frequency shininess is removed; (2) global structures are changed (last two edits); and (3) normals are enhanced, leading to a completely different leathered brick material. Please zoom-in to see fine-scale controls.}
\label{fig:editing}
\end{figure}

Fig.~\ref{fig:main-results} shows our inverse procedural modeling results on different materials. For each example, inputs are albedo map, normal map and roughness map only. We decompose these material maps using our hierarchical segmentation methods and visualize the computed labeled mask maps. Given material maps and computed hierarchical mask maps, we generate procedural materials and output albedo, normal and roughness maps (Sec. \ref{sec: mat-proc}). The results show that our pipeline can reproduce well a variety of stochastic and regular materials. We see that our method recovers both large scale patterns and fine-grained details thanks to our sub-material decomposition approach. The general structure and texture appearance are not perfectly registered with the original input because our model is a procedural approximation of it. We aim at reproducing its appearance rather than a pixel-perfect match~\cite{Deschaintre18}, allowing us to preserve the material global appearance while sampling a new realization of it or editing it, as shown in Fig.~\ref{fig:editing}.
\subsection{Natural Images as Input}
Although our pipeline was designed to take material maps as input, it can also work on natural and flash images, benefiting from existing material acquisition methods~\cite{Deschaintre18, Deschaintre19, DDB20, Gao19, Guo20}. Providing a set of captured images, we first apply a material acquisition method to generate material maps, and use the generated SVBRDF maps with our method. Fig. \ref{fig:image-input-results} shows examples of picture(s) as input. We convert the input natural image(s) to SVBRDF maps using state-of-the-art material acquisition methods \cite{Deschaintre19, DDB20, Guo20}.As opposed to "artist-designed" SVBRDF maps or manually post-processed SVBRDF maps, automatically generated SVBRDF maps are not perfectly clean. Different maps can be noisy and exhibit shading and color variations due to incomplete lighting removal. Normal maps in particular can represent strong height variations even within a single sub-material. This makes it challenging to recover an exact match to the input image. Furthermore, irregularities often occur in acquired SVBRDF which should be regular. We show in Fig.~\ref{fig:image-input-results} that our pipeline is capable of matching acquired SVBRDF overall appearance, and also of enforcing better regularity.

\begin{figure} 
	\centering
	\addtolength{\tabcolsep}{-3pt}
    \begin{tabular}{ccll}
    \includegraphics[width=0.14\textwidth]{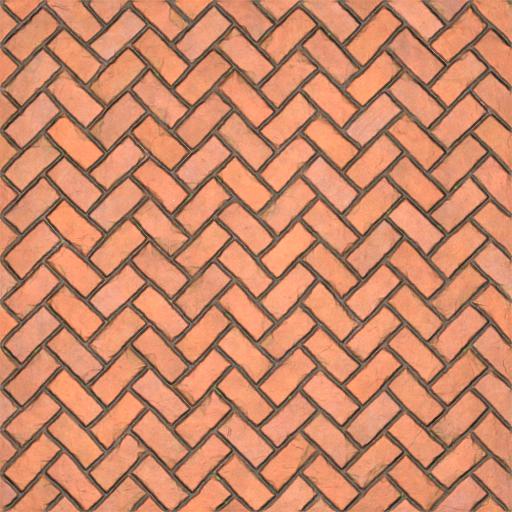} & \multicolumn{3}{c}{} \\
    Input & \multicolumn{3}{c}{}                   \\
    \includegraphics[width=0.14\textwidth]{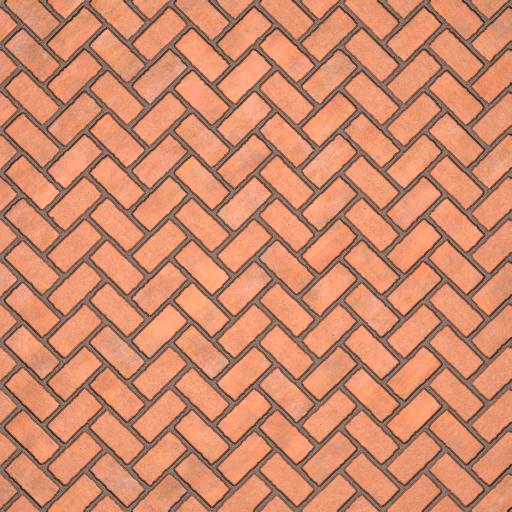} & \multicolumn{3}{c}{\multirow[t]{3}{*}{\includegraphics[width=0.308\textwidth]{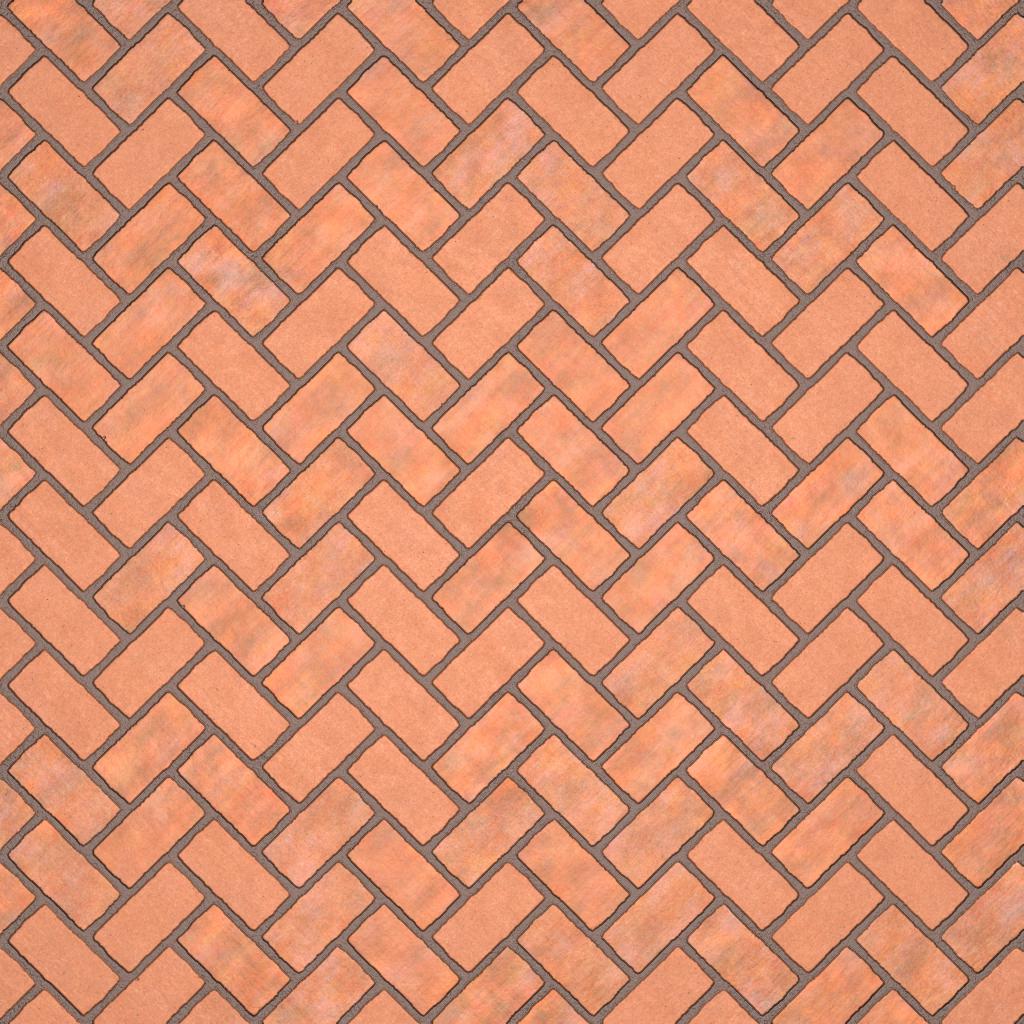}}}                   \\
    Ours & \multicolumn{3}{c}{$2\times$ upscale}                 
    \end{tabular}
    \vspace{-10pt}
\caption{Material super-resolution. We show our model can generate higher-resolution textures but without changing its scale. This is achieved by only procedurally upscaling mask maps which controls the global spatial distributions of sub-materials.}
\label{fig:superres}
\end{figure}
\subsection{Application}
\subsubsection{Procedural Material Editing}
Once proceduralized using our method, materials can be further edited. Fig. \ref{fig:editing} shows examples of operations that we can apply on our results. We can freely edit the way mask maps and noise models are combined, the large scale structures (e.g patterns and distributions) and fine-scale details (e.g. fine normals and roughness). Once generated, our procedural model allows users to get interactive feedback on each of their edits. As opposed to \cite{Shi20, hu2019}, we do not rely on pre-defined material graph, allowing artists to use our method to generate a small tune-able and extendable material graph to start a new design. Furthermore, Hu et al.~\shortcite{hu2019} relied on a style transfer like post-process step to better match the input texture appearance, limiting the editability of their final results.

\subsubsection{High-resolution Material Generation}
As a fully procedural model, materials converted by our pipeline can be expanded to arbitrary size. Global structures can be reproduced and extended using the procedural mask map with PPTBF and random sampling, while the appearance of sub-materials can be losslessly synthesized to higher resolution using procedural noise models. Fig. \ref{fig:high-res} shows examples of high-resolution material synthesis. The resolution of the result material and rendered images is $2048\times1024$. We also show an example of material super-resolution in Fig. \ref{fig:superres} where the resolution of the input SVBRDF is $512\times512$ and we double its size in each dimension by procedurally upsampling its mask maps. Such operation will not affect its global scale, providing a super-resolved material rather than synthesize a larger scale one as demonstrated in Fig.~\ref{fig:high-res}. More high-resolution materials generated by our method are shown in the supplemental material.

\begin{figure*} 
	\centering
	\addtolength{\tabcolsep}{-4pt}
	\begin{tabular}{ccccccc}
		\includegraphics[width=0.09\textwidth]{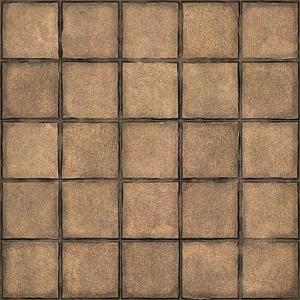} &
		\includegraphics[width=0.15\textwidth]{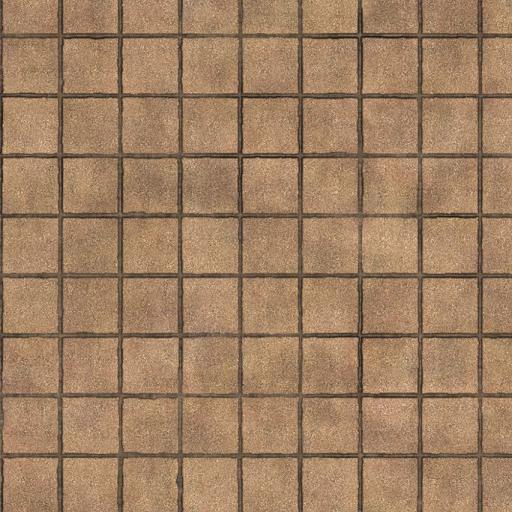}&
		\includegraphics[width=0.15\textwidth]{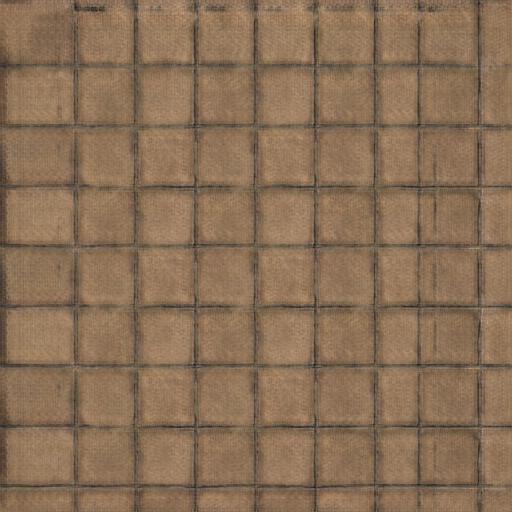} &
		\includegraphics[width=0.15\textwidth]{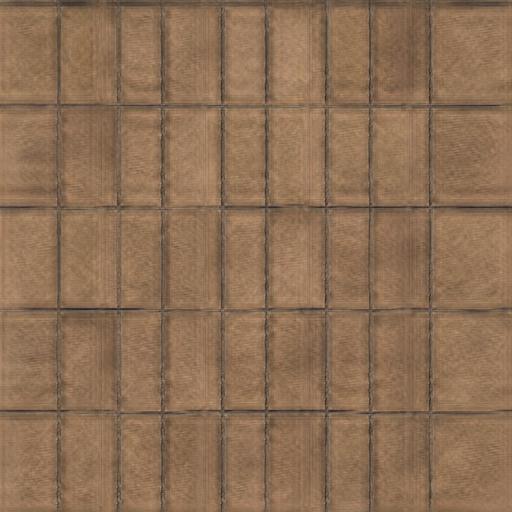} &
		\includegraphics[width=0.15\textwidth]{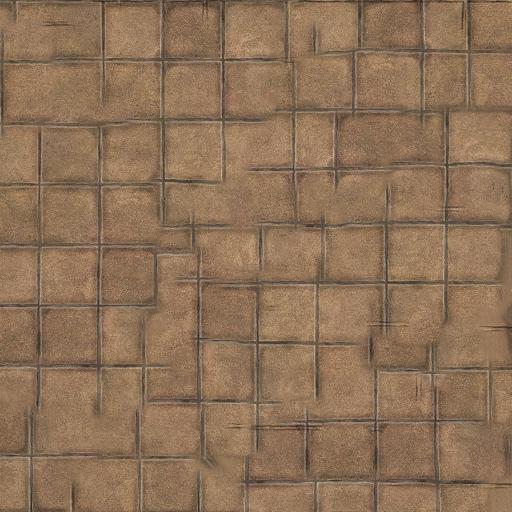} &
		\includegraphics[width=0.15\textwidth]{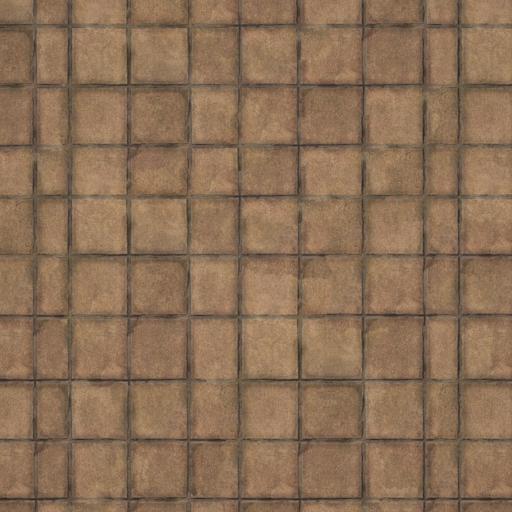} \\
		\includegraphics[width=0.09\textwidth]{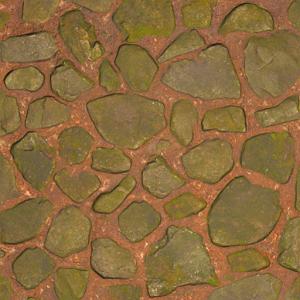}&
		\includegraphics[width=0.15\textwidth]{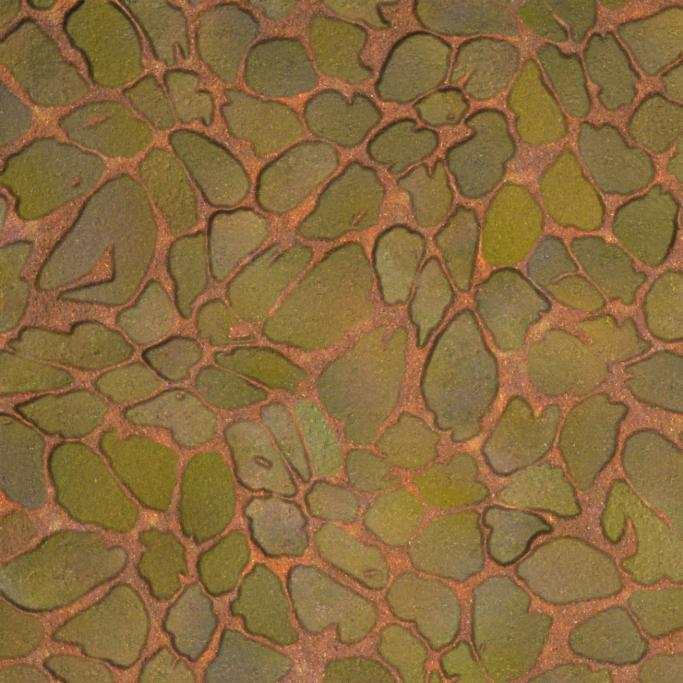}&
		\includegraphics[width=0.15\textwidth]{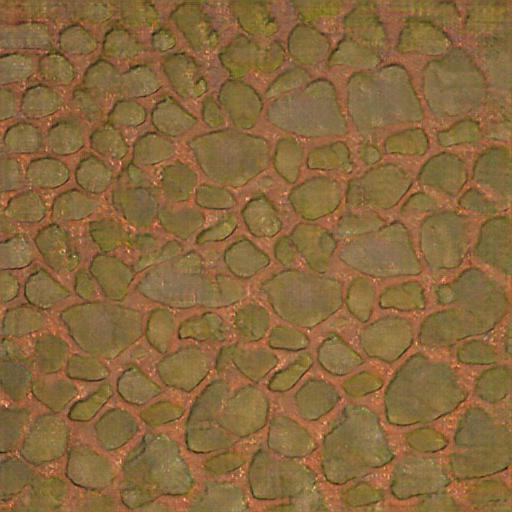} &
		\includegraphics[width=0.15\textwidth]{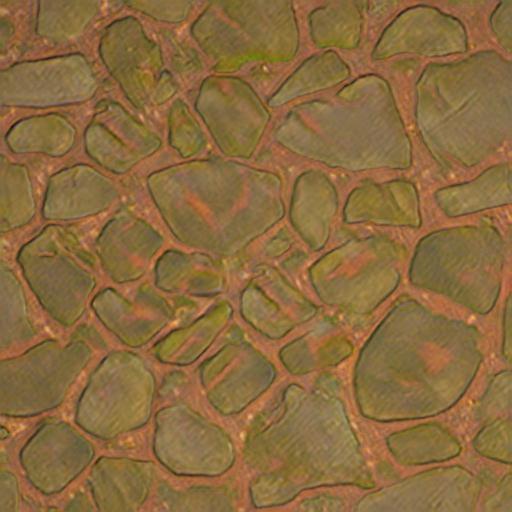} &
		\includegraphics[width=0.15\textwidth]{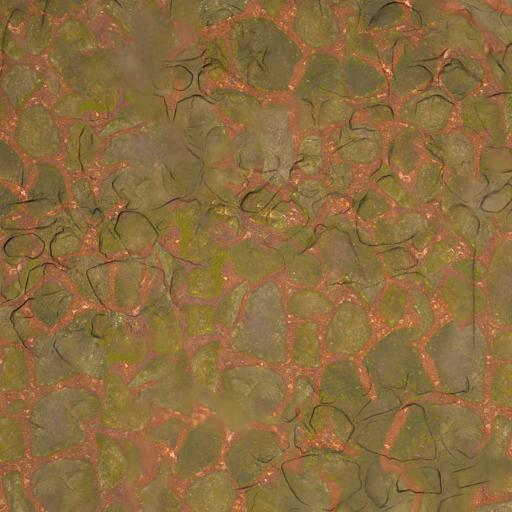} &
		\includegraphics[width=0.15\textwidth]{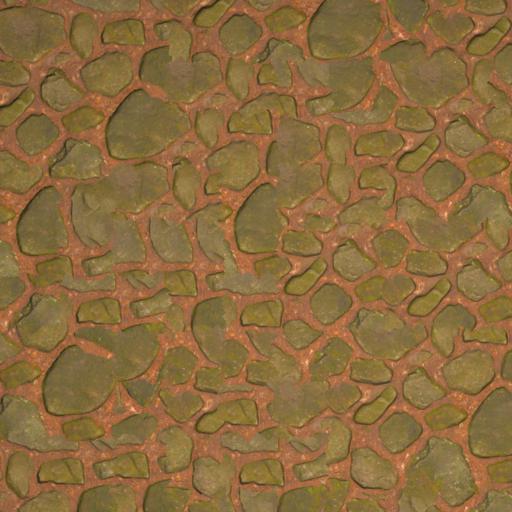} \\
		\includegraphics[width=0.09\textwidth]{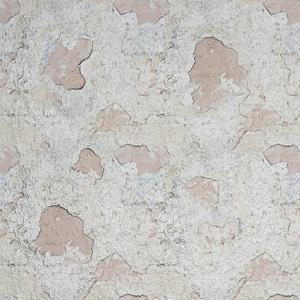}&
		\includegraphics[width=0.15\textwidth]{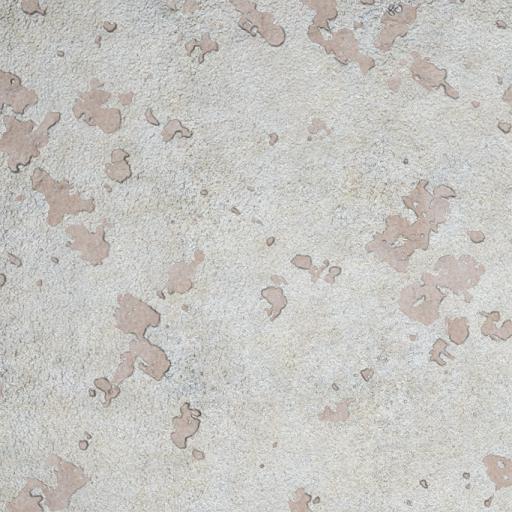}&
		\includegraphics[width=0.15\textwidth]{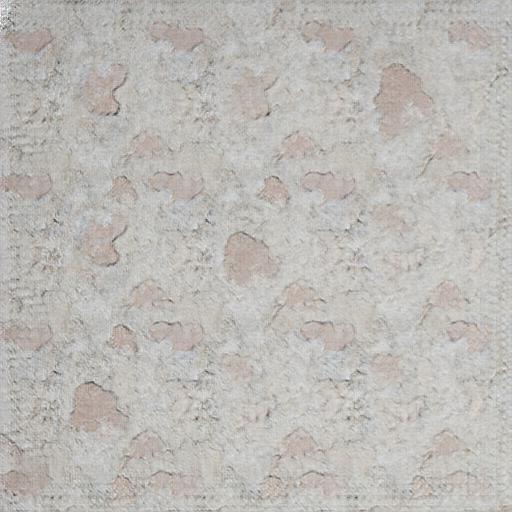} &
		\includegraphics[width=0.15\textwidth]{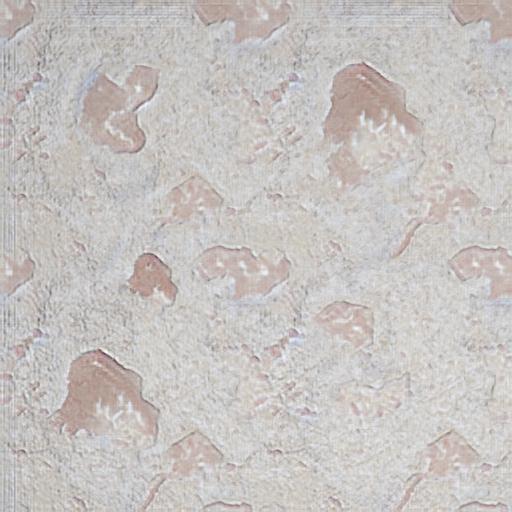} &
		\includegraphics[width=0.15\textwidth]{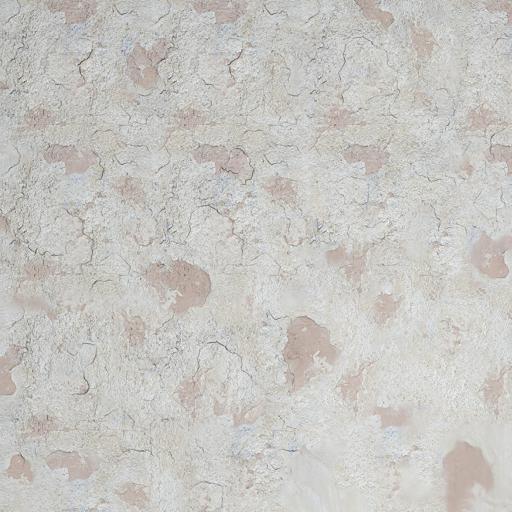} &
		\includegraphics[width=0.15\textwidth]{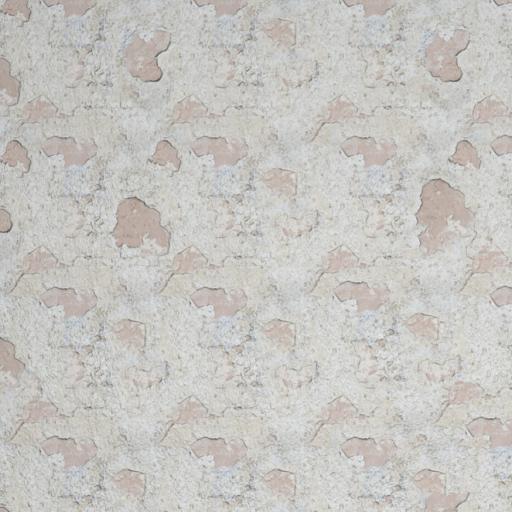} \\
		\scalebox{0.8}{Input} & \scalebox{0.8}{Ours} & \scalebox{0.8}{Adversarial Expansion} & 
		\scalebox{0.8}{InGAN} & \scalebox{0.8}{Self-tuning} & \scalebox{0.8}{Quilting} \\
	\end{tabular}
\vspace{-10pt}
\caption{Comparison of our method with example-based texture synthesis methods on SVBRDF maps. We generalize these methods to process multi-channel SVBRDF maps. We show our method; InGAN \cite{InGAN}; Non-stationary Texture Synthesis by Adversarial Expansion \cite{TexSyn18}; Self-tuning Texture Optimization \cite{SelfTuning}; Image Quilting \cite{quilting}. Images shown here are rendered by GGX shading model.}
\label{fig:comp-texture-syn}
\end{figure*}
\begin{figure} 
	\centering
	\addtolength{\tabcolsep}{-4pt}
	\begin{tabular}{ccccc}
		\includegraphics[width=0.115\textwidth]{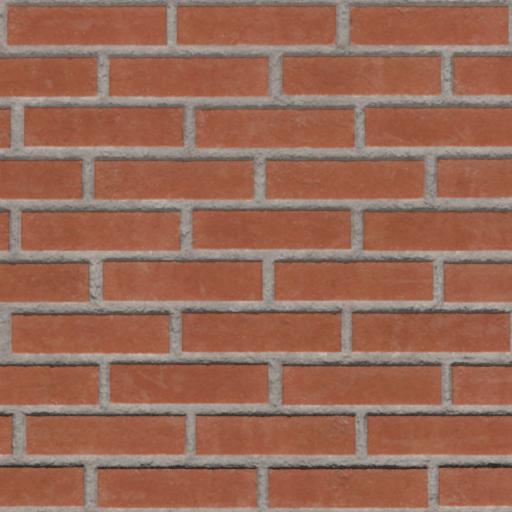} &
		\includegraphics[width=0.115\textwidth]{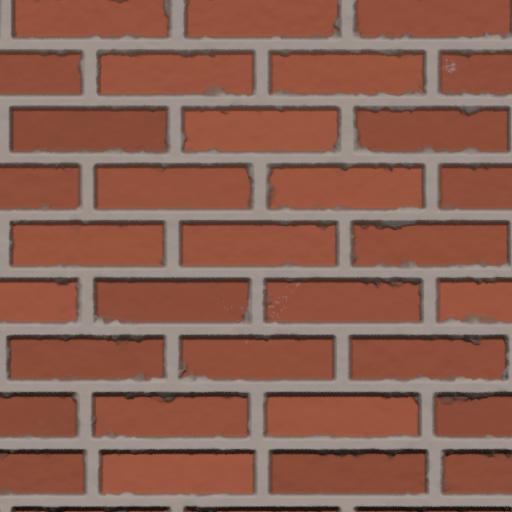} &
		\includegraphics[width=0.115\textwidth]{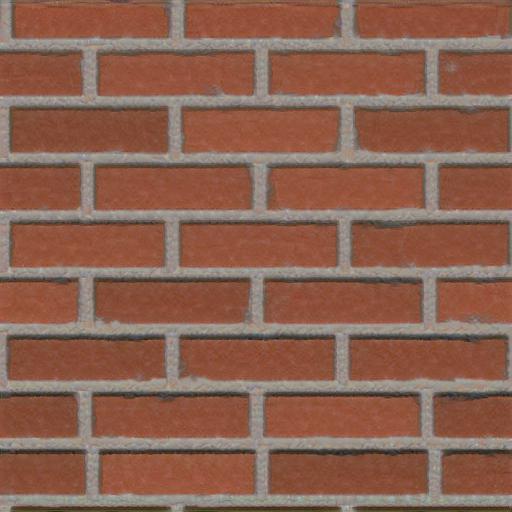} &
		\includegraphics[width=0.115\textwidth]{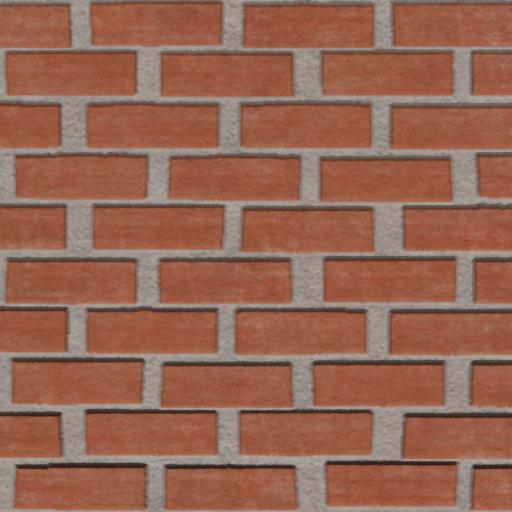} \\
		\includegraphics[width=0.115\textwidth]{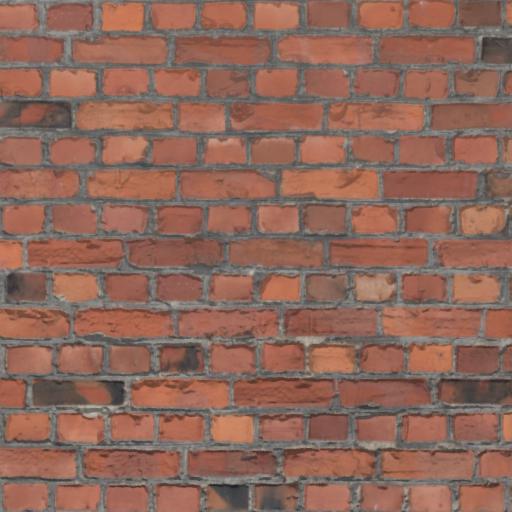} &
		\includegraphics[width=0.115\textwidth]{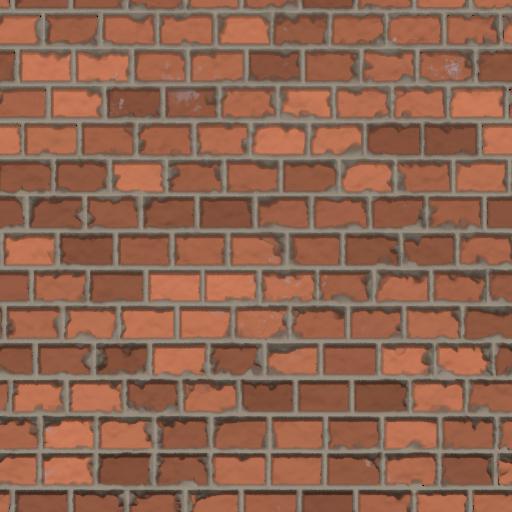} &
		\includegraphics[width=0.115\textwidth]{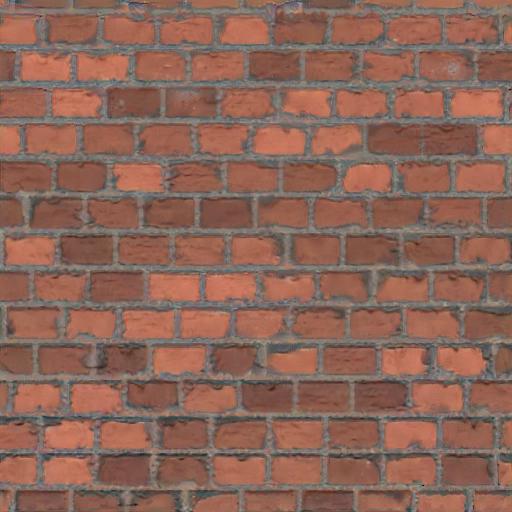} &
		\includegraphics[width=0.115\textwidth]{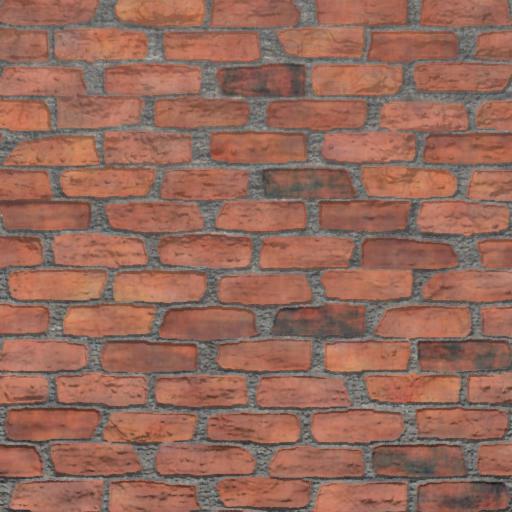} \\
		\includegraphics[width=0.115\textwidth]{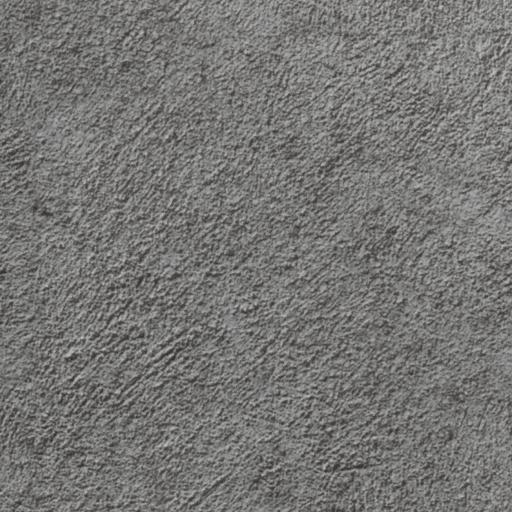} &
		\includegraphics[width=0.115\textwidth]{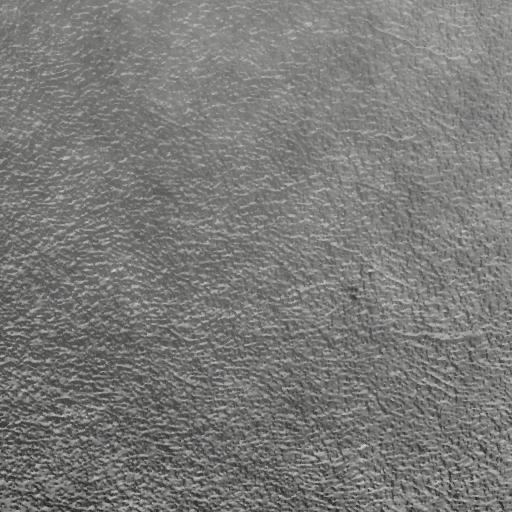} &
		\includegraphics[width=0.115\textwidth]{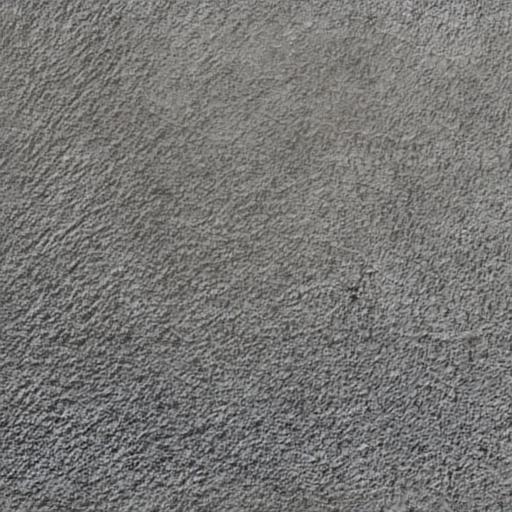} &
		\includegraphics[width=0.115\textwidth]{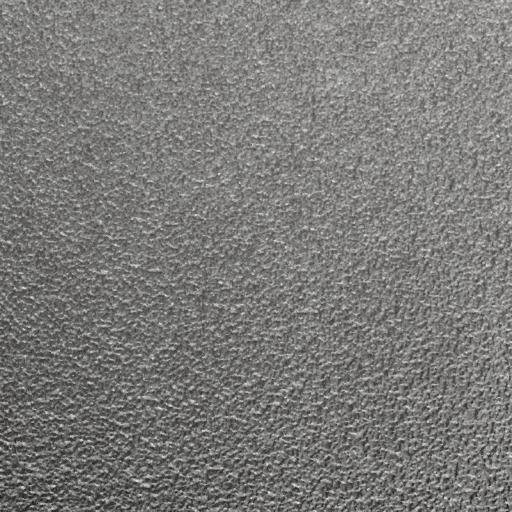} \\		\includegraphics[width=0.115\textwidth]{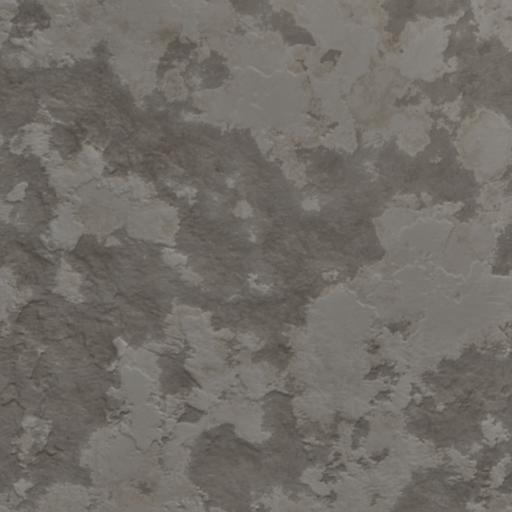} &
		\includegraphics[width=0.115\textwidth]{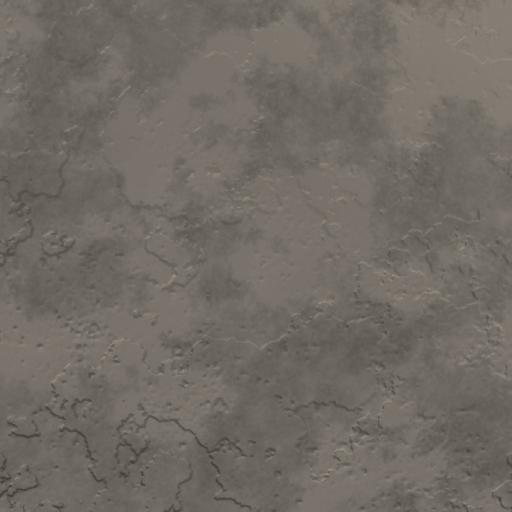} &
		\includegraphics[width=0.115\textwidth]{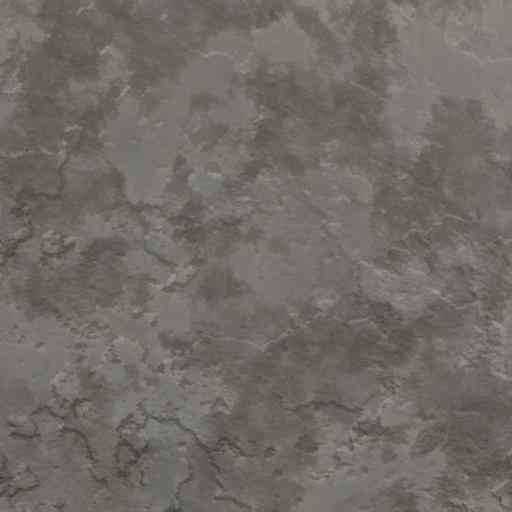} &
		\includegraphics[width=0.115\textwidth]{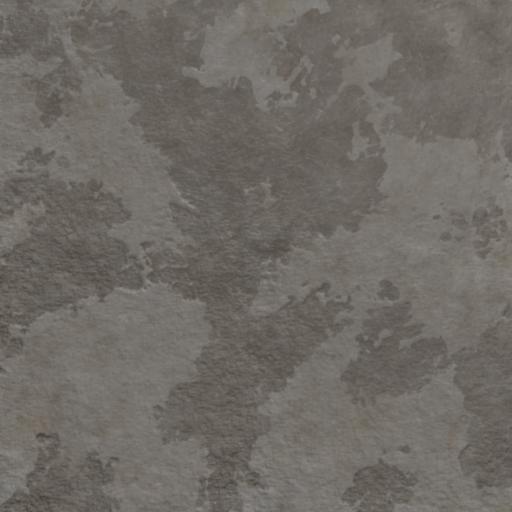} \\
        Input & Hu et al. (a) & Hu et al. (b) & Ours \\
	\end{tabular}
\vspace{-10pt}
\caption{Comparison of our method to \cite{hu2019}. The second column shows predicted procedural results by \cite{hu2019} while the third column is their style augmented results (non-procedural and no edibility). In contrast to their method, our pipeline generates fully procedural materials without a pre-existing material graph as an auxiliary input. The images are rendered using Blender with diffuse reflectance to match \cite{hu2019}.}
\label{fig:comp-hu2019}
\end{figure}
\begin{figure} 
	\centering
	\addtolength{\tabcolsep}{-4pt}
	\begin{tabular}{ccccc}
		\includegraphics[width=0.115\textwidth]{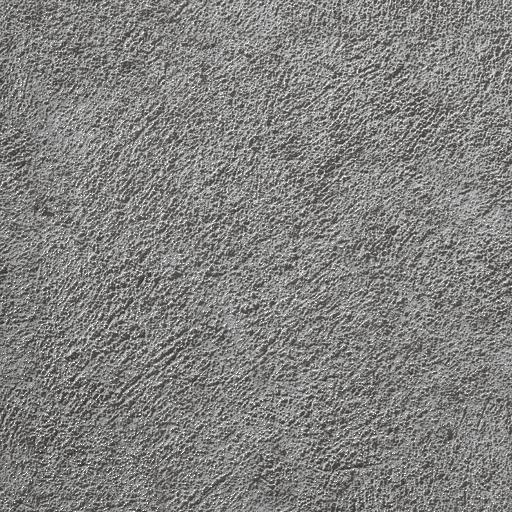} &
		\includegraphics[width=0.115\textwidth]{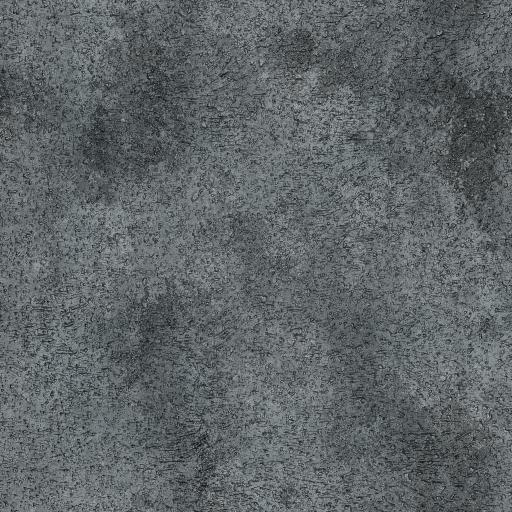} &
		\includegraphics[width=0.115\textwidth]{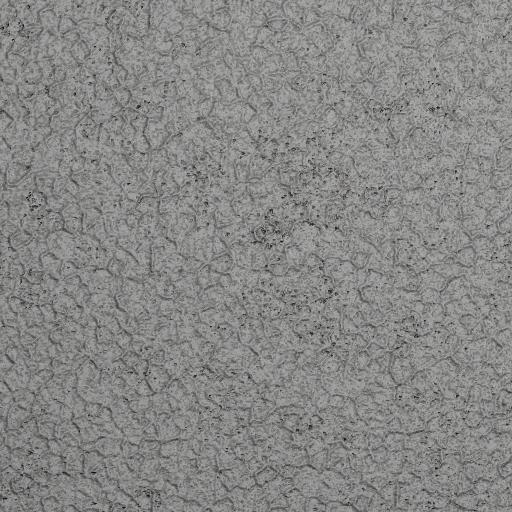} &
		\includegraphics[width=0.115\textwidth]{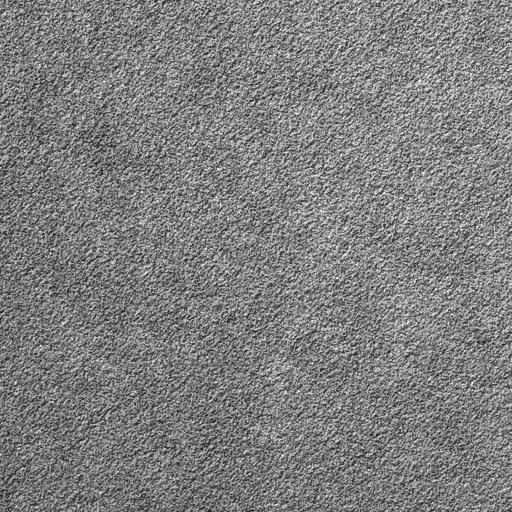} \\
		\includegraphics[width=0.115\textwidth]{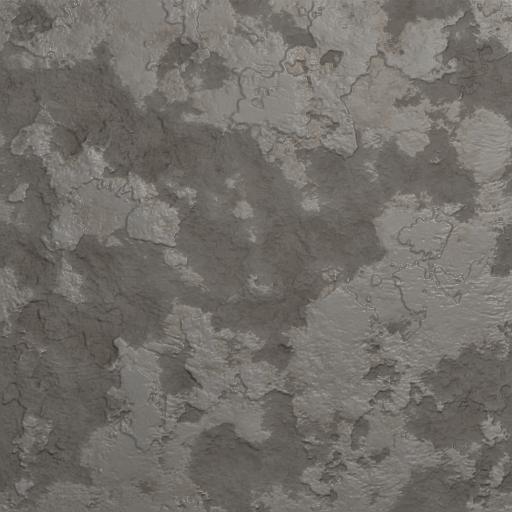} &
		\includegraphics[width=0.115\textwidth]{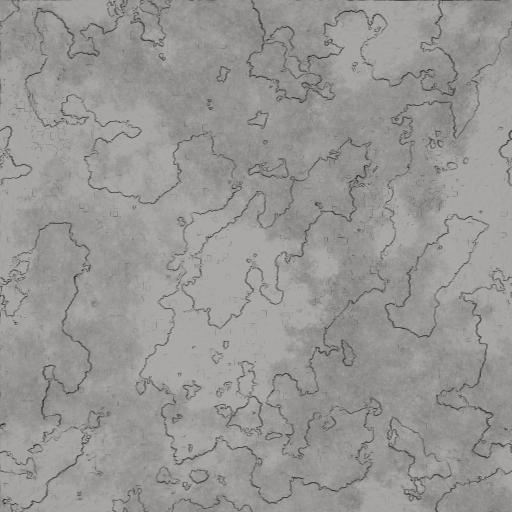} &
		\includegraphics[width=0.115\textwidth]{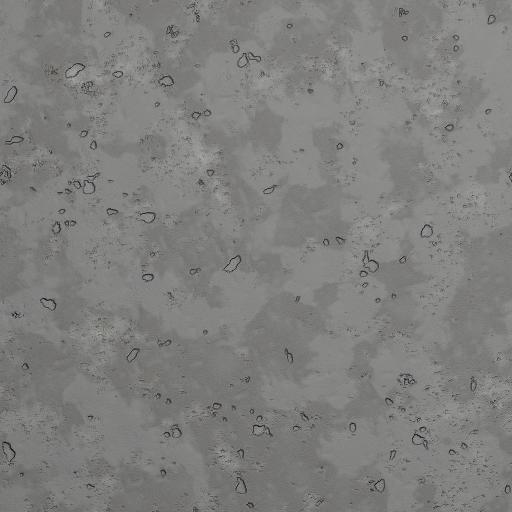} &
		\includegraphics[width=0.115\textwidth]{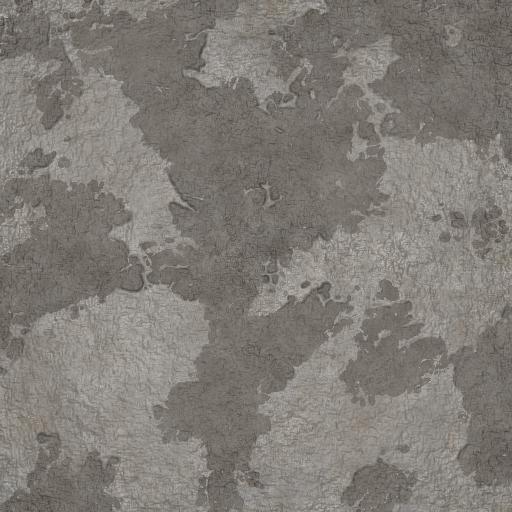} \\
		Input & Default & MATch & Ours \\
	\end{tabular}
\vspace{-10pt}
\caption{Comparison of our method to MATch~\shortcite{Shi20}. We use Hu et al. ~\shortcite{hu2019}'s framework to select a close Substance model (Default), and use MATch to optimize its appearance. As the MATch framework does not handle discrete parameters and requires good initialization, it  can generate poor output if the initialization and the discrete parameters are not hand-tuned. This can give poor results for bricks as shown in the supplemental material. Materials are rendered using the GGX shading model.}
\label{fig:comp-match}
\end{figure}
\subsection{Comparison with Prior Work}
We compare our results to previous work in inverse procedural modeling and in texture synthesis. Since we take the output of a material acquisition method as input to our pipeline, we do not compare to material acquisition methods.
\subsubsection{Inverse Procedural Material Modeling}
We compare our approach with state-of-the-art inverse procedural material modeling frameworks \cite{hu2019, Shi20} in Figs.~\ref{fig:comp-hu2019}~\&~\ref{fig:comp-match}. \emph{Both are built on a collection of pre-defined material graphs and rely on model selection with parameter estimation}. While Hu et al. directly applies neural network to predict parameters, Shi et al. (MATch) optimize parameters by end-to-end differential rendering. We use the model selection scheme of Hu et al. to chose input substance models for both methods. \emph{Despite not using pre-existing material graphs}, we show that our pipeline well reproduces  material maps with regular and stochastic patterns e.g. brick and stucco, similar to the examples shown in \cite{hu2019}. Hu et al. additionally propose a post-processing step (see Hu et al. (b) in Fig. \ref{fig:comp-hu2019}) to enrich the details, but \new{this step is done on the rendered images (instead of material maps)} and loses its procedural aspect. 

Direct comparison to (MATch)~\shortcite{Shi20} is difficult because the optimization framework of MATch \new{strongly} depends on the quality of initialization. Indeed, the MATch framework cannot optimize discrete parameters. \new{This limitation requires the scale of the structured target to be the same as the selected procedural material. The user is then required to manually fine-tune discrete parameters in Substance Designer. Additionally, only a subset of the Substance Engine was made differentiable in MATch, limiting the pool of compatible procedural graphs.} In Fig.~\ref{fig:comp-match} and the supplemental material we attempt to provide good initialization and discrete parameters for MATch. 

\subsubsection{Texture Synthesis}
We also compare our approach with texture synthesis methods as they share the ability of resolution extension our method allows. We experiment with several state-of-the-art example-based texture synthesis methods and generalize them to accept material maps as input. We stack all material maps together to build a high-dimensional texture map where each texel encodes albedo values, normal directions and roughness values. Loss functions are computed on each material map and averaged. For self-tuning texture optimization \cite{SelfTuning}, its generalization to multi-channel material maps is not trivial and we therefore run their algorithm separately on each material map. Fig. \ref{fig:comp-texture-syn} shows these comparisons in which the input resolution is $300\times300$ and the output resolution is $512\times512$. We see that our method is capable of extending both structured and stochastic materials and does not suffer from artifacts seen in traditional texture synthesis approaches such as structure error. Finally, and most significantly, compared with these other texture synthesis approaches, our method enables editing and fast synthesis of larger resolution without necessarily augmenting the scale.

\section{Limitations}
\subsection{Texture Taxonomy for Proceduralization}

\begin{figure}
	\centering
	\includegraphics[width=0.4\textwidth]{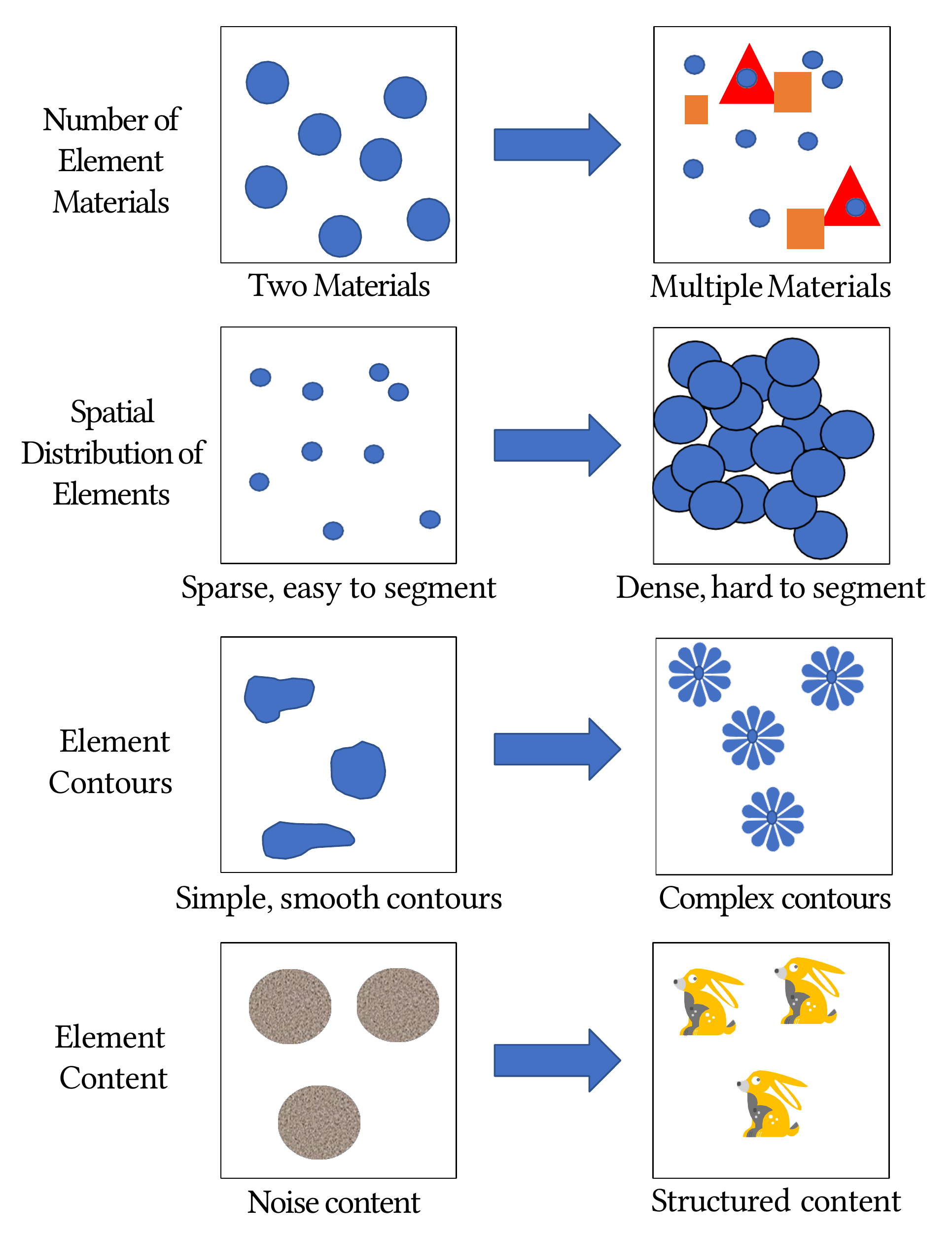} 
\vspace{-10pt}
\caption{Illustration of the four dimensions of variation of element collections in the space of material textures that we model.}
\label{fig:taxonomy}
\end{figure}

To characterize the limitations of our work, we describe the space of possible materials we attempt to model. We consider materials as collections of elements. The space of different element collections has four dimensions defining the complexity of the proceduralization of a material, illustrated in Fig.~\ref{fig:taxonomy}. %
The first axis is the number of different material element types in the texture. Both the types of elements and their spatial distribution relative to one another need to be modeled.
The second axis is the nature of the spatial distribution, whether the elements on the texture are sparsely or densely positioned, with overlap for example. Dense spatial distribution are harder to segment using masks. The third axis is complexity of the contour of the spatial mask. The more complex, the more difficult it is for PPTBF or any procedural approach to represent them faithfully. The fourth axis is the texture within each element, whether it can easily be represented as  noise or is more structured and semantically meaningful. 

Our general pipeline that segments SVBRDF and then estimates parameters for each element type, applies to the complete space we have described. However, our current implementation is limited to the less complex end of each of the last three axes. %
Our segmentation method is limited in that it cannot appropriately segment dense overlapping material. As stated, PPTBF performs well with irregular simple contours but not with complex contours. Finally, for the fourth axis we have not developed a method to proceduralize semantically meaningful elements such as the bunny.

\begin{figure} 
	\centering
	\addtolength{\tabcolsep}{-4pt}
	\begin{tabular}{cccccc}
		\includegraphics[width=0.09\textwidth]{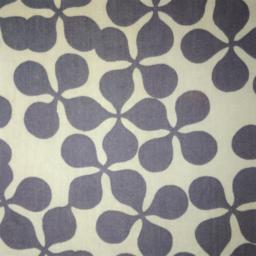} &
		\includegraphics[width=0.09\textwidth]{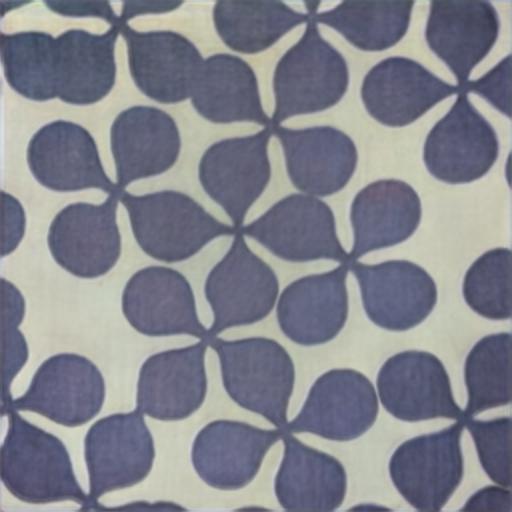} &
		\includegraphics[width=0.09\textwidth]{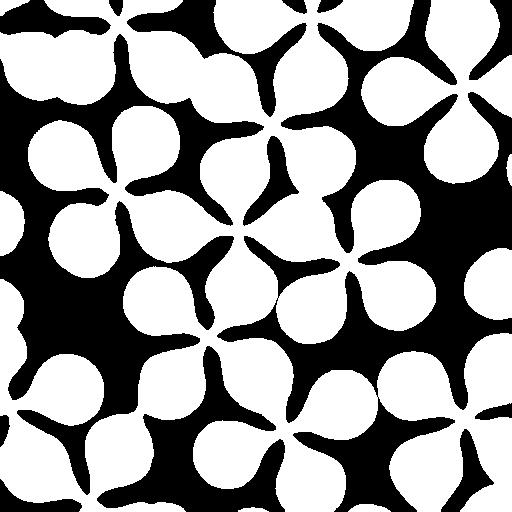} &
		\includegraphics[width=0.09\textwidth]{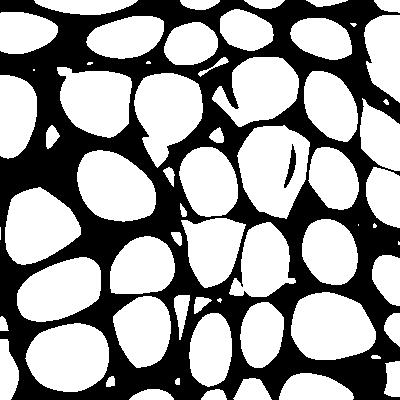} &
		\includegraphics[width=0.09\textwidth]{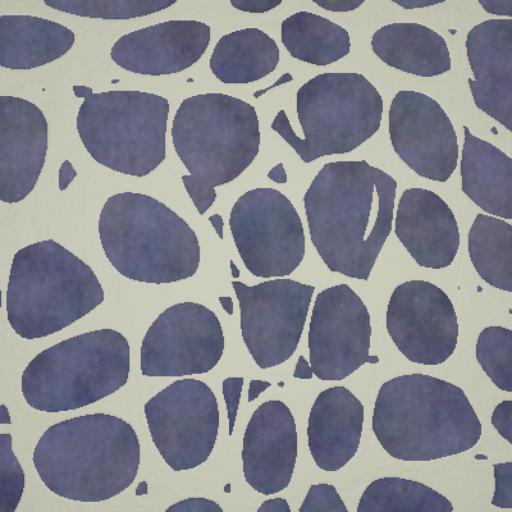} \\
		\includegraphics[width=0.09\textwidth]{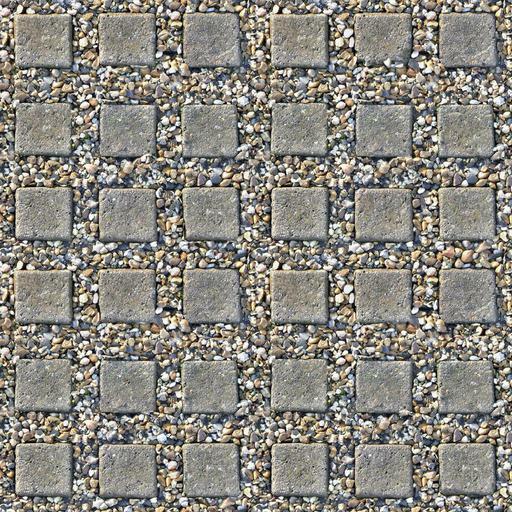} &
		\includegraphics[width=0.09\textwidth]{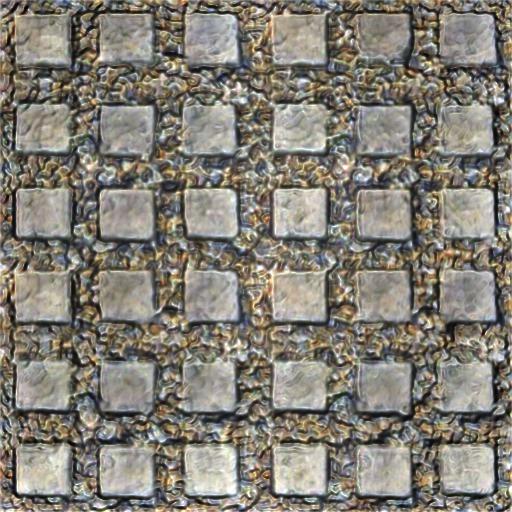} &
		\includegraphics[width=0.09\textwidth]{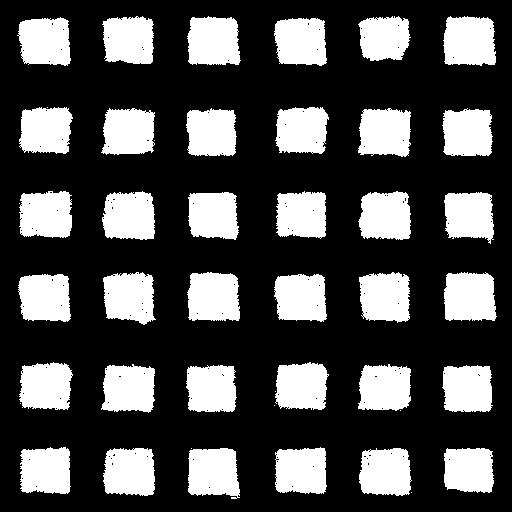} &
		\includegraphics[width=0.09\textwidth]{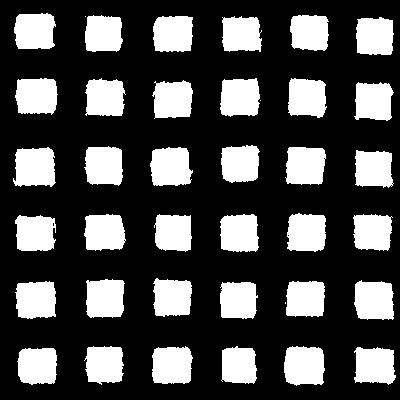} &
		\includegraphics[width=0.09\textwidth]{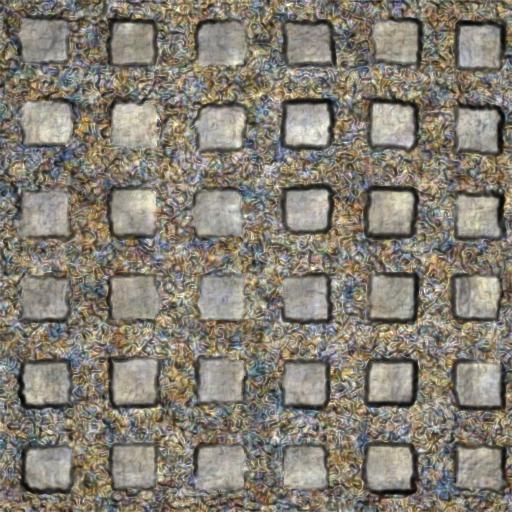} \\
		\scalebox{0.8}{Input} & 
		\scalebox{0.8}{Render("GT")} & 
		\scalebox{0.8}{Segmented} &
		\scalebox{0.8}{Our Procedural} & 
		\scalebox{0.8}{Our Render}
	\end{tabular}
\vspace{-10pt}
\caption{Failure cases \new{on semantically-shaped pattern and densely-overlapped material}. The input images are a captured image (first row) and an arbitrary texture (second row). The "Ground Truth" render is generated using \cite{DDB20} results. The first row shows a case where our PPTBF-based procedural mask map fails to reproduce the structure of the segmented mask map, missing the semantic "flower" shape. In the second row, one of the sub-material contains dense, small, overlapping pebbles which cannot be well reproduced by our multi-layer noise model or easily segmented. As a result, individual pebbles cannot be distinguished in our rendered result. Please zoom in to see details.}
\label{fig:failure}
\end{figure}
\begin{figure} 
	\centering
	\addtolength{\tabcolsep}{-4pt}
	\begin{tabular}{cccccc}
	    \raisebox{25pt}{\scalebox{0.9}{\rotatebox[origin=c]{90}{Input}}} &
		\includegraphics[width=0.11\textwidth]{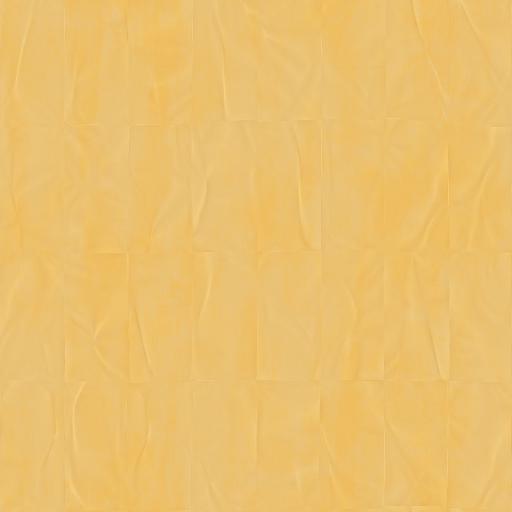} &
		\includegraphics[width=0.11\textwidth]{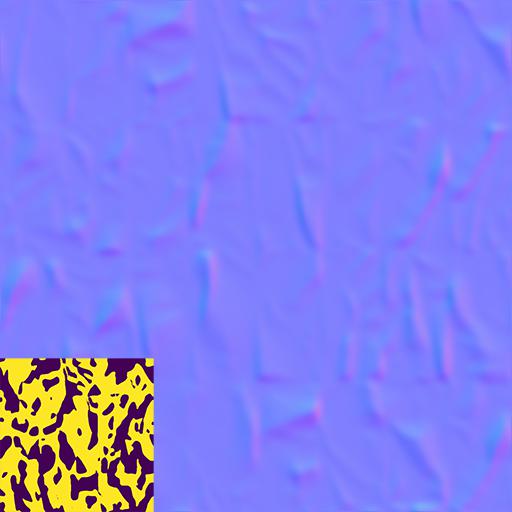} &
		\includegraphics[width=0.11\textwidth]{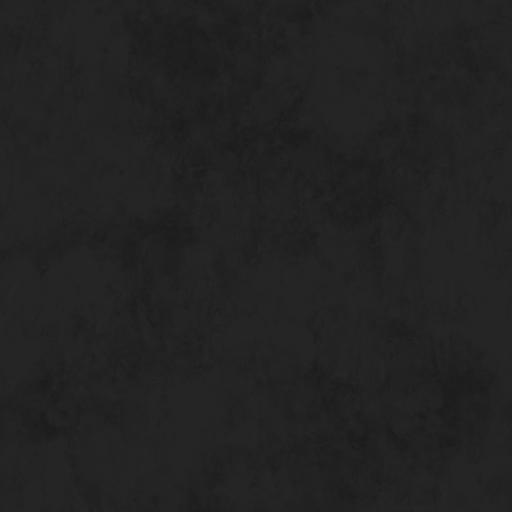} &
		\includegraphics[width=0.11\textwidth]{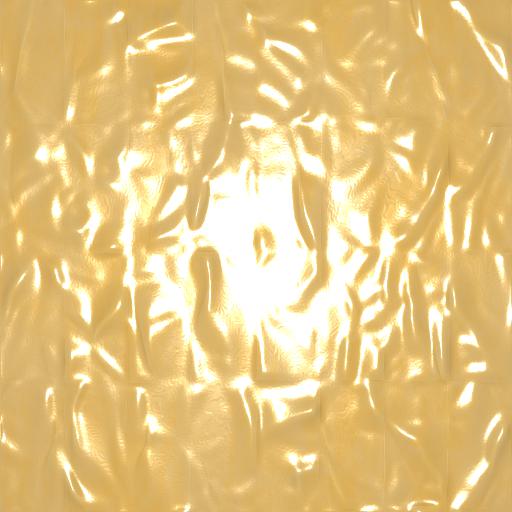} \\
		
		\raisebox{25pt}{\scalebox{0.9}{\rotatebox[origin=c]{90}{Ours (Proc.)}}} &
		\includegraphics[width=0.11\textwidth]{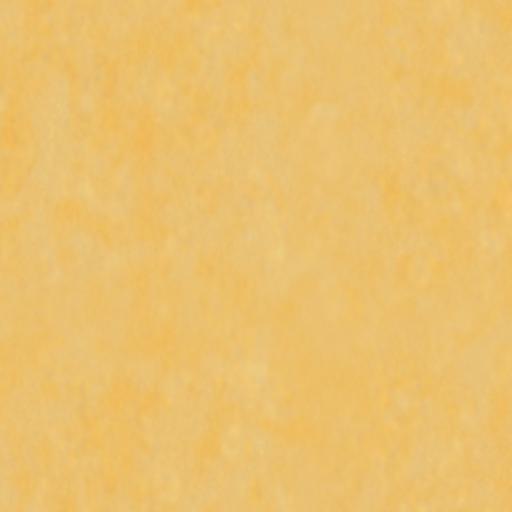} &
		\includegraphics[width=0.11\textwidth]{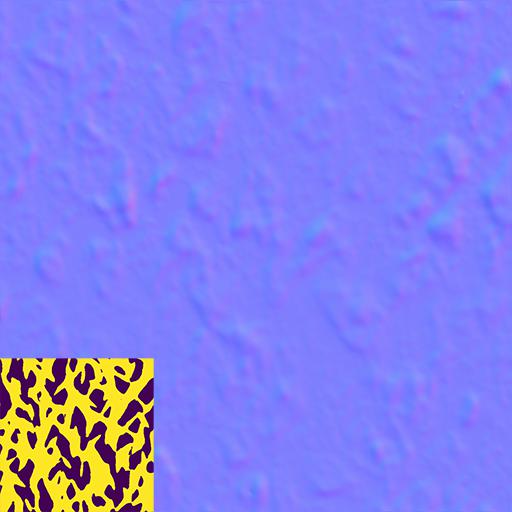} &
		\includegraphics[width=0.11\textwidth]{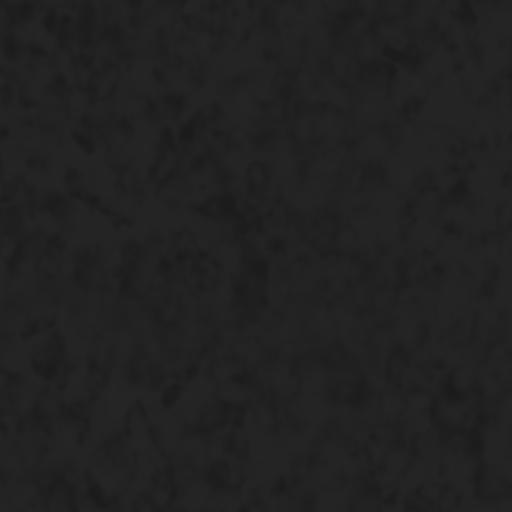} &
		\includegraphics[width=0.11\textwidth]{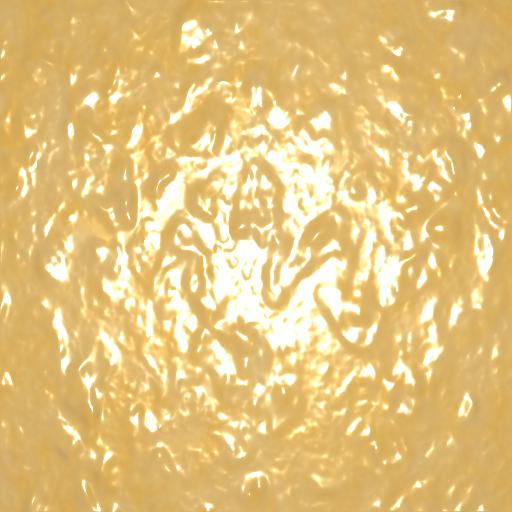} \\
		
	    \raisebox{25pt}{\scalebox{0.9}{\rotatebox[origin=c]{90}{Input}}} &
		\includegraphics[width=0.11\textwidth]{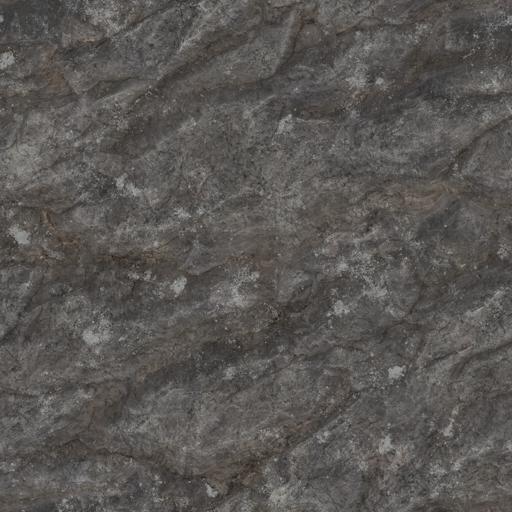} &
		\includegraphics[width=0.11\textwidth]{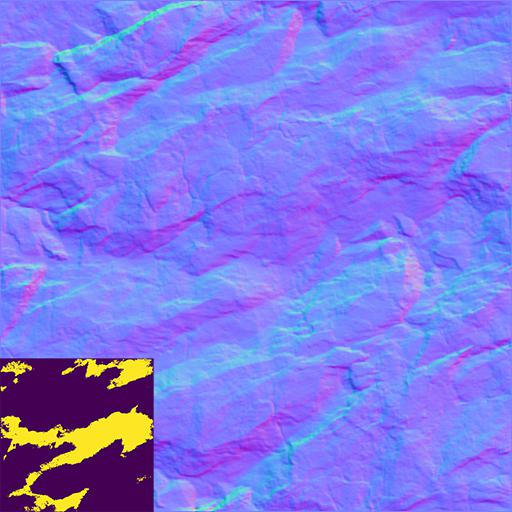} &
		\includegraphics[width=0.11\textwidth]{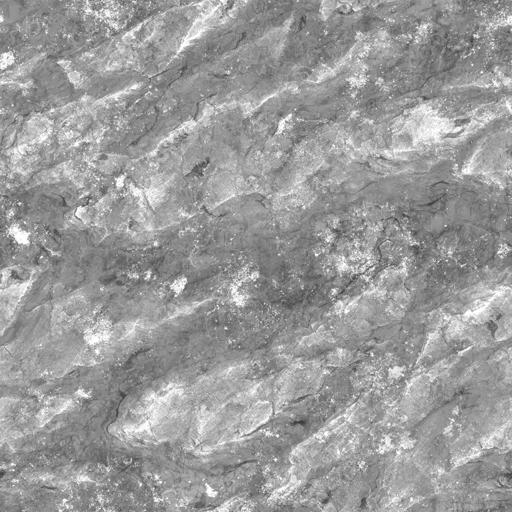} &
		\includegraphics[width=0.11\textwidth]{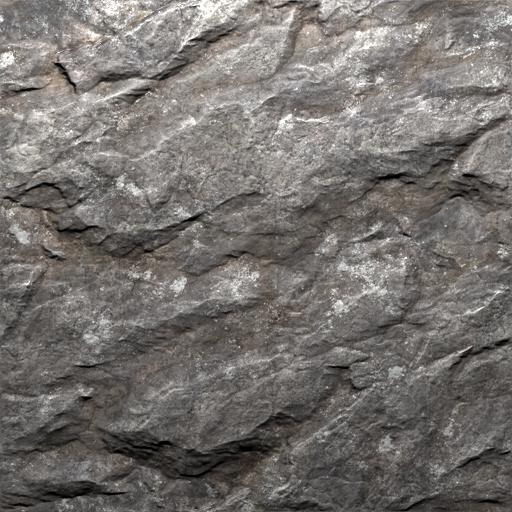} \\
		
		\raisebox{25pt}{\scalebox{0.9}{\rotatebox[origin=c]{90}{Ours (Proc.)}}} &
		\includegraphics[width=0.11\textwidth]{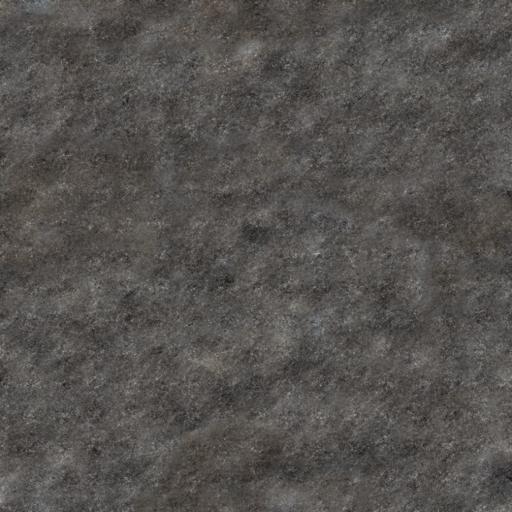} &
		\includegraphics[width=0.11\textwidth]{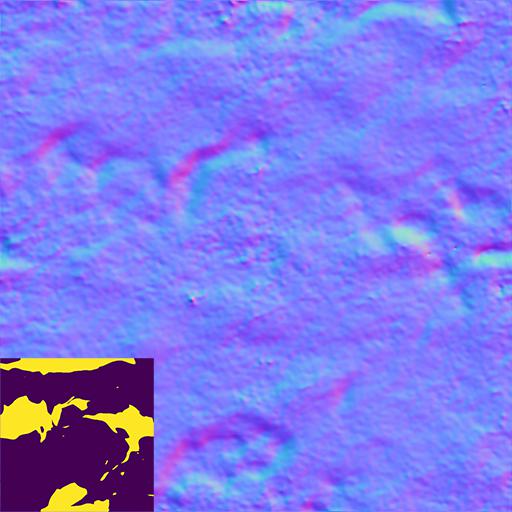} &
		\includegraphics[width=0.11\textwidth]{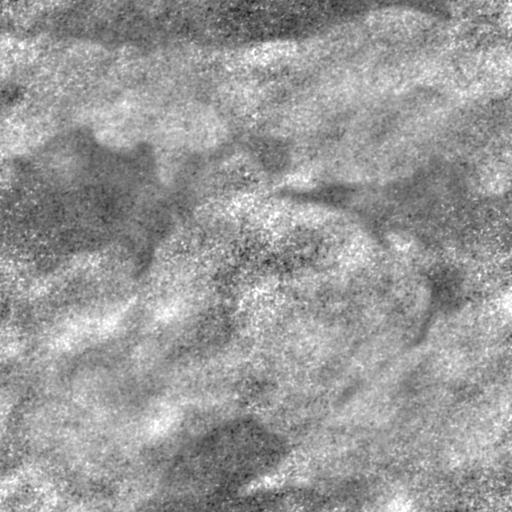} &
		\includegraphics[width=0.11\textwidth]{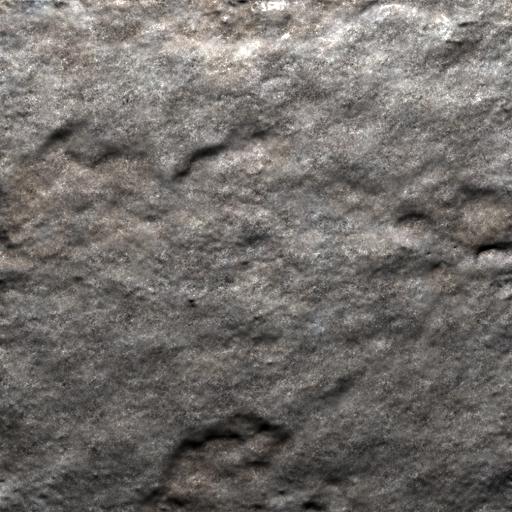} \\
		
		&
		\scalebox{1}{Albedo} & 
		\scalebox{1}{Normal} & 
		\scalebox{1}{Roughness} &
		\scalebox{1}{Render}\\
	\end{tabular}
\vspace{-10pt}
\caption{\new{Failure cases on material exemplars with highly-structured variations. We show additional failure cases on extremely structured normal variations which cannot be either 1) identified by segmentation or 2) recovered by noise models and Gaussian kernels on the transition regions of adjacent sub-materials. Insets on normal map show segmentation map and our procedural map. For better visualization, the first material is rendered with a central light while the second material is rendered with a top side light.}}
\label{fig:failure2}
\end{figure}

\subsection{Specific Limitation Examples} 
Our results show that our pipeline is able to proceduralize different types of materials with multiple SVBRDF maps e.g. albedo, normal and roughness. In Fig.~\ref{fig:failure} we show failure cases that result from the general limitations we just described.
The combination of complex contours and a semantically meaningful structure results in the failure shown in the top row of Fig.~\ref{fig:failure}. The complex contours form a "flower" shape that is not preserved by the PPTBF mask. The presence of a sub-material that is itself composed of a densely packed set of elements results in the failure shown in the bottom row of Fig.~\ref{fig:failure}. Neither a noise pattern nor our segmentation method can capture the arrangement of the small pebbles forming the fill material between the square elements. \new{Finer segmentation could produce better results, but the overlapping partial shapes would still be difficult to match using PPTBF. In some cases, such as the second example of Fig.~\ref{fig:comp-texture-syn}, the retrieved procedural mask will have small differences in the contours compared to the original. Adding a loss more sensitive to contours could improve such cases.}

\new{Finally, as our procedural material model is built upon multi-scale Gaussian noise models and Gaussian filters, it poorly reproduces highly-structured variations e.g. extremely-strong and directional normal variations seen as Fig. \ref{fig:failure2}. In these cases, the segmentation of the height map is also ambiguous, leading to less faithful procedural reconstruction.}
\section{Future work}
\label{sec:future}
In this work, we propose the first complete pipeline for inverse procedural modelling of general materials and highlight the challenges of each steps. We describe here interesting future work to enable better proceduralization. 

\textbf{Modelling.} New \new{generative} approaches such as deep texture or material generation~\cite{Henzler20, henzler21neuralmaterial} could help better reproduce complex material appearances that cannot be simply represented by noise models, \new{while sacrificing some control}. %

\textbf{Segmentation.} Our pipeline relies on user input, allowing user control and specification of the artistically important elements in the material. Nevertheless, a fully automatic segmentation would provide a faster approach and allow for quickly proceduralization of a large amount of materials. Current methods~\cite{cho2016, xu2017, bell2015, deeplabv3plus2018} are not geared toward material segmentation and fail on the materials we consider.%

\textbf{Recomposition.}  Complex details and patterns might be lost during procedural modeling as our method relies on noise and mask fitting. Introducing advanced differentiable filters and generator -- such as ones in Substance Designer and MATch~\cite{Shi20} -- to modify the appearance of procedural noise maps and binary maps, would allow to represent a wider range of appearances, further narrowing the gap between input and generated procedural materials.

\section{Conclusion}
We present the first pipeline for semi-automatic material proceduralization. Given a set of input material maps, our pipeline decomposes them into a tree of sub-materials and corresponding binary mask maps. We model the local appearance of sub-materials by procedural noise models and proceduralize binary mask map to reproduce global distribution of sub-materials.%

Compared with previous work \cite{hu2019, Shi20}, our pipeline does not rely on any predefined material graphs. %
With our approach, combined with the state-of-the-art material acquisition methods, we enable convenient creation of high quality spatially varying procedural materials. %
We take a first step toward general material proceduralization and hope our work and highlighted challenges will inspire future research.

\section*{Acknowledgment}
This work was funded by NSF \#2007283, and was supported by Dr. Abhijeet Ghosh with his EPSRC Early Career Fellowship (EP/N006259/1)

\bibliographystyle{ACM-Reference-Format}
\bibliography{bibliography} 


\begin{thebibliography}{64}


\ifx \showCODEN    \undefined \def \showCODEN     #1{\unskip}     \fi
\ifx \showDOI      \undefined \def \showDOI       #1{#1}\fi
\ifx \showISBNx    \undefined \def \showISBNx     #1{\unskip}     \fi
\ifx \showISBNxiii \undefined \def \showISBNxiii  #1{\unskip}     \fi
\ifx \showISSN     \undefined \def \showISSN      #1{\unskip}     \fi
\ifx \showLCCN     \undefined \def \showLCCN      #1{\unskip}     \fi
\ifx \shownote     \undefined \def \shownote      #1{#1}          \fi
\ifx \showarticletitle \undefined \def \showarticletitle #1{#1}   \fi
\ifx \showURL      \undefined \def \showURL       {\relax}        \fi
\providecommand\bibfield[2]{#2}
\providecommand\bibinfo[2]{#2}
\providecommand\natexlab[1]{#1}
\providecommand\showeprint[2][]{arXiv:#2}

\bibitem[\protect\citeauthoryear{Adobe}{Adobe}{2021}]%
        {SubstanceDes}
\bibfield{author}{\bibinfo{person}{Adobe}.} \bibinfo{year}{2021}\natexlab{}.
\newblock \bibinfo{title}{Substance Designer}.
\newblock
\newblock
\newblock
\shownote{https://www.substance3d.com/.}


\bibitem[\protect\citeauthoryear{Akl, Yaacoub, Donias, Da~Costa, and
  Germain}{Akl et~al\mbox{.}}{2018}]%
        {akl18}
\bibfield{author}{\bibinfo{person}{Adib Akl}, \bibinfo{person}{Charles
  Yaacoub}, \bibinfo{person}{Marc Donias}, \bibinfo{person}{Jean-Pierre
  Da~Costa}, {and} \bibinfo{person}{Christian Germain}.}
  \bibinfo{year}{2018}\natexlab{}.
\newblock \showarticletitle{A survey of exemplar-based texture synthesis
  methods}.
\newblock \bibinfo{journal}{\emph{Computer Vision and Image Understanding}}
  \bibinfo{volume}{172} (\bibinfo{year}{2018}), \bibinfo{pages}{12--24}.
\newblock


\bibitem[\protect\citeauthoryear{Aksoy, Oh, Paris, Pollefeys, and
  Matusik}{Aksoy et~al\mbox{.}}{2018}]%
        {aksoy2018}
\bibfield{author}{\bibinfo{person}{Ya{\u{g}}iz Aksoy},
  \bibinfo{person}{Tae-Hyun Oh}, \bibinfo{person}{Sylvain Paris},
  \bibinfo{person}{Marc Pollefeys}, {and} \bibinfo{person}{Wojciech Matusik}.}
  \bibinfo{year}{2018}\natexlab{}.
\newblock \showarticletitle{Semantic soft segmentation}.
\newblock \bibinfo{journal}{\emph{ACM Trans. Graph.}} \bibinfo{volume}{37},
  \bibinfo{number}{4}, Article \bibinfo{articleno}{72} (\bibinfo{year}{2018}),
  \bibinfo{numpages}{13}~pages.
\newblock


\bibitem[\protect\citeauthoryear{Aksoy, Ozan~Aydin, and Pollefeys}{Aksoy
  et~al\mbox{.}}{2017}]%
        {aksoy2017}
\bibfield{author}{\bibinfo{person}{Yagiz Aksoy}, \bibinfo{person}{Tunc
  Ozan~Aydin}, {and} \bibinfo{person}{Marc Pollefeys}.}
  \bibinfo{year}{2017}\natexlab{}.
\newblock \showarticletitle{Designing effective inter-pixel information flow
  for natural image matting}. In \bibinfo{booktitle}{\emph{Proceedings of the
  IEEE Conference on Computer Vision and Pattern Recognition (CVPR)}}.
  \bibinfo{pages}{29--37}.
\newblock


\bibitem[\protect\citeauthoryear{An and Pellacini}{An and Pellacini}{2008}]%
        {An08}
\bibfield{author}{\bibinfo{person}{Xiaobo An} {and} \bibinfo{person}{Fabio
  Pellacini}.} \bibinfo{year}{2008}\natexlab{}.
\newblock \showarticletitle{AppProp: All-Pairs Appearance-Space Edit
  Propagation}.
\newblock \bibinfo{journal}{\emph{ACM Trans. Graph.}} \bibinfo{volume}{27},
  \bibinfo{number}{3} (\bibinfo{date}{Aug.} \bibinfo{year}{2008}),
  \bibinfo{pages}{1–9}.
\newblock
\showISSN{0730-0301}
\urldef\tempurl%
\url{https://doi.org/10.1145/1360612.1360639}
\showDOI{\tempurl}


\bibitem[\protect\citeauthoryear{authors}{authors}{2016}]%
        {gpyopt2016}
\bibfield{author}{\bibinfo{person}{The~GPyOpt authors}.}
  \bibinfo{year}{2016}\natexlab{}.
\newblock \bibinfo{title}{{GPyOpt}: A Bayesian Optimization framework in
  python}.
\newblock \bibinfo{howpublished}{\url{http://github.com/SheffieldML/GPyOpt}}.
\newblock


\bibitem[\protect\citeauthoryear{Barnes, Shechtman, Finkelstein, and
  Goldman}{Barnes et~al\mbox{.}}{2009}]%
        {PatchMatch}
\bibfield{author}{\bibinfo{person}{Connelly Barnes}, \bibinfo{person}{Eli
  Shechtman}, \bibinfo{person}{Adam Finkelstein}, {and} \bibinfo{person}{Dan~B
  Goldman}.} \bibinfo{year}{2009}\natexlab{}.
\newblock \showarticletitle{PatchMatch: A Randomized Correspondence Algorithm
  for Structural Image Editing}.
\newblock \bibinfo{journal}{\emph{ACM Trans. Graph.}} \bibinfo{volume}{28},
  \bibinfo{number}{3}, Article \bibinfo{articleno}{24} (\bibinfo{date}{July}
  \bibinfo{year}{2009}), \bibinfo{numpages}{11}~pages.
\newblock
\showISSN{0730-0301}
\urldef\tempurl%
\url{https://doi.org/10.1145/1531326.1531330}
\showDOI{\tempurl}


\bibitem[\protect\citeauthoryear{Bell, Upchurch, Snavely, and Bala}{Bell
  et~al\mbox{.}}{2015}]%
        {bell2015}
\bibfield{author}{\bibinfo{person}{Sean Bell}, \bibinfo{person}{Paul Upchurch},
  \bibinfo{person}{Noah Snavely}, {and} \bibinfo{person}{Kavita Bala}.}
  \bibinfo{year}{2015}\natexlab{}.
\newblock \showarticletitle{Material recognition in the wild with the materials
  in context database}. In \bibinfo{booktitle}{\emph{Proceedings of the IEEE
  Conference on Computer Vision and Pattern Recognition (CVPR)}}.
  \bibinfo{pages}{3479--3487}.
\newblock


\bibitem[\protect\citeauthoryear{Bergmann, Jetchev, and Vollgraf}{Bergmann
  et~al\mbox{.}}{2017}]%
        {Bergmann2017}
\bibfield{author}{\bibinfo{person}{Urs Bergmann}, \bibinfo{person}{Nikolay
  Jetchev}, {and} \bibinfo{person}{Roland Vollgraf}.}
  \bibinfo{year}{2017}\natexlab{}.
\newblock \showarticletitle{Learning Texture Manifolds with the Periodic
  Spatial {GAN}}. In \bibinfo{booktitle}{\emph{Proceedings of the 34th
  International Conference on Machine Learning}}
  \emph{(\bibinfo{series}{Proceedings of Machine Learning Research},
  Vol.~\bibinfo{volume}{70})}, \bibfield{editor}{\bibinfo{person}{Doina Precup}
  {and} \bibinfo{person}{Yee~Whye Teh}} (Eds.). \bibinfo{publisher}{PMLR},
  \bibinfo{pages}{469--477}.
\newblock
\urldef\tempurl%
\url{http://proceedings.mlr.press/v70/bergmann17a.html}
\showURL{%
\tempurl}


\bibitem[\protect\citeauthoryear{Boss, Jampani, Kim, Lensch, and Kautz}{Boss
  et~al\mbox{.}}{2020}]%
        {Boss20}
\bibfield{author}{\bibinfo{person}{Mark Boss}, \bibinfo{person}{Varun Jampani},
  \bibinfo{person}{Kihwan Kim}, \bibinfo{person}{Hendrik~P.A. Lensch}, {and}
  \bibinfo{person}{Jan Kautz}.} \bibinfo{year}{2020}\natexlab{}.
\newblock \showarticletitle{Two-Shot Spatially-Varying BRDF and Shape
  Estimation}. In \bibinfo{booktitle}{\emph{Proceedings of the IEEE/CVF
  Conference on Computer Vision and Pattern Recognition (CVPR)}}.
  \bibinfo{pages}{3982--3991}.
\newblock


\bibitem[\protect\citeauthoryear{Chen, Zhu, Papandreou, Schroff, and Adam}{Chen
  et~al\mbox{.}}{2018}]%
        {deeplabv3plus2018}
\bibfield{author}{\bibinfo{person}{Liang-Chieh Chen}, \bibinfo{person}{Yukun
  Zhu}, \bibinfo{person}{George Papandreou}, \bibinfo{person}{Florian Schroff},
  {and} \bibinfo{person}{Hartwig Adam}.} \bibinfo{year}{2018}\natexlab{}.
\newblock \showarticletitle{Encoder-Decoder with Atrous Separable Convolution
  for Semantic Image Segmentation}. In \bibinfo{booktitle}{\emph{Proceedings of
  the European Conference on Computer Vision (ECCV)}}.
  \bibinfo{pages}{801--818}.
\newblock


\bibitem[\protect\citeauthoryear{Chen, Li, and Tang}{Chen
  et~al\mbox{.}}{2013}]%
        {chen2013}
\bibfield{author}{\bibinfo{person}{Qifeng Chen}, \bibinfo{person}{Dingzeyu Li},
  {and} \bibinfo{person}{Chi-Keung Tang}.} \bibinfo{year}{2013}\natexlab{}.
\newblock \showarticletitle{KNN matting}.
\newblock \bibinfo{journal}{\emph{IEEE Transactions on Pattern Analysis and
  Machine Intelligence}} \bibinfo{volume}{35}, \bibinfo{number}{9}
  (\bibinfo{year}{2013}), \bibinfo{pages}{2175--2188}.
\newblock


\bibitem[\protect\citeauthoryear{Cho, Tai, and Kweon}{Cho
  et~al\mbox{.}}{2016}]%
        {cho2016}
\bibfield{author}{\bibinfo{person}{Donghyeon Cho}, \bibinfo{person}{Yu-Wing
  Tai}, {and} \bibinfo{person}{Inso Kweon}.} \bibinfo{year}{2016}\natexlab{}.
\newblock \showarticletitle{Natural image matting using deep convolutional
  neural networks}. In \bibinfo{booktitle}{\emph{Proceedings of European
  Conference on Computer Vision (ECCV)}}. Springer, \bibinfo{pages}{626--643}.
\newblock


\bibitem[\protect\citeauthoryear{Cimpoi, Maji, Kokkinos, and Vedaldi}{Cimpoi
  et~al\mbox{.}}{2016}]%
        {cimpoi2016}
\bibfield{author}{\bibinfo{person}{Mircea Cimpoi}, \bibinfo{person}{Subhransu
  Maji}, \bibinfo{person}{Iasonas Kokkinos}, {and} \bibinfo{person}{Andrea
  Vedaldi}.} \bibinfo{year}{2016}\natexlab{}.
\newblock \showarticletitle{Deep filter banks for texture recognition,
  description, and segmentation}.
\newblock \bibinfo{journal}{\emph{International Journal of Computer Vision}}
  \bibinfo{volume}{118}, \bibinfo{number}{1} (\bibinfo{year}{2016}),
  \bibinfo{pages}{65--94}.
\newblock


\bibitem[\protect\citeauthoryear{Deschaintre, Aittala, Durand, Drettakis, and
  Bousseau}{Deschaintre et~al\mbox{.}}{2018}]%
        {Deschaintre18}
\bibfield{author}{\bibinfo{person}{Valentin Deschaintre},
  \bibinfo{person}{Miika Aittala}, \bibinfo{person}{Fr\'edo Durand},
  \bibinfo{person}{George Drettakis}, {and} \bibinfo{person}{Adrien Bousseau}.}
  \bibinfo{year}{2018}\natexlab{}.
\newblock \showarticletitle{Single-Image SVBRDF Capture with a Rendering-Aware
  Deep Network}.
\newblock \bibinfo{journal}{\emph{ACM Trans. Graph.}} \bibinfo{volume}{37},
  \bibinfo{number}{4}, Article \bibinfo{articleno}{128} (\bibinfo{date}{Aug}
  \bibinfo{year}{2018}), \bibinfo{numpages}{15}~pages.
\newblock


\bibitem[\protect\citeauthoryear{Deschaintre, Aittala, Durand, Drettakis, and
  Bousseau}{Deschaintre et~al\mbox{.}}{2019}]%
        {Deschaintre19}
\bibfield{author}{\bibinfo{person}{Valentin Deschaintre},
  \bibinfo{person}{Miika Aittala}, \bibinfo{person}{Fr\'edo Durand},
  \bibinfo{person}{George Drettakis}, {and} \bibinfo{person}{Adrien Bousseau}.}
  \bibinfo{year}{2019}\natexlab{}.
\newblock \showarticletitle{Flexible SVBRDF Capture with a Multi-Image Deep
  Network}.
\newblock \bibinfo{journal}{\emph{Computer Graphics Forum (Proceedings of the
  Eurographics Symposium on Rendering)}} \bibinfo{volume}{38},
  \bibinfo{number}{4} (\bibinfo{date}{July} \bibinfo{year}{2019}),
  \bibinfo{pages}{1--13}.
\newblock


\bibitem[\protect\citeauthoryear{Deschaintre, Drettakis, and
  Bousseau}{Deschaintre et~al\mbox{.}}{2020}]%
        {DDB20}
\bibfield{author}{\bibinfo{person}{Valentin Deschaintre},
  \bibinfo{person}{George Drettakis}, {and} \bibinfo{person}{Adrien Bousseau}.}
  \bibinfo{year}{2020}\natexlab{}.
\newblock \showarticletitle{Guided Fine-Tuning for Large-Scale Material
  Transfer}.
\newblock \bibinfo{journal}{\emph{Computer Graphics Forum (Proceedings of the
  Eurographics Symposium on Rendering)}} \bibinfo{volume}{39},
  \bibinfo{number}{4} (\bibinfo{year}{2020}), \bibinfo{pages}{91--105}.
\newblock
\urldef\tempurl%
\url{http://www-sop.inria.fr/reves/Basilic/2020/DDB20}
\showURL{%
\tempurl}


\bibitem[\protect\citeauthoryear{Deschaintre, Lin, and Ghosh}{Deschaintre
  et~al\mbox{.}}{2021}]%
        {Deschaintre21}
\bibfield{author}{\bibinfo{person}{Valentin Deschaintre},
  \bibinfo{person}{Yiming Lin}, {and} \bibinfo{person}{Abhijeet Ghosh}.}
  \bibinfo{year}{2021}\natexlab{}.
\newblock \showarticletitle{Deep polarization imaging for 3D shape and SVBRDF
  acquisition}. In \bibinfo{booktitle}{\emph{Proceedings of the IEEE/CVF
  Conference on Computer Vision and Pattern Recognition (CVPR)}}.
\newblock


\bibitem[\protect\citeauthoryear{Dong}{Dong}{2019}]%
        {dong19}
\bibfield{author}{\bibinfo{person}{Yue Dong}.} \bibinfo{year}{2019}\natexlab{}.
\newblock \showarticletitle{Deep appearance modeling: A survey}.
\newblock \bibinfo{journal}{\emph{Visual Informatics}} \bibinfo{volume}{3},
  \bibinfo{number}{2} (\bibinfo{year}{2019}), \bibinfo{pages}{59 -- 68}.
\newblock
\showISSN{2468-502X}
\urldef\tempurl%
\url{https://doi.org/10.1016/j.visinf.2019.07.003}
\showDOI{\tempurl}


\bibitem[\protect\citeauthoryear{Efros and Freeman}{Efros and Freeman}{2001}]%
        {quilting}
\bibfield{author}{\bibinfo{person}{Alexei~A. Efros} {and}
  \bibinfo{person}{William~T. Freeman}.} \bibinfo{year}{2001}\natexlab{}.
\newblock \showarticletitle{Image Quilting for Texture Synthesis and Transfer}.
  In \bibinfo{booktitle}{\emph{Proceedings of the 28th Annual Conference on
  Computer Graphics and Interactive Techniques}}
  \emph{(\bibinfo{series}{SIGGRAPH '01})}. \bibinfo{publisher}{Association for
  Computing Machinery}, \bibinfo{address}{New York, NY, USA},
  \bibinfo{pages}{341–346}.
\newblock
\showISBNx{158113374X}
\urldef\tempurl%
\url{https://doi.org/10.1145/383259.383296}
\showDOI{\tempurl}


\bibitem[\protect\citeauthoryear{Efros and Leung}{Efros and Leung}{1999}]%
        {Efros1999}
\bibfield{author}{\bibinfo{person}{Alexei~A Efros} {and}
  \bibinfo{person}{Thomas~K Leung}.} \bibinfo{year}{1999}\natexlab{}.
\newblock \showarticletitle{Texture synthesis by non-parametric sampling}. In
  \bibinfo{booktitle}{\emph{Proceedings of the IEEE International Conference on
  Computer Vision (ICCV)}}, Vol.~\bibinfo{volume}{2}. IEEE,
  \bibinfo{pages}{1033--1038}.
\newblock


\bibitem[\protect\citeauthoryear{Foo}{Foo}{1997}]%
        {Foo97}
\bibfield{author}{\bibinfo{person}{Sing~Choong Foo}.}
  \bibinfo{year}{1997}\natexlab{}.
\newblock \bibinfo{title}{A Gonioreflectometer For Measuring The Bidirectional
  Reflectance Of Material For Use In Illumination Computation}.
\newblock \bibinfo{howpublished}{Doctoral Dissertation, Cornell University}.
\newblock


\bibitem[\protect\citeauthoryear{{Galerne}, {Gousseau}, and {Morel}}{{Galerne}
  et~al\mbox{.}}{2011}]%
        {Galerne2011}
\bibfield{author}{\bibinfo{person}{B. {Galerne}}, \bibinfo{person}{Y.
  {Gousseau}}, {and} \bibinfo{person}{J. {Morel}}.}
  \bibinfo{year}{2011}\natexlab{}.
\newblock \showarticletitle{Random Phase Textures: Theory and Synthesis}.
\newblock \bibinfo{journal}{\emph{IEEE Transactions on Image Processing}}
  \bibinfo{volume}{20}, \bibinfo{number}{1} (\bibinfo{year}{2011}),
  \bibinfo{pages}{257--267}.
\newblock


\bibitem[\protect\citeauthoryear{Galerne, Lagae, Lefebvre, and
  Drettakis}{Galerne et~al\mbox{.}}{2012}]%
        {Galerne2012}
\bibfield{author}{\bibinfo{person}{Bruno Galerne}, \bibinfo{person}{Ares
  Lagae}, \bibinfo{person}{Sylvain Lefebvre}, {and} \bibinfo{person}{George
  Drettakis}.} \bibinfo{year}{2012}\natexlab{}.
\newblock \showarticletitle{Gabor Noise by Example}.
\newblock \bibinfo{journal}{\emph{ACM Trans. Graph.}} \bibinfo{volume}{31},
  \bibinfo{number}{4}, Article \bibinfo{articleno}{73} (\bibinfo{date}{July}
  \bibinfo{year}{2012}), \bibinfo{numpages}{9}~pages.
\newblock
\showISSN{0730-0301}
\urldef\tempurl%
\url{https://doi.org/10.1145/2185520.2185569}
\showDOI{\tempurl}


\bibitem[\protect\citeauthoryear{Galerne, Leclaire, and Moisan}{Galerne
  et~al\mbox{.}}{2017}]%
        {Galerne2017}
\bibfield{author}{\bibinfo{person}{B. Galerne}, \bibinfo{person}{A. Leclaire},
  {and} \bibinfo{person}{L. Moisan}.} \bibinfo{year}{2017}\natexlab{}.
\newblock \showarticletitle{Texton Noise}.
\newblock \bibinfo{journal}{\emph{Computer Graphics Forum}}
  \bibinfo{volume}{36}, \bibinfo{number}{8} (\bibinfo{year}{2017}),
  \bibinfo{pages}{205--218}.
\newblock
\urldef\tempurl%
\url{https://doi.org/10.1111/cgf.13073}
\showDOI{\tempurl}
\showeprint{https://onlinelibrary.wiley.com/doi/pdf/10.1111/cgf.13073}


\bibitem[\protect\citeauthoryear{Gao, Li, Dong, Peers, Xu, and Tong}{Gao
  et~al\mbox{.}}{2019}]%
        {Gao19}
\bibfield{author}{\bibinfo{person}{Duan Gao}, \bibinfo{person}{Xiao Li},
  \bibinfo{person}{Yue Dong}, \bibinfo{person}{Pieter Peers},
  \bibinfo{person}{Kun Xu}, {and} \bibinfo{person}{Xin Tong}.}
  \bibinfo{year}{2019}\natexlab{}.
\newblock \showarticletitle{Deep Inverse Rendering for High-resolution SVBRDF
  Estimation from an Arbitrary Number of Images}.
\newblock \bibinfo{journal}{\emph{ACM Trans. Graph.}} \bibinfo{volume}{38},
  \bibinfo{number}{4}, Article \bibinfo{articleno}{134} (\bibinfo{date}{July}
  \bibinfo{year}{2019}), \bibinfo{numpages}{15}~pages.
\newblock
\showISSN{0730-0301}
\urldef\tempurl%
\url{https://doi.org/10.1145/3306346.3323042}
\showDOI{\tempurl}


\bibitem[\protect\citeauthoryear{Gatys, Ecker, and Bethge}{Gatys
  et~al\mbox{.}}{2016}]%
        {gatys2016}
\bibfield{author}{\bibinfo{person}{Leon~A Gatys}, \bibinfo{person}{Alexander~S
  Ecker}, {and} \bibinfo{person}{Matthias Bethge}.}
  \bibinfo{year}{2016}\natexlab{}.
\newblock \showarticletitle{Image style transfer using convolutional neural
  networks}. In \bibinfo{booktitle}{\emph{Proceedings of the IEEE Conference on
  Computer Vision and Pattern Recognition (CVPR)}}.
  \bibinfo{pages}{2414--2423}.
\newblock


\bibitem[\protect\citeauthoryear{Gilet, Sauvage, Vanhoey, Dischler, and
  Ghazanfarpour}{Gilet et~al\mbox{.}}{2014}]%
        {Gilet2014}
\bibfield{author}{\bibinfo{person}{Guillaume Gilet}, \bibinfo{person}{Basile
  Sauvage}, \bibinfo{person}{Kenneth Vanhoey}, \bibinfo{person}{Jean-Michel
  Dischler}, {and} \bibinfo{person}{Djamchid Ghazanfarpour}.}
  \bibinfo{year}{2014}\natexlab{}.
\newblock \showarticletitle{Local Random-Phase Noise for Procedural Texturing}.
\newblock \bibinfo{journal}{\emph{ACM Trans. Graph.}} \bibinfo{volume}{33},
  \bibinfo{number}{6}, Article \bibinfo{articleno}{195} (\bibinfo{date}{Nov.}
  \bibinfo{year}{2014}), \bibinfo{numpages}{11}~pages.
\newblock
\showISSN{0730-0301}
\urldef\tempurl%
\url{https://doi.org/10.1145/2661229.2661249}
\showDOI{\tempurl}


\bibitem[\protect\citeauthoryear{Guarnera, Guarnera, Ghosh, Denk, and
  Glencross}{Guarnera et~al\mbox{.}}{2016}]%
        {Guarnera16}
\bibfield{author}{\bibinfo{person}{Dar'ya Guarnera},
  \bibinfo{person}{Giuseppe~Claudio Guarnera}, \bibinfo{person}{Abhijeet
  Ghosh}, \bibinfo{person}{Cornelia Denk}, {and} \bibinfo{person}{Mashhuda
  Glencross}.} \bibinfo{year}{2016}\natexlab{}.
\newblock \showarticletitle{{BRDF Representation and Acquisition}}.
\newblock \bibinfo{journal}{\emph{Computer Graphics Forum}}
  \bibinfo{volume}{35}, \bibinfo{number}{2} (\bibinfo{year}{2016}),
  \bibinfo{pages}{625--650}.
\newblock


\bibitem[\protect\citeauthoryear{Guehl, Allègre, Dischler, Benes, and
  Galin}{Guehl et~al\mbox{.}}{2020}]%
        {Guehl20}
\bibfield{author}{\bibinfo{person}{Pascal Guehl}, \bibinfo{person}{Remi
  Allègre}, \bibinfo{person}{Jean-Michel Dischler}, \bibinfo{person}{Bedrich
  Benes}, {and} \bibinfo{person}{Eric Galin}.} \bibinfo{year}{2020}\natexlab{}.
\newblock \showarticletitle{{Semi-Procedural Textures Using Point Process
  Texture Basis Functions}}.
\newblock \bibinfo{journal}{\emph{Computer Graphics Forum}}
  \bibinfo{volume}{39}, \bibinfo{number}{4} (\bibinfo{year}{2020}),
  \bibinfo{pages}{159--171}.
\newblock
\showISSN{1467-8659}
\urldef\tempurl%
\url{https://doi.org/10.1111/cgf.14061}
\showDOI{\tempurl}


\bibitem[\protect\citeauthoryear{Guingo, Sauvage, Dischler, and Cani}{Guingo
  et~al\mbox{.}}{2017}]%
        {Guingo2017}
\bibfield{author}{\bibinfo{person}{Geoffrey Guingo}, \bibinfo{person}{Basile
  Sauvage}, \bibinfo{person}{Jean-Michel Dischler}, {and}
  \bibinfo{person}{Marie-Paule Cani}.} \bibinfo{year}{2017}\natexlab{}.
\newblock \showarticletitle{{Bi-Layer Textures: a Model for Synthesis and
  Deformation of Composite Textures}}.
\newblock \bibinfo{journal}{\emph{Computer Graphics Forum}}
  (\bibinfo{year}{2017}), \bibinfo{pages}{111--122}.
\newblock
\showISSN{1467-8659}
\urldef\tempurl%
\url{https://doi.org/10.1111/cgf.13229}
\showDOI{\tempurl}


\bibitem[\protect\citeauthoryear{Guo, Ha\v{s}an, Yan, and Zhao}{Guo
  et~al\mbox{.}}{2020a}]%
        {GuoBayesian20}
\bibfield{author}{\bibinfo{person}{Yu Guo}, \bibinfo{person}{Milo\v{s}
  Ha\v{s}an}, \bibinfo{person}{Lingqi Yan}, {and} \bibinfo{person}{Shuang
  Zhao}.} \bibinfo{year}{2020}\natexlab{a}.
\newblock \showarticletitle{A Bayesian Inference Framework for Procedural
  Material Parameter Estimation}.
\newblock \bibinfo{journal}{\emph{Computer Graphics Forum}}
  \bibinfo{volume}{39}, \bibinfo{number}{7} (\bibinfo{year}{2020}),
  \bibinfo{pages}{255 -- 266}.
\newblock


\bibitem[\protect\citeauthoryear{Guo, Smith, Ha\v{s}an, Sunkavalli, and
  Zhao}{Guo et~al\mbox{.}}{2020b}]%
        {Guo20}
\bibfield{author}{\bibinfo{person}{Yu Guo}, \bibinfo{person}{Cameron Smith},
  \bibinfo{person}{Milo\v{s} Ha\v{s}an}, \bibinfo{person}{Kalyan Sunkavalli},
  {and} \bibinfo{person}{Shuang Zhao}.} \bibinfo{year}{2020}\natexlab{b}.
\newblock \showarticletitle{MaterialGAN: Reflectance Capture Using a Generative
  SVBRDF Model}.
\newblock \bibinfo{journal}{\emph{ACM Trans. Graph.}} \bibinfo{volume}{39},
  \bibinfo{number}{6}, Article \bibinfo{articleno}{254} (\bibinfo{date}{Nov.}
  \bibinfo{year}{2020}), \bibinfo{numpages}{13}~pages.
\newblock
\showISSN{0730-0301}
\urldef\tempurl%
\url{https://doi.org/10.1145/3414685.3417779}
\showDOI{\tempurl}


\bibitem[\protect\citeauthoryear{Heeger and Bergen}{Heeger and Bergen}{1995}]%
        {Heeger1995}
\bibfield{author}{\bibinfo{person}{David~J Heeger} {and}
  \bibinfo{person}{James~R Bergen}.} \bibinfo{year}{1995}\natexlab{}.
\newblock \showarticletitle{Pyramid-based texture analysis/synthesis}. In
  \bibinfo{booktitle}{\emph{Proceedings of the 22nd Annual Conference on
  Computer Graphics and Interactive Techniques (SIGGRAPH)}}.
  \bibinfo{pages}{229--238}.
\newblock


\bibitem[\protect\citeauthoryear{Heitz and Neyret}{Heitz and Neyret}{2018}]%
        {Heitz2018}
\bibfield{author}{\bibinfo{person}{Eric Heitz} {and} \bibinfo{person}{Fabrice
  Neyret}.} \bibinfo{year}{2018}\natexlab{}.
\newblock \showarticletitle{High-Performance By-Example Noise Using a
  Histogram-Preserving Blending Operator}.
\newblock \bibinfo{journal}{\emph{ACM Trans. Graph.}} \bibinfo{volume}{1},
  \bibinfo{number}{2}, Article \bibinfo{articleno}{31} (\bibinfo{date}{Aug.}
  \bibinfo{year}{2018}), \bibinfo{numpages}{25}~pages.
\newblock
\urldef\tempurl%
\url{https://doi.org/10.1145/3233304}
\showDOI{\tempurl}


\bibitem[\protect\citeauthoryear{Henzler, Deschaintre, Mitra, and
  Ritschel}{Henzler et~al\mbox{.}}{2021}]%
        {henzler21neuralmaterial}
\bibfield{author}{\bibinfo{person}{Philipp Henzler}, \bibinfo{person}{Valentin
  Deschaintre}, \bibinfo{person}{Niloy~J Mitra}, {and} \bibinfo{person}{Tobias
  Ritschel}.} \bibinfo{year}{2021}\natexlab{}.
\newblock \showarticletitle{Generative Modelling of BRDF Textures from Flash
  Images}.
\newblock \bibinfo{journal}{\emph{ACM Trans Graph (Proc. SIGGRAPH Asia)}}
  \bibinfo{volume}{40}, \bibinfo{number}{6} (\bibinfo{year}{2021}).
\newblock


\bibitem[\protect\citeauthoryear{Henzler, Mitra, and Ritschel}{Henzler
  et~al\mbox{.}}{2020}]%
        {Henzler20}
\bibfield{author}{\bibinfo{person}{Philipp Henzler}, \bibinfo{person}{Niloy~J
  Mitra}, {and} \bibinfo{person}{Tobias Ritschel}.}
  \bibinfo{year}{2020}\natexlab{}.
\newblock \showarticletitle{Learning a Neural 3D Texture Space from 2D
  Exemplars}. In \bibinfo{booktitle}{\emph{Proceedings of the IEEE/CVF
  Conference on Computer Vision and Pattern Recognition (CVPR)}}.
  \bibinfo{pages}{8356 -- 8364}.
\newblock


\bibitem[\protect\citeauthoryear{Hu, Dorsey, and Rushmeier}{Hu
  et~al\mbox{.}}{2019}]%
        {hu2019}
\bibfield{author}{\bibinfo{person}{Yiwei Hu}, \bibinfo{person}{Julie Dorsey},
  {and} \bibinfo{person}{Holly Rushmeier}.} \bibinfo{year}{2019}\natexlab{}.
\newblock \showarticletitle{A Novel Framework for Inverse Procedural Texture
  Modeling}.
\newblock \bibinfo{journal}{\emph{ACM Trans. Graph.}} \bibinfo{volume}{38},
  \bibinfo{number}{6}, Article \bibinfo{articleno}{186} (\bibinfo{date}{Nov.}
  \bibinfo{year}{2019}), \bibinfo{numpages}{14}~pages.
\newblock
\showISSN{0730-0301}
\urldef\tempurl%
\url{https://doi.org/10.1145/3355089.3356516}
\showDOI{\tempurl}


\bibitem[\protect\citeauthoryear{Kaspar, Neubert, Lischinski, Pauly, and
  Kopf}{Kaspar et~al\mbox{.}}{2015}]%
        {SelfTuning}
\bibfield{author}{\bibinfo{person}{Alexandre Kaspar}, \bibinfo{person}{Boris
  Neubert}, \bibinfo{person}{Dani Lischinski}, \bibinfo{person}{Mark Pauly},
  {and} \bibinfo{person}{Johannes Kopf}.} \bibinfo{year}{2015}\natexlab{}.
\newblock \showarticletitle{Self Tuning Texture Optimization}.
\newblock \bibinfo{journal}{\emph{Comput. Graph. Forum}} \bibinfo{volume}{34},
  \bibinfo{number}{2} (\bibinfo{date}{May} \bibinfo{year}{2015}),
  \bibinfo{pages}{349–359}.
\newblock
\showISSN{0167-7055}
\urldef\tempurl%
\url{https://doi.org/10.1111/cgf.12565}
\showDOI{\tempurl}


\bibitem[\protect\citeauthoryear{Khan}{Khan}{2014}]%
        {khan2014survey}
\bibfield{author}{\bibinfo{person}{Muhammad~Waseem Khan}.}
  \bibinfo{year}{2014}\natexlab{}.
\newblock \showarticletitle{A survey: image segmentation techniques}.
\newblock \bibinfo{journal}{\emph{International Journal of Future Computer and
  Communication}} \bibinfo{volume}{3}, \bibinfo{number}{2}
  (\bibinfo{year}{2014}), \bibinfo{pages}{89}.
\newblock


\bibitem[\protect\citeauthoryear{Lawrence, Ben-Artzi, DeCoro, Matusik, Pfister,
  Ramamoorthi, and Rusinkiewicz}{Lawrence et~al\mbox{.}}{2006}]%
        {Lawrence06}
\bibfield{author}{\bibinfo{person}{Jason Lawrence}, \bibinfo{person}{Aner
  Ben-Artzi}, \bibinfo{person}{Christopher DeCoro}, \bibinfo{person}{Wojciech
  Matusik}, \bibinfo{person}{Hanspeter Pfister}, \bibinfo{person}{Ravi
  Ramamoorthi}, {and} \bibinfo{person}{Szymon Rusinkiewicz}.}
  \bibinfo{year}{2006}\natexlab{}.
\newblock \showarticletitle{Inverse Shade Trees for Non-Parametric Material
  Representation and Editing}.
\newblock \bibinfo{journal}{\emph{ACM Transactions on Graphics (Proc.
  SIGGRAPH)}} \bibinfo{volume}{25}, \bibinfo{number}{3} (\bibinfo{date}{July}
  \bibinfo{year}{2006}).
\newblock


\bibitem[\protect\citeauthoryear{Lepage and Lawrence}{Lepage and
  Lawrence}{2011}]%
        {Lepage11}
\bibfield{author}{\bibinfo{person}{Daniel Lepage} {and} \bibinfo{person}{Jason
  Lawrence}.} \bibinfo{year}{2011}\natexlab{}.
\newblock \showarticletitle{Material Matting}.
\newblock \bibinfo{journal}{\emph{ACM Trans. Graph.}} \bibinfo{volume}{30},
  \bibinfo{number}{6} (\bibinfo{date}{Dec.} \bibinfo{year}{2011}),
  \bibinfo{pages}{1–10}.
\newblock
\showISSN{0730-0301}
\urldef\tempurl%
\url{https://doi.org/10.1145/2070781.2024178}
\showDOI{\tempurl}


\bibitem[\protect\citeauthoryear{Levin, Rav-Acha, and Lischinski}{Levin
  et~al\mbox{.}}{2008}]%
        {levin2008}
\bibfield{author}{\bibinfo{person}{Anat Levin}, \bibinfo{person}{Alex
  Rav-Acha}, {and} \bibinfo{person}{Dani Lischinski}.}
  \bibinfo{year}{2008}\natexlab{}.
\newblock \showarticletitle{Spectral matting}.
\newblock \bibinfo{journal}{\emph{IEEE Transactions on Pattern Analysis and
  Machine Intelligence}} \bibinfo{volume}{30}, \bibinfo{number}{10}
  (\bibinfo{year}{2008}), \bibinfo{pages}{1699--1712}.
\newblock


\bibitem[\protect\citeauthoryear{Li, Dong, Peers, and Tong}{Li
  et~al\mbox{.}}{2017}]%
        {li17}
\bibfield{author}{\bibinfo{person}{Xiao Li}, \bibinfo{person}{Yue Dong},
  \bibinfo{person}{Pieter Peers}, {and} \bibinfo{person}{Xin Tong}.}
  \bibinfo{year}{2017}\natexlab{}.
\newblock \showarticletitle{Modeling Surface Appearance from a Single
  Photograph using Self-augmented Convolutional Neural Networks}.
\newblock \bibinfo{journal}{\emph{ACM Trans. Graph.}} \bibinfo{volume}{36},
  \bibinfo{number}{4}, Article \bibinfo{articleno}{45} (\bibinfo{year}{2017}),
  \bibinfo{numpages}{11}~pages.
\newblock


\bibitem[\protect\citeauthoryear{Li, Sunkavalli, and Chandraker}{Li
  et~al\mbox{.}}{2018a}]%
        {li_kal18}
\bibfield{author}{\bibinfo{person}{Zhengqin Li}, \bibinfo{person}{Kalyan
  Sunkavalli}, {and} \bibinfo{person}{Manmohan Chandraker}.}
  \bibinfo{year}{2018}\natexlab{a}.
\newblock \showarticletitle{Materials for Masses: {SVBRDF} Acquisition with a
  Single Mobile Phone Image}.
\newblock \bibinfo{journal}{\emph{Proceedings of the European Conference on
  Computer Vision (ECCV)}} (\bibinfo{year}{2018}), \bibinfo{pages}{72--87}.
\newblock


\bibitem[\protect\citeauthoryear{Li, Xu, Ramamoorthi, Sunkavalli, and
  Chandraker}{Li et~al\mbox{.}}{2018b}]%
        {Li18b}
\bibfield{author}{\bibinfo{person}{Zhengqin Li}, \bibinfo{person}{Zexiang Xu},
  \bibinfo{person}{Ravi Ramamoorthi}, \bibinfo{person}{Kalyan Sunkavalli},
  {and} \bibinfo{person}{Manmohan Chandraker}.}
  \bibinfo{year}{2018}\natexlab{b}.
\newblock \showarticletitle{Learning to Reconstruct Shape and Spatially-varying
  Reflectance from a Single Image}.
\newblock \bibinfo{journal}{\emph{ACM Trans. Graph.}} \bibinfo{volume}{37},
  \bibinfo{number}{6}, Article \bibinfo{articleno}{269} (\bibinfo{date}{Dec.}
  \bibinfo{year}{2018}), \bibinfo{numpages}{11}~pages.
\newblock


\bibitem[\protect\citeauthoryear{Minaee, Boykov, Porikli, Plaza, Kehtarnavaz,
  and Terzopoulos}{Minaee et~al\mbox{.}}{2020}]%
        {minaee20}
\bibfield{author}{\bibinfo{person}{Shervin Minaee}, \bibinfo{person}{Yuri
  Boykov}, \bibinfo{person}{Fatih Porikli}, \bibinfo{person}{Antonio Plaza},
  \bibinfo{person}{Nasser Kehtarnavaz}, {and} \bibinfo{person}{Demetri
  Terzopoulos}.} \bibinfo{year}{2020}\natexlab{}.
\newblock \bibinfo{title}{Image Segmentation Using Deep Learning: A Survey}.
\newblock
\newblock
\showeprint[arxiv]{2001.05566}~[cs.CV]


\bibitem[\protect\citeauthoryear{Niklasson, Mordvintsev, Randazzo, and
  Levin}{Niklasson et~al\mbox{.}}{2021}]%
        {Niklasson21}
\bibfield{author}{\bibinfo{person}{Eyvind Niklasson},
  \bibinfo{person}{Alexander Mordvintsev}, \bibinfo{person}{Ettore Randazzo},
  {and} \bibinfo{person}{Michael Levin}.} \bibinfo{year}{2021}\natexlab{}.
\newblock \showarticletitle{Self-Organising Textures}.
\newblock \bibinfo{journal}{\emph{Distill}} (\bibinfo{year}{2021}).
\newblock
\urldef\tempurl%
\url{https://doi.org/10.23915/distill.00027.003}
\showDOI{\tempurl}
\newblock
\shownote{https://distill.pub/selforg/2021/textures.}


\bibitem[\protect\citeauthoryear{Pellacini and Lawrence}{Pellacini and
  Lawrence}{2007}]%
        {Pellacini07}
\bibfield{author}{\bibinfo{person}{Fabio Pellacini} {and}
  \bibinfo{person}{Jason Lawrence}.} \bibinfo{year}{2007}\natexlab{}.
\newblock \showarticletitle{AppWand: Editing Measured Materials Using
  Appearance-Driven Optimization}.
\newblock \bibinfo{journal}{\emph{ACM Trans. Graph.}} \bibinfo{volume}{26},
  \bibinfo{number}{3} (\bibinfo{date}{July} \bibinfo{year}{2007}),
  \bibinfo{pages}{54–es}.
\newblock
\showISSN{0730-0301}
\urldef\tempurl%
\url{https://doi.org/10.1145/1276377.1276444}
\showDOI{\tempurl}


\bibitem[\protect\citeauthoryear{P\'{e}rez, Gangnet, and Blake}{P\'{e}rez
  et~al\mbox{.}}{2003}]%
        {Perez2003}
\bibfield{author}{\bibinfo{person}{Patrick P\'{e}rez}, \bibinfo{person}{Michel
  Gangnet}, {and} \bibinfo{person}{Andrew Blake}.}
  \bibinfo{year}{2003}\natexlab{}.
\newblock \showarticletitle{Poisson Image Editing}.
\newblock \bibinfo{journal}{\emph{ACM Trans. Graph.}} \bibinfo{volume}{22},
  \bibinfo{number}{3} (\bibinfo{date}{July} \bibinfo{year}{2003}),
  \bibinfo{pages}{313–318}.
\newblock
\showISSN{0730-0301}
\urldef\tempurl%
\url{https://doi.org/10.1145/882262.882269}
\showDOI{\tempurl}


\bibitem[\protect\citeauthoryear{Raad, Davy, Desolneux, and Morel}{Raad
  et~al\mbox{.}}{2018}]%
        {raad18}
\bibfield{author}{\bibinfo{person}{Lara Raad}, \bibinfo{person}{Axel Davy},
  \bibinfo{person}{Agn{\`e}s Desolneux}, {and} \bibinfo{person}{Jean-Michel
  Morel}.} \bibinfo{year}{2018}\natexlab{}.
\newblock \showarticletitle{A survey of exemplar-based texture synthesis}.
\newblock \bibinfo{journal}{\emph{Annals of Mathematical Sciences and
  Applications}} \bibinfo{volume}{3}, \bibinfo{number}{1}
  (\bibinfo{year}{2018}), \bibinfo{pages}{89--148}.
\newblock


\bibitem[\protect\citeauthoryear{Rosenberger, Cohen-Or, and
  Lischinski}{Rosenberger et~al\mbox{.}}{2009}]%
        {Rosenberger2009}
\bibfield{author}{\bibinfo{person}{Amir Rosenberger}, \bibinfo{person}{Daniel
  Cohen-Or}, {and} \bibinfo{person}{Dani Lischinski}.}
  \bibinfo{year}{2009}\natexlab{}.
\newblock \showarticletitle{Layered Shape Synthesis: Automatic Generation of
  Control Maps for Non-Stationary Textures}.
\newblock \bibinfo{journal}{\emph{ACM Trans. Graph.}} \bibinfo{volume}{28},
  \bibinfo{number}{5}, Article \bibinfo{articleno}{107} (\bibinfo{date}{Dec.}
  \bibinfo{year}{2009}), \bibinfo{numpages}{9}~pages.
\newblock
\showISSN{0730-0301}
\urldef\tempurl%
\url{https://doi.org/10.1145/1618452.1618453}
\showDOI{\tempurl}


\bibitem[\protect\citeauthoryear{Shi, Li, Ha{\v s}an, Sunkavalli, Boubekeur,
  Mech, and Matusik}{Shi et~al\mbox{.}}{2020}]%
        {Shi20}
\bibfield{author}{\bibinfo{person}{Liang Shi}, \bibinfo{person}{Beichen Li},
  \bibinfo{person}{Milo{\v s} Ha{\v s}an}, \bibinfo{person}{Kalyan Sunkavalli},
  \bibinfo{person}{Tamy Boubekeur}, \bibinfo{person}{Radomir Mech}, {and}
  \bibinfo{person}{Wojciech Matusik}.} \bibinfo{year}{2020}\natexlab{}.
\newblock \showarticletitle{MATch: Differentiable Material Graphs for
  Procedural Material Capture}.
\newblock \bibinfo{journal}{\emph{ACM Trans. Graph.}} \bibinfo{volume}{39},
  \bibinfo{number}{6}, Article \bibinfo{articleno}{196} (\bibinfo{date}{Dec.}
  \bibinfo{year}{2020}), \bibinfo{numpages}{15}~pages.
\newblock


\bibitem[\protect\citeauthoryear{Shocher, Bagon, Isola, and Irani}{Shocher
  et~al\mbox{.}}{2019}]%
        {InGAN}
\bibfield{author}{\bibinfo{person}{Assaf Shocher}, \bibinfo{person}{Shai
  Bagon}, \bibinfo{person}{Phillip Isola}, {and} \bibinfo{person}{Michal
  Irani}.} \bibinfo{year}{2019}\natexlab{}.
\newblock \showarticletitle{InGAN: Capturing and Retargeting the "DNA" of a
  Natural Image}. In \bibinfo{booktitle}{\emph{The IEEE International
  Conference on Computer Vision (ICCV)}}.
\newblock


\bibitem[\protect\citeauthoryear{Simonyan and Zisserman}{Simonyan and
  Zisserman}{2015}]%
        {vgg19}
\bibfield{author}{\bibinfo{person}{Karen Simonyan} {and}
  \bibinfo{person}{Andrew Zisserman}.} \bibinfo{year}{2015}\natexlab{}.
\newblock \showarticletitle{Very Deep Convolutional Networks for Large-Scale
  Image Recognition}. In \bibinfo{booktitle}{\emph{3rd International Conference
  on Learning Representations, {ICLR} 2015, San Diego, CA, USA, May 7-9, 2015,
  Conference Track Proceedings}}, \bibfield{editor}{\bibinfo{person}{Yoshua
  Bengio} {and} \bibinfo{person}{Yann LeCun}} (Eds.).
\newblock
\urldef\tempurl%
\url{http://arxiv.org/abs/1409.1556}
\showURL{%
\tempurl}


\bibitem[\protect\citeauthoryear{{\v{S}}t'ava, Bene{\v{s}}, M{\v{e}}ch, Aliaga,
  and Kri{\v{s}}tof}{{\v{S}}t'ava et~al\mbox{.}}{2010}]%
        {vst2010}
\bibfield{author}{\bibinfo{person}{Ondrej {\v{S}}t'ava},
  \bibinfo{person}{Bedrich Bene{\v{s}}}, \bibinfo{person}{Radomir M{\v{e}}ch},
  \bibinfo{person}{Daniel~G Aliaga}, {and} \bibinfo{person}{Peter
  Kri{\v{s}}tof}.} \bibinfo{year}{2010}\natexlab{}.
\newblock \showarticletitle{Inverse procedural modeling by automatic generation
  of l-systems}. In \bibinfo{booktitle}{\emph{Computer Graphics Forum}},
  Vol.~\bibinfo{volume}{29}. Wiley Online Library, \bibinfo{pages}{665--674}.
\newblock


\bibitem[\protect\citeauthoryear{Telea}{Telea}{2004}]%
        {telea2004}
\bibfield{author}{\bibinfo{person}{Alexandru Telea}.}
  \bibinfo{year}{2004}\natexlab{}.
\newblock \showarticletitle{An image inpainting technique based on the fast
  marching method}.
\newblock \bibinfo{journal}{\emph{Journal of Graphics Tools}}
  \bibinfo{volume}{9}, \bibinfo{number}{1} (\bibinfo{year}{2004}),
  \bibinfo{pages}{23--34}.
\newblock


\bibitem[\protect\citeauthoryear{Ulyanov, Lebedev, Vedaldi, and
  Lempitsky}{Ulyanov et~al\mbox{.}}{2016}]%
        {Ulyanov2016}
\bibfield{author}{\bibinfo{person}{Dmitry Ulyanov}, \bibinfo{person}{Vadim
  Lebedev}, \bibinfo{person}{Andrea Vedaldi}, {and} \bibinfo{person}{Victor
  Lempitsky}.} \bibinfo{year}{2016}\natexlab{}.
\newblock \showarticletitle{Texture Networks: Feed-Forward Synthesis of
  Textures and Stylized Images}. In \bibinfo{booktitle}{\emph{Proceedings of
  the 33rd International Conference on International Conference on Machine
  Learning - Volume 48}} (New York, NY, USA)
  \emph{(\bibinfo{series}{ICML'16})}. \bibinfo{publisher}{JMLR.org},
  \bibinfo{pages}{1349–1357}.
\newblock


\bibitem[\protect\citeauthoryear{Virtanen, Gommers, Oliphant, Haberland, Reddy,
  Cournapeau, Burovski, Peterson, Weckesser, Bright, {van der Walt}, Brett,
  Wilson, Millman, Mayorov, Nelson, Jones, Kern, Larson, Carey, Polat, Feng,
  Moore, {VanderPlas}, Laxalde, Perktold, Cimrman, Henriksen, Quintero, Harris,
  Archibald, Ribeiro, Pedregosa, {van Mulbregt}, and {SciPy 1.0
  Contributors}}{Virtanen et~al\mbox{.}}{2020}]%
        {scipy2020}
\bibfield{author}{\bibinfo{person}{Pauli Virtanen}, \bibinfo{person}{Ralf
  Gommers}, \bibinfo{person}{Travis~E. Oliphant}, \bibinfo{person}{Matt
  Haberland}, \bibinfo{person}{Tyler Reddy}, \bibinfo{person}{David
  Cournapeau}, \bibinfo{person}{Evgeni Burovski}, \bibinfo{person}{Pearu
  Peterson}, \bibinfo{person}{Warren Weckesser}, \bibinfo{person}{Jonathan
  Bright}, \bibinfo{person}{St{\'e}fan~J. {van der Walt}},
  \bibinfo{person}{Matthew Brett}, \bibinfo{person}{Joshua Wilson},
  \bibinfo{person}{K.~Jarrod Millman}, \bibinfo{person}{Nikolay Mayorov},
  \bibinfo{person}{Andrew R.~J. Nelson}, \bibinfo{person}{Eric Jones},
  \bibinfo{person}{Robert Kern}, \bibinfo{person}{Eric Larson},
  \bibinfo{person}{C~J Carey}, \bibinfo{person}{{\.I}lhan Polat},
  \bibinfo{person}{Yu Feng}, \bibinfo{person}{Eric~W. Moore},
  \bibinfo{person}{Jake {VanderPlas}}, \bibinfo{person}{Denis Laxalde},
  \bibinfo{person}{Josef Perktold}, \bibinfo{person}{Robert Cimrman},
  \bibinfo{person}{Ian Henriksen}, \bibinfo{person}{E.~A. Quintero},
  \bibinfo{person}{Charles~R. Harris}, \bibinfo{person}{Anne~M. Archibald},
  \bibinfo{person}{Ant{\^o}nio~H. Ribeiro}, \bibinfo{person}{Fabian Pedregosa},
  \bibinfo{person}{Paul {van Mulbregt}}, {and} \bibinfo{person}{{SciPy 1.0
  Contributors}}.} \bibinfo{year}{2020}\natexlab{}.
\newblock \showarticletitle{{{SciPy} 1.0: Fundamental Algorithms for Scientific
  Computing in Python}}.
\newblock \bibinfo{journal}{\emph{Nature Methods}}  \bibinfo{volume}{17}
  (\bibinfo{year}{2020}), \bibinfo{pages}{261--272}.
\newblock
\urldef\tempurl%
\url{https://doi.org/10.1038/s41592-019-0686-2}
\showDOI{\tempurl}


\bibitem[\protect\citeauthoryear{Welch}{Welch}{1967}]%
        {Welch1967}
\bibfield{author}{\bibinfo{person}{P. Welch}.} \bibinfo{year}{1967}\natexlab{}.
\newblock \showarticletitle{The use of fast Fourier transform for the
  estimation of power spectra: A method based on time averaging over short,
  modified periodograms}.
\newblock \bibinfo{journal}{\emph{IEEE Transactions on Audio and
  Electroacoustics}} \bibinfo{volume}{15}, \bibinfo{number}{2}
  (\bibinfo{year}{1967}), \bibinfo{pages}{70--73}.
\newblock
\urldef\tempurl%
\url{https://doi.org/10.1109/TAU.1967.1161901}
\showDOI{\tempurl}


\bibitem[\protect\citeauthoryear{Wie, Lefebvre, Kwatra, and Turk}{Wie
  et~al\mbox{.}}{2009}]%
        {wie09}
\bibfield{author}{\bibinfo{person}{Li-Yi Wie}, \bibinfo{person}{Sylvain
  Lefebvre}, \bibinfo{person}{Vivek Kwatra}, {and} \bibinfo{person}{Greg
  Turk}.} \bibinfo{year}{2009}\natexlab{}.
\newblock \showarticletitle{{State of the Art in Example-based Texture
  Synthesis}}. In \bibinfo{booktitle}{\emph{Eurographics 2009 - State of the
  Art Reports}}, \bibfield{editor}{\bibinfo{person}{M.~Pauly} {and}
  \bibinfo{person}{G.~Greiner}} (Eds.). \bibinfo{publisher}{The Eurographics
  Association}.
\newblock
\urldef\tempurl%
\url{https://doi.org/10.2312/egst.20091063}
\showDOI{\tempurl}


\bibitem[\protect\citeauthoryear{Xu, Price, Cohen, and Huang}{Xu
  et~al\mbox{.}}{2017}]%
        {xu2017}
\bibfield{author}{\bibinfo{person}{Ning Xu}, \bibinfo{person}{Brian Price},
  \bibinfo{person}{Scott Cohen}, {and} \bibinfo{person}{Thomas Huang}.}
  \bibinfo{year}{2017}\natexlab{}.
\newblock \showarticletitle{Deep image matting}. In
  \bibinfo{booktitle}{\emph{Proceedings of the IEEE Conference on Computer
  Vision and Pattern Recognition (CVPR)}}. \bibinfo{pages}{2970--2979}.
\newblock


\bibitem[\protect\citeauthoryear{Zhou, Zhu, Bai, Lischinski, Cohen-Or, and
  Huang}{Zhou et~al\mbox{.}}{2018}]%
        {TexSyn18}
\bibfield{author}{\bibinfo{person}{Yang Zhou}, \bibinfo{person}{Zhen Zhu},
  \bibinfo{person}{Xiang Bai}, \bibinfo{person}{Dani Lischinski},
  \bibinfo{person}{Daniel Cohen-Or}, {and} \bibinfo{person}{Hui Huang}.}
  \bibinfo{year}{2018}\natexlab{}.
\newblock \showarticletitle{Non-stationary Texture Synthesis by Adversarial
  Expansion}.
\newblock \bibinfo{journal}{\emph{ACM Transactions on Graphics (Proc.
  SIGGRAPH)}} \bibinfo{volume}{37}, \bibinfo{number}{4} (\bibinfo{year}{2018}).
\newblock


\bibitem[\protect\citeauthoryear{{Zhou Wang}, {Bovik}, {Sheikh}, and
  {Simoncelli}}{{Zhou Wang} et~al\mbox{.}}{2004}]%
        {ssim}
\bibfield{author}{\bibinfo{person}{{Zhou Wang}}, \bibinfo{person}{A.~C.
  {Bovik}}, \bibinfo{person}{H.~R. {Sheikh}}, {and} \bibinfo{person}{E.~P.
  {Simoncelli}}.} \bibinfo{year}{2004}\natexlab{}.
\newblock \showarticletitle{Image quality assessment: from error visibility to
  structural similarity}.
\newblock \bibinfo{journal}{\emph{IEEE Transactions on Image Processing}}
  \bibinfo{volume}{13}, \bibinfo{number}{4} (\bibinfo{year}{2004}),
  \bibinfo{pages}{600--612}.
\newblock
\urldef\tempurl%
\url{https://doi.org/10.1109/TIP.2003.819861}
\showDOI{\tempurl}


\end{thebibliography}

\end{document}